\documentclass[12pt]{article}

\usepackage[utf8]{inputenc}
\usepackage[dvipsnames]{xcolor}
\usepackage{bbm}
\pdfoutput=24
\usepackage{amsmath}
\usepackage{float}
\DeclareMathOperator*{\minimize}{minimize}
\usepackage[linesnumbered,ruled]{algorithm2e}

\usepackage{esvect}
\usepackage[toc,page]{appendix}
\usepackage{leftidx}
\usepackage{color}	
\usepackage{framed, color}
\usepackage{multirow}
\usepackage{pdfpages}
\usepackage{multicol}
\usepackage{wrapfig,lipsum,booktabs}
\usepackage{relsize}
\usepackage{mathtools,hyperref}
\usepackage{xparse}
\usepackage{amsthm}

\hypersetup{
    colorlinks=true,
    linkcolor=cyan,
    filecolor=cyan,
    urlcolor=red,
    citecolor=red
}
\DeclarePairedDelimiter{\ceil}{\lceil}{\rceil}
\DeclarePairedDelimiter\floor{\lfloor}{\rfloor}

\usepackage{cleveref}
\usepackage{commath}
\usepackage{enumitem}
\usepackage{moreenum}
\usepackage{amssymb}
\renewcommand{\Bbb}{\mathbb}

\let\emptyset\varnothing

\crefformat{section}{\S#2#1#3} 
\crefformat{subsection}{\S#2#1#3}
\crefformat{subsubsection}{\S#2#1#3}

\definecolor{mgreen}{RGB}{25,147,100}
\definecolor{shadecolor}{rgb}{1,.8,.1}
\definecolor{shadecolor2}{RGB}{245,237,0}
\definecolor{orange}{RGB}{255,137,20}
\definecolor{orange}{RGB}{245,37,100}

\usepackage{pgfplots}
\usetikzlibrary{patterns}
\usepackage{mdframed}
\usepackage{adjustbox}
\usepackage{tcolorbox}
\usepackage{tikz,ifthen,fp,calc}
\usepackage{caption}
\usepackage{subcaption}
\usetikzlibrary{plotmarks}
\usepackage{graphicx}

\theoremstyle{plain}
\newtheorem{theorem}{Theorem}[section]
\newtheorem{prop}{Proposition}[section]
\newtheorem{corr}{Corollary}[section]
\theoremstyle{definition}
\newtheorem{definition}{Definition}[section]
\newtheorem{assumption}{Assumption}[section]

\theoremstyle{definition}
\newtheorem{remark}{Remark}[section]

\theoremstyle{example}
\newtheorem{example}{Example}[section]

\usepackage[english]{babel}
\usepackage{babel,blindtext}

\usepackage{fullpage}
\usepackage{amsfonts}
\usepackage{lscape}
\usepackage{bbm}

\usepackage{todonotes}
\usepackage{cite}
\usepackage{verbatim}
\usepackage{bm}

\usepackage[margin=1in]{geometry}
\providecommand{\keywords}[1]{\textbf{\textit{Keywords:---}} #1}

\usepackage[T1]{fontenc}
\usepackage[utf8]{inputenc}
\usepackage{authblk}
\usepackage{cite}

\title{\textbf{From classical to modern opinion dynamics}}

\author[1]{Hossein Noorazar \thanks{h.noorazar@gmail.com}}
\author[2]{Kevin R. Vixie\thanks{vixie@speakeasy.net}}
\author[3]{Arghavan Talebanpour\thanks{arghavantalebanpour@gmail.com}}
\author[4]{Yunfeng Hu\thanks{yunfeng.hu90@gmail.com}}
\affil[1]{Washington State University}
\affil[2]{Department of Mathematics and Statistics, Washington State University}
\affil[3]{Department of Mechanical and Material Engineering, Washington State University}
\affil[1]{Washington State University}

\date{}

\providecommand{\keywords}[1]{\textbf{\textit{Keywords:}} #1}

\begin{document}

\maketitle

\begin{abstract}
In this age of Facebook, Instagram, and Twitter, 
there is rapidly growing interest in understanding 
network-enabled opinion dynamics in large groups 
of autonomous agents. The phenomena of opinion 
polarization, the spread of propaganda and fake news, 
and the manipulation of sentiment is of interest to large 
numbers of organizations and people. Whether it is the more 
nefarious players such as foreign governments that are 
attempting to sway elections or it is more open and above 
board, such as researchers who want to make large groups 
of people aware of helpful innovations, what is at stake is often significant.

In this paper, we review opinion dynamics including the extensions of many 
classical models as well as some new models that deepen understanding. 
For example, we look at models that track the evolution of an 
individual’s power, that include noise, and that feature sequentially dependent topics, to name a few.

While the first papers studying opinion dynamics 
appeared over 60 years ago, there is still a 
great deal of room for innovation and exploration. 
We believe that the political climate and the extraordinary 
(even unprecedented) events in the sphere of politics 
in the last few years will inspire new interest and new ideas.

It is our aim to help those interested researchers understand what has already 
been explored in a significant portion of the field of opinion dynamics. We believe that in 
doing this, it will become clear that there is still much to be done.

\end{abstract}

\tableofcontents

\section{Introduction}\label{intro}

The problem with studying human relational dynamics mathematically is that as soon as we present such phenomena in a way that is amenable to mathematical analysis, we have stripped away much of the nuance and, more importantly, the complexity that exist in the real world. Of course, this is also a benefit, since it is impossible to represent such systems in their full complexity. Therefore, the art of modeling consists in performing this reduction in a way that leaves something of value intact.

However, carefully crafted models can provide useful insights into how real human systems work.

In our case, we want to understand the dynamics of opinions in groups
of people who interact with each other and a context of information --
what causes people to change opinions and groups that support or
oppose an opinion to gain or lose influence and power? This is, of
course a very old topic of interest. As long as there has been groups
of people whose opinions differed (and mattered), this has been of
interest.

In this paper we review a partial cross section of the mathematical
approaches to answering these questions. 
While these models are all
radically simplified representation of the social and economic systems, 
what has been done is at least provacative and
interesting. The idea that we can mathematically model and study the
evolution of opinions is not a new one - research using a mathematical
perspective dates back at least as far as John R. P. French's 1956
paper \emph{A Formal Theory of Social Power}\cite{FRENCH1956}. More recently, but still
many decades old, is the work of DeGroot in 1974\cite{DeGroot1974}
and Friedkin in 1986\cite{friedkin1986formal}.

\subsection{Mathematical Representation}\label{sec:model-represent}
When modeling opinions and their dynamics we must, 
at a minimum, represent the opinions that
are held as well as the means by which people interact, both influencing 
and being influenced. We also need to choose how we represent time. 
\begin{description}
\item[Opinion Spaces] Opinions can be represented 
by both discrete variables as well as by 
continuous representations. For example, 
in a two-party election, we might 
represent opinions by a discrete 
variable, candidate $A$ or
  candidate $B$ (Fig.~\ref{fig:binarySp}). In contrast, 
  while designing a new product, we might care 
  about the distribution of prices that customers 
  are willing to pay, which we might 
  represent by real numbers in the 
  interval $[0,1]$ (Fig.~\ref{fig:zeroOne}), $0$ representing some minimal amount and $1$ the maximal
  potential price. It is possible that one opinion be presented via an ordered pair (Fig.~\ref{fig:square}). If one wants
  to choose a favorite color based on combination of red, green and blue
  then the opinion space can be the unit cube (Fig.~\ref{fig:cubeOp}).
  \emph{In this paper, we focus our attention on models
  of opinion spaces as intervals in $\Bbb{R}$.}

\begin{figure}[h!]
\begin{subfigure}{.5\textwidth}
  \centering
  \includegraphics[width=.8\linewidth]{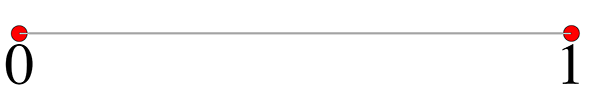}
  \caption{\small{binary opinion space}}
   \label{fig:binarySp}
\end{subfigure}
\begin{subfigure}{.5\textwidth}
  \centering
  \includegraphics[width=.8\linewidth]{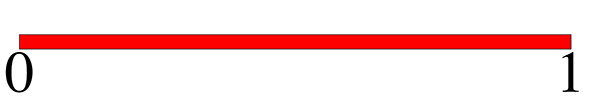}
  \caption{\small{continuous 1D opinion space}}
  \label{fig:zeroOne}
\end{subfigure}
\begin{subfigure}{.5\textwidth}
  \centering
  \includegraphics[width=.7\linewidth]{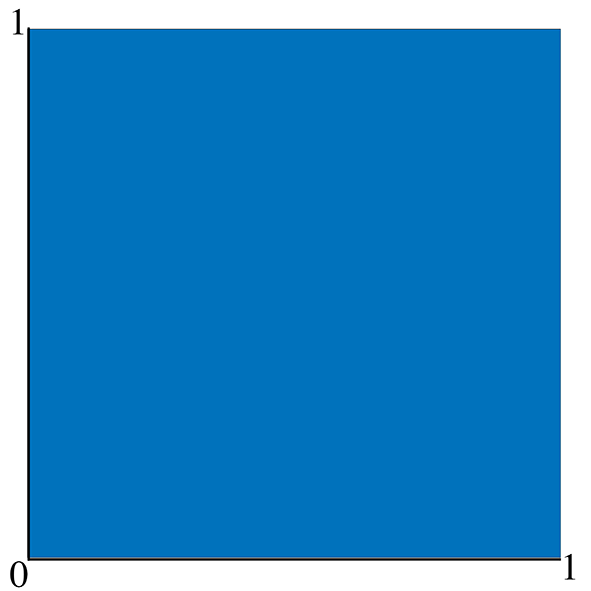}
  \caption{\small{2D opinion space}}
  \label{fig:square}
\end{subfigure}
\begin{subfigure}{.5\textwidth}
  \centering
  \includegraphics[width=.8\linewidth]{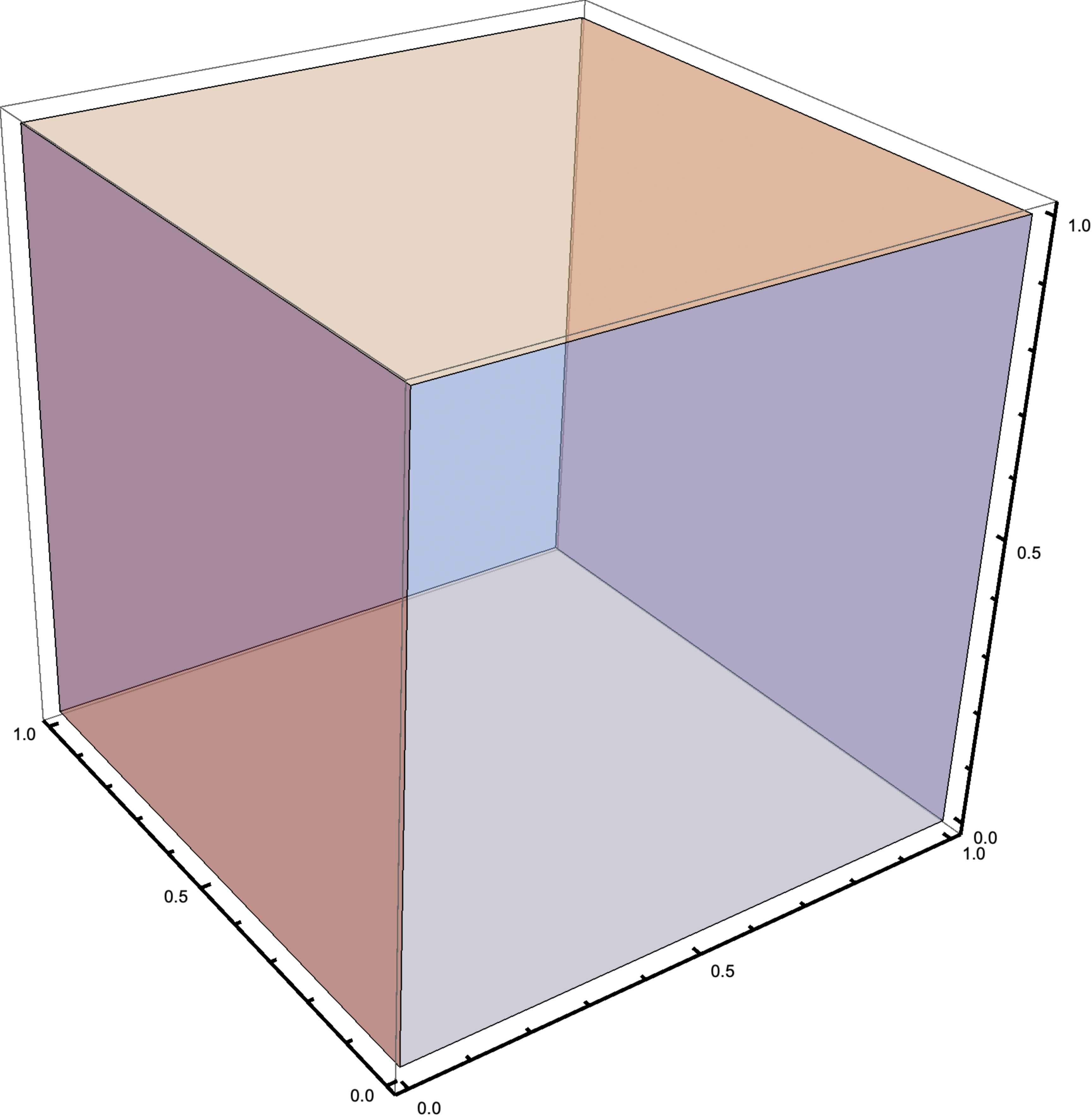}
  \caption{\small{a cubical opinion space}}
  \label{fig:cubeOp}
\end{subfigure}
\caption{opinion space examples. In Fig.~\ref{fig:binarySp} the opinions are binary either 0 or 1,
in Fig.~\ref{fig:zeroOne} opinions are continuous anywhere between 0 and 1, and Fig.~\ref{fig:square}
is a two dimensional opinion space, it could also be a triangle or a simplex.}
\label{fig:opSpacePlt}
\end{figure}

\item[Interactions] A natural starting place for the representation of
  interactions is a network, with a node for each person (we will call
  them agents) and an edge, representing pairwise interactions between
  each pair of agents. If we have $N$ people each with an opinion,
  then there will be $\frac{N*(N-1)}{2}$ pairs of people and possible
  interactions, assuming we focus on pairwise interactions. The result
  is a network with $N$ nodes and $\frac{N*(N-1)}{2}$ possible edges,
  each perhaps with a weight or even two weights for influence if
  there is an asymmetry in persuasiveness. If a pair of agents have
  different influences on each other then the relations are defined by
  directed edges (Fig.~\ref{fig:directedInteractions}) and in this case the complete-graph
  will have $N * (N-1)$ directed edges.
  Of course, the agents can
  also be media entities, in which case it is clear that there would
  typically be an asymmetry in influence where a directed
  graph can be used (Fig.~\ref{fig:directedInteractions}).
\begin{figure}[httb!]
\begin{subfigure}{.5\textwidth}
  \centering
  \includegraphics[width=1\linewidth]{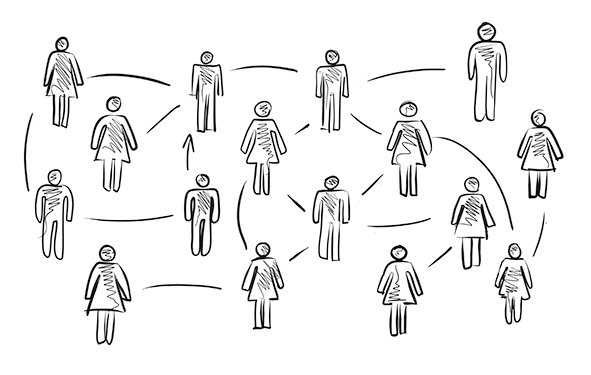}
  \caption{undirected network} 
   \label{fig:interactions}
\end{subfigure}
\begin{subfigure}{.5\textwidth}
  \centering
  \includegraphics[width=1\linewidth]{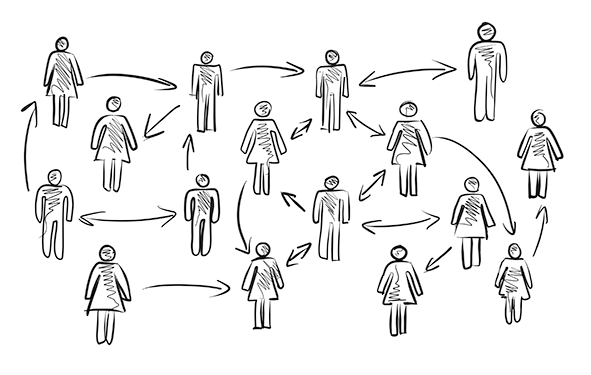}
  \caption{directed network}
  \label{fig:directedInteractions}
\end{subfigure}
\caption{The interaction network can be undirected or directed. In an undirected graph
relationships is assumed to be symmetric and bidirectional. In opinion dynamics an arrow from 
Alice to Bob means Alice puts some weights on Bob's opinion, i.e. she listens to her.}
\label{fig:interactionPlots}
\end{figure}
\item[Time] We can model time as continuous (Fig.~\ref{fig:continuousTime}), but this is usually not
the choice. Therefore, \emph{in each of the models we review in this paper,
time is discrete: $t = 0, 1, 2, \dots$} (Fig.~\ref{fig:discreteTime}).
\begin{figure}[httb!]
\begin{subfigure}{.5\textwidth}
  \centering
  \includegraphics[width=1\linewidth]{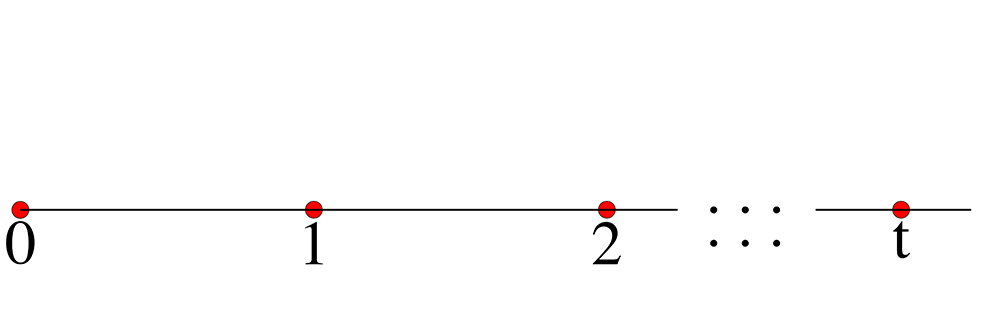}
  \caption{discrete time steps} 
   \label{fig:discreteTime}
\end{subfigure}
\begin{subfigure}{.5\textwidth}
  \centering
  \includegraphics[width=1\linewidth]{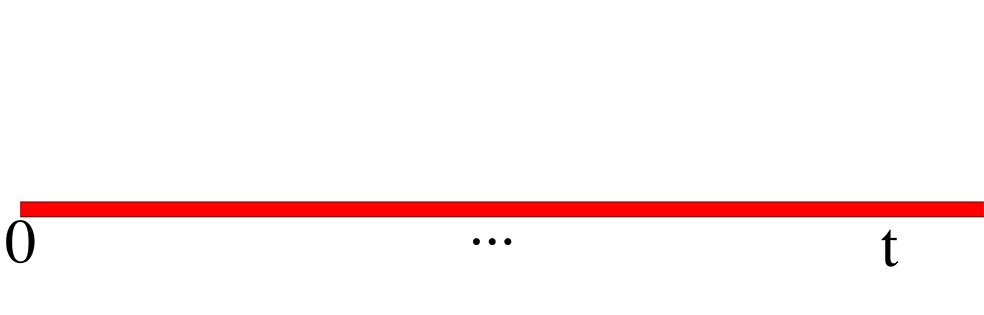}
  \caption{continuous time}
  \label{fig:continuousTime}
\end{subfigure}
\caption{The time in the opinion dynamics can be 
discrete, e.g. the DeGroot model~\cite{DeGroot1974}, or continuous
e.g. the Altafini model~\cite{Altafini2012}.}
\end{figure}
\end{description}
\subsection{Outline of the Paper}\label{sec:outline}

We begin by providing some definitions (Sec.~\ref{DefAndNota})
that are needed to either establish
the models or present of the results. Then we briefly 
present the classical models of opinion dynamics (Sec.~\ref{ClassicalModels}) 
for readers who are new to the field, and to refresh 
the memories of readers who have some knowledge 
of them; these classical models will be the foundations 
for the extensions presented later in the paper.

In Sec.~\ref{degrootianModelSec}, we study the simple DeGroot model and
its variants which permits the use of advanced linear algebra insights
and tools. In this model, the opinions of all participants
are usually updated simultaneously. In Sec.~\ref{BCMSection},
we consider the bounded confidence models. Then, in Sec.~\ref{FurtherWorkQuestions}, 
models that include repulsive forces are discussed. Other
works that we did not cover in detail will also be touched upon in this section and we will make suggestion for future research directions.

The paper is organized based on the DeGroot model and its extensions and the bounded confidence models and their extensions. 
Moreover, not all concepts are
applied to all the models. For example, a repulsive behavior has not been applied
to the HK model. Therefore, Sec.~\ref{degrootianModelSec} includes a subsection
devoted to repulsive forces but the 
Sec.~\ref{BCMSection} does not.

Hence, the organization of the paper cannot be solely based on
models or based on concepts. The final section 
includes novel models and some of influential 
works that did not fit 
into the framework of well-known models.

\subsection{Note on terminology}\label{sec:term}

Before diving into the definitions we would like to mention that there
is no convergence on terminology in the literature.  Many of the model
names are text dependent, i.e. different models may be known by the
same name in different articles. For example, most authors by
``bounded confidence model'' refer to the DW
model~\cite{Deffuant2000}, however, any model with a confidence radius
beyond which agents ignore each other can be referred to as a bounded
confidence model.  The Hegselmann-Krause version of the bounded confidence
model is given by:
\begin{equation}\label{thisEq}
o_i^{(t+1)} = \frac{1}{|N_i^{(t)}|} \sum_{j \in N_i^{(t)} }o_j^{(t)} 
\end{equation}

\noindent where
$N_i^{(t)} = \{ j | |o_i^{(t)} - o_j^{(t)}| \le r_i \}$ is the set of
agents whose opinions fall in the confidence interval of agent $i$ at
time $t$, i.e. hold close-enough opinions.  A general framework is
established in Hegselmann-Krause's work and then different directions are
studied, all of which are referred to as HK model in different works.
Proskurnikov\cite{Proskurnikov2018} refers to the model given by
Eq.~(\ref{thisEq}) as the HK model, and not bounded confidence
model. It is noteworthy that in the DW model the update rule is a
function of the difference of opinions of the pair, however, in
Eq.~\ref{thisEq} the update is independent of differences.

In the equation above, agent $i$ weighs its own opinion and that of
other nodes equally. Fu et al.~\cite{Fu2015} modifies it so that agent $i$
can weigh its own opinion freely. He uses the term ``stubborn'' for
agents who never change their opinions; however, in other works
stubbornness refers to agents who incorporate their ``initial''
opinion~\cite{Tian2019}, partially, to their current opinion. Hence, a more accurate terminology could be the use of
\emph{fully} and \emph{partially} stubborn. A fully stubborn agent
can be considered as a node that spreads a fixed opinion,
regardless of that of others - much like a source of propaganda
intended to forcibly shift the opinion of a population.
\section{Basic definitions, notation and classical models}\label{sec:definitions}
In this section we define notations and concepts that will be 
used globally in this paper. Local variables that are 
applicable only in specific situations will be 
defined within the appropriate sections.

The purpose of this section is to avoid misunderstanding 
and introduce ideas and terminology from a variety of
fields. We need the definitions to present the results. 
For example, Thm.~\ref{thm4graphdef} 
uses graph properties and Prop.~\ref{DeGrootConvergence80} 
uses properties of the weight matrix to state relevant results. This section also clarifies
confusing terminologies for first-time readers, e.g., consensus versus convergence.
 Moreover, there are terms about which the community of researchers 
 is not yet in total agreement,
 for example \emph{leaders} vs \emph{stubborn-agents}. We will present all models
 with a unified terminology (see~\ref{DefAndNota}).

\subsection{Definitions and Notations}\label{DefAndNota}
\begin{definition}{}
A graph $\mathcal{G}$ is an object consisting of two sets, 
$V$ and $E$, denoted by an ordered pair $\mathcal{G} = (V, E)$, 
where $V$ is a finite nonempty set where each of its
 members is called a \emph{vertex} or \emph{node}, 
 and $E$ is a subset of $V \times V$ where each of its 
 elements is called an \emph{edge}. An edge connecting 
 a given vertex to itself is called a \emph{loop}.
 In a social network a vertex is an agent, 
 such as a person. 
An edge represents the relationship between
 two agents.
\end{definition}
\begin{definition}{}
A path is a sequence of vertices
from one node to the next using the edges.
\end{definition}

For example the sequence $v_0, v_1, v_2, \dots, v_{k-1}, v_k$ is a path from $v_0$ to $v_k$,
where $v_i$ and $v_{i+1}$, for $i \in \{0, 1, 2, \dots, k-1\}$, are connected by an edge.
\begin{definition}{}
A \emph{connected} graph is a graph in which there is a path between all pairs of nodes.
\end{definition}

In this work we use the terms graph and network 
interchangeably; vertices represent agents; and 
edges represent connections between individual agents.
\begin{definition}{}
The number of edges connected to a vertex is 
called the \emph{degree} of that vertex.
\end{definition}
\begin{definition}{}
If the edges in the set $E$ given in the definition 
above are unordered, i.e. unoriented,  
such that $e = (x,y) = (y,x) = \{x,y \}$, then 
the graph is an \emph{undirected graph}; 
otherwise, it is a \emph{directed} graph or a \emph{digraph}.
 In a directed graph, relationships can be unilateral,
such as the relationship between a judge and the person being sentenced,
or between a teacher and a student who will receive a grade.
\end{definition}
\begin{definition}{}
A \emph{weighted} graph has a weight assigned to each edge.
\end{definition}

The weights associated with edges in a graph 
can represent various factors such as geographical 
distance, the probability of interaction between 
the two agents related by a given 
edge, or the influence an agent has over another 
agent. For example, the weights 
assigned to edges are similar to 
powers assigned to relationships; for example,
the power of a teacher over a student 
tends to be much greater than the power the 
student has over the teacher.
The assignment and values of weights are 
dependent on the goals of the model.
\begin{definition}{}
Two vertices $v_i$ and $v_j$ are said to be
 \emph{adjacent} if there is an edge connecting them. 
 This definition gives rise to the \emph{adjacency matrix} 
$\mathbf{A} = \mathbf{A}(\mathcal{G}) = (a_{ij}) = (\mathbf{A}_{ij})$. In an unweighted graph 
\begin{equation*}
\mathbf{A}_{ij}  =
\begin{cases} 1 & \text{if $(v_i, v_j) \in E$,} \\
0 &\text{o.w.} \\
\end{cases}
\end{equation*}
In a directed graph we might 
have $a_{ij} \neq a_{ji}$. 
Some graphs are weighted such that 
the weight assigned to each edge 
represents the influence of each agent on the other. 
(The weights also could represent the probability or the 
frequency of interactions, or other variables.)
In this case the adjacency matrix 
can also be referred to by \emph{weight matrix} or 
\emph{influence matrix}. In this paper we
use the terms adjacency matrix or 
weight matrix or influence matrix interchangeably, and
while each entry of the matrix will 
be denoted by $w_{ij}$ the matrix itself
still will be denoted by $\mathbf{A}$.
\begin{equation*}
\mathbf{A}_{ij}  = 
\begin{cases} w_{ij} & \text{if $(v_i, v_j) \in E$,} \\
0 &\text{o.w.} \\
\end{cases}
\end{equation*}
\end{definition}
\noindent For a weighted graph, the indicator of the 
influence matrix $\mathbf{A}$ is a matrix whose 
entries are 1 whenever $\mathbf{A}_{ij} > 0$ 
and zero whenever $\mathbf{A}_{ij} = 0$. 
We use $\mathbf{A}$ for both whenever 
it does not cause confusion.
\begin{remark}
Similar to $\mathbf{A}(\mathcal{G})$, which is the 
adjacency matrix induced by graph $\mathcal{G}$,
$\mathcal{G}(\mathbf{A})$ is a graph induced by $\mathbf{A}$.
\end{remark}
\begin{definition}{}
A \emph{full graph} or a \emph{complete graph} 
is a graph in which all nodes are adjacent to each other.
\end{definition}
\begin{definition}{}
The set of all possible (numerical) 
opinions, denoted by $\mathcal{O}$, 
is called the \textit{opinion space}. 
Examples of opinion spaces are $\{0, 1\}$ 
for binary opinions, $[0, 1]$, $[-1, 1]$, simplices in $\mathbb{R}$, etc.
\end{definition}
\begin{definition}{}
Define the indicator function by 
\begin{equation*}
\mathbbm{1_\mathcal{A}}(x) = 
\begin{cases} 1 & \text{if $x \in \mathcal{A}$,} \\
0 & \text{if $x \not\in \mathcal{A}$}  \\
\end{cases}
\end{equation*}
for a given set $\mathcal{A}$.
\end{definition}
\begin{example}{}
Let $x= -0.2$ and $\mathcal{A} = [-1, 1]$, 
then, since $ x = -0.2 \in [-1, 1] = \mathcal{A}$ 
we have $\mathbbm{1_\mathcal{A}}(x)=1$. 
One can use the following notation as well: 
$\mathbbm{1}(x \in \mathcal{A})=1$.
\end{example}

Please note the same idea can be defined when 
$\mathcal{A}$ is a condition, and $\mathbbm{1_\mathcal{A}}(x)=1$
whenever the condition $\mathcal{A}$ is met.
\begin{definition}{}
Opinions held by agents are defined as:
\begin{itemize}
\item The opinion of agent $i$ at time $t$ is $o_i^{(t)}$. 

\item Let $\mathbf{o}^{(t)} = [o_1^{(t)}, o_2^{(t)}, \cdots, o_N^{(t)}]$ 
be the vector of opinions of all agents; 
this is also referred to as a \emph{profile} 
in some articles. $N$ will denote the population of the network.

\item The set of neighbors of node 
$i$ is denoted by $N_i$ when 
including itself, and by $N_{\bar i}$ 
when excluding itself; this notation 
allows graphs to contain self-loops, 
and to use or exclude the opinion of agent $i$ in updates.
 \end{itemize}
\end{definition}

As mentioned before some of the terminology in the field is not
standardized. 
For example, Dong et al.~\cite{Dong2017} defines a leader as an
agent who is connected, directly or indirectly, to all other agents
and Dietrich et al.~\cite{Dietrich2018} defines a leader as an agent who
does not change its opinion whatsoever. Yet, the latter definition is used
to describe a {\it fully-stubborn} agent in some other texts. We will use
unified definitions for these cases.

\begin{definition}{}
A connected-agent is an 
agent that is connected directly
 or indirectly to all other agents. We will denote it by CA.
 In the case of digraphs there is a path from all agents to
 the connected agent.
\end{definition}

\begin{definition}{}
A fully-stubborn agent is an agent that
does not change its opinion at all. A partially-stubborn
agent is an agent that incorporates its initial opinion in subsequent 
 updates, but is open to change. Stubbornness is denoted by $1-d_i$, where 
$d_i$ is the measurement of susceptibility to influence.
If $1-d_i = 0$, then the agent $i$ is called 
non-stubborn; if $0 < 1-d_i < 1$, then the 
agent is called partially-stubborn; and if 
$1-d_i = 1$, then the agent is fully-stubborn.
\end{definition}

\begin{definition}{}
A fully-stubborn agent that is connected directly via an edge to all other agents is labeled as {media}.
\end{definition}

The terms leader, fully-stubborn agent or media can be used in
the contexts in which agents purposefully
steer or manipulate other agents
toward consensus or even more specifically 
toward a pre-determined opinion.
\begin{definition}{}
Let $\mathbf{A}$ be a given matrix. The $t$-th power of the matrix is 
denoted by $\mathbf{A}^t = \underbrace{A \times A \times \cdots \times A}_\text{t times} $.
\end{definition}

\begin{definition}{}
For a dynamically changing adjacency matrix, the adjacency matrix 
at time $t$ is given by $\mathbf{A}^{(t)}$. 
\end{definition}

\begin{definition}{}
The vector $\mathbf{e}_k$ is 
a vector with 1 in its $k^{th}$ position
and zeros elsewhere.
\end{definition}
\begin{definition}{}
The \emph{consensus} value 
is the opinion shared by all agents 
and is denoted by $o^* \in \mathcal{O}$. 
 \end{definition}

\begin{definition}{}
Convergence is defined as an equilibrium state, 
which may or may not be 
the consensus state. The equilibrium 
state is denoted by $\mathbf{O}^*$.
 \end{definition}

If the equilibrium state coincides with
consensus, then all its elements are identical; $\mathbf{O}^* = \left [ o^*, o^*, \dots, o^* \right]$.
\begin{definition}{}
Polarization refers to existence of two distinct groups that are not necessarily at opposite 
 extremes. 
\end{definition}
\begin{definition}{}
A bounded confidence model is a model in which
agents ignore other agents whose opinions 
are too far from their own and
take into account the opinions 
of agents that are close enough to their own.
The region within which an agent 
considers other opinions in is 
called the confidence interval of the agent. In Fig.~\ref{fig:confidenceIntervalDefPlot}
the blue intervals are confidence 
intervals of agents where other
opinions can be considered during an interaction.
\begin{figure}[httb!]
\begin{subfigure}{.5\textwidth}
  \centering
  \includegraphics[width=1\linewidth]{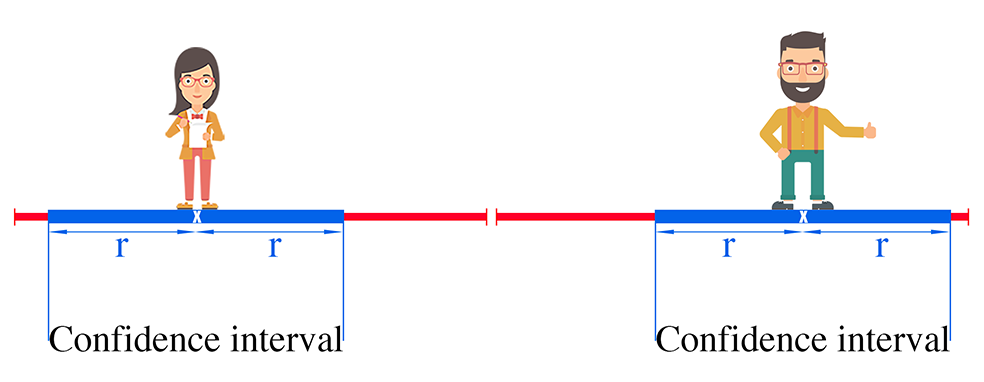}
  \caption{homogeneous and symmetric} 
   \label{fig:homo_sym}
\end{subfigure}
\begin{subfigure}{.5\textwidth}
  \centering
  \includegraphics[width=1\linewidth]{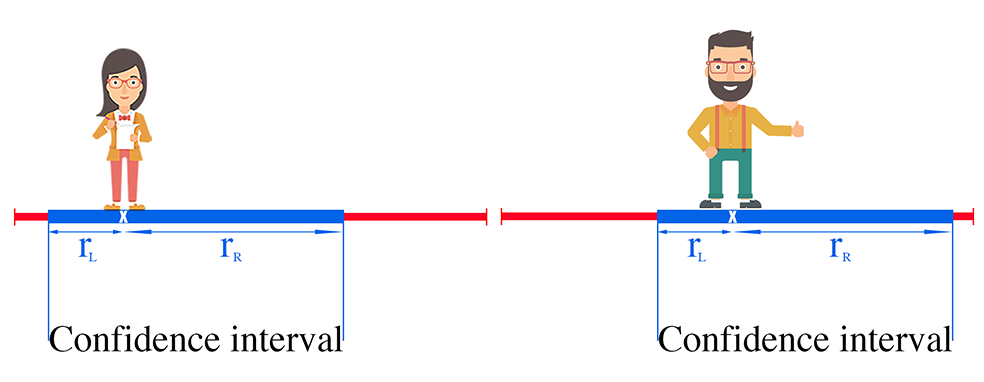}
  \caption{homogeneous and asymmetric}
  \label{fig:homoAsymConf}
\end{subfigure}
\begin{subfigure}{.5\textwidth}
  \centering
  \includegraphics[width=1\linewidth]{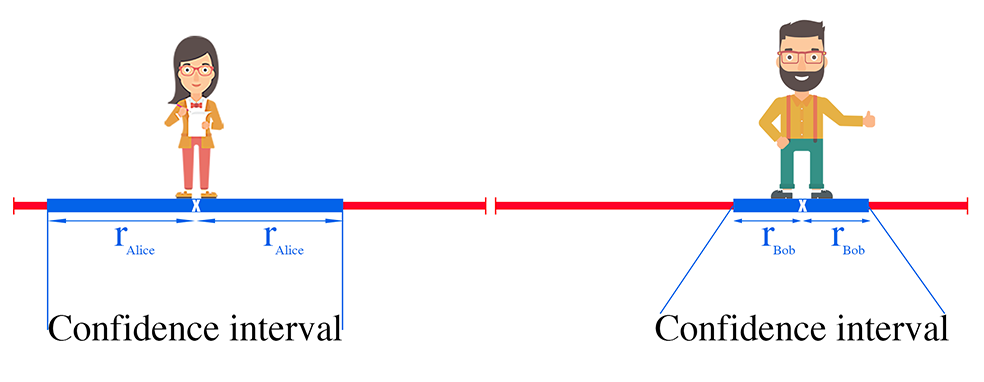}
  \caption{heterogeneous and asymmetric}
  \label{fig:hetroSymm}
\end{subfigure}
\begin{subfigure}{.5\textwidth}
  \centering
  \includegraphics[width=1\linewidth]{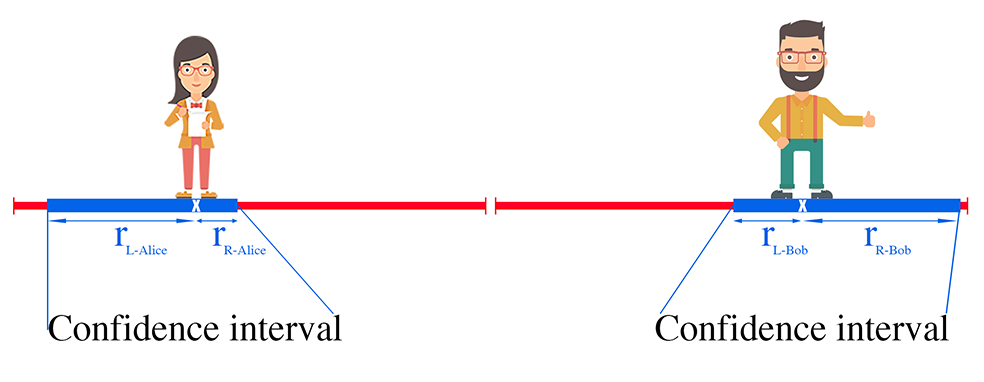}
  \caption{heterogeneous and asymmetric}
  \label{fig:hetroAsymConf}
\end{subfigure}
\caption{confidence interval 
types. \tiny{In these figures the opinion 
space, e.g. $[0, 1]$, 
is in red and the blue 
lines define the area 
where an interaction 
partner's opinion can fall and be 
accepted by Alice or Bob. An agent's opinion is
 shown by a white ``X''. 
In \ref{fig:homo_sym} both agents 
 have the same confidence interval, hence the system 
 is homogeneous (in terms of the confidence interval, 
 not the learning rate) and the intervals are symmetric. In~\ref{fig:homoAsymConf}
both agents have the same confidence interval, hence 
the system is homogeneous, but the intervals are asymmetric — 
agents are more accepting of opinions that are closer to 1, i.e. more than the 
agents' current opinion. In figures~\ref{fig:hetroSymm} 
and~\ref{fig:homoAsymConf} we see Bob's and 
Alice's intervals are different from each other, i.e.
 the system is heterogeneous. In the former case, the intervals are 
 symmetric, while in the latter they are asymmetric.}}
 \label{fig:confidenceIntervalDefPlot}
\end{figure}
If all agents have the same symmetric confidence interval, the interval's radius is denoted by $r$. 
If the confidence levels differ for left and right, they 
are denoted by $r_l$ and $r_r$, respectively. 
Cases in which all agents enjoy identical confidence 
intervals are referred to as \emph{homogenous} models. 
If each agent has its own confidence radius, it is 
denoted by $r_i$ in the case of symmetric confidence 
intervals. If the confidence levels are different for left 
and right, they are denoted by $r_{il}$ and $r_{ir}$, respectively; 
such models are referred to as \emph{heterogenous} systems.
\end{definition}
\textbf{Note}
\emph{Homogeneous} is used for two purposes: 1. to indicate 
that all agents share the same confidence interval, or 2. to 
indicate that they share the same learning rate, i.e. the same step 
size coefficient in updating an opinion due to an interaction. 
Likewise, heterogeneity can be used for two purposes. 
Note that in a given system, agents can be simultaneously 
homogeneous with respect to confidence interval and 
heterogeneous with respect to learning rate, and all other 
combinations of homogeneity/heterogeneity.

To refresh the memory of readers, we briefly list the 
well-known models that have been the foundation 
of opinion dynamics and other scientists' works. 
Then we delve into the most recently published 
modifications of these models.
\subsection{Classical models}\label{ClassicalModels}
\begin{enumerate}
\item The DeGroot\cite{DeGroot1974} model, 
a simple averaging scheme that defines a 
linear system that can be analyzed by classical linear algebra, is given by 
\begin{equation}\label{eq:DeGroot}
  \tag{DeGroot}
\mathbf{o}^{(t)} = \mathbf{A o}^{(t-1)} = \mathbf{A}^2 \mathbf{o}^{(t-2)}  = \cdots = \mathbf{A}^t \mathbf{o}^{(0)}
\end{equation}
where $\mathbf{A}$ is a row stochastic matrix 
of weights. In the original paper‚ stochastic 
indicated row stochastic and doubly stochastic 
referred to a matrix whose row sums and column 
sums each add up to 1. The matrix entries are the 
weights that a given agent puts on other agents' 
opinions, i.e. the weights determine how much 
a given agent is influenced by any other agent. 
The updates in this model are synchronous. 
The stochasticity of the adjacency matrix 
means that the weights 
each agent puts on all of its friends/neighbors, 
including itself, add up to 1 or 100\%. 
So, everything an agent learns
is calculated exactly by the amount he trusts his friends and himself.
\item The Friedkin-Johnsen (FJ) model, which was 
introduced in 1990\cite{FJmodel1990} and 1999\cite{ FJmodel}, is given by
(the FJ model is extension of the DeGroot model that includes stubborn agents):
\begin{equation}\label{eq:FJmodelEq}
\tag{FJ}
\mathbf{o}^{(t+1)} = \mathbf{D} \mathbf{A o}^{(t)} + ( \mathbf{I} -  \mathbf{D}) \mathbf{o}^{(0)}
\end{equation}
where $\mathbf{D} = diag([d_1, d_2, \dots, d_N])$ 
with entries that specify the susceptibility of individual 
agents to influence, i.e., $(1-d_i)$ is the level of 
stubbornness of agent $i$, and $\mathbf{A}$ is a 
row stochastic matrix. In this model, agents are attached 
to their initial opinions, and are referred to as \emph{stubborn} 
agents. If the adjacency matrix with entries that are 
influence weights is not symmetric then we can 
assume its associated graph is a digraph. Updates of the system are synchronous.
\item A well-known bounded confidence model is
introduced by Deffuant and Weisbuch~\cite{Deffuant2000},
commonly referred to as the DW model, is given by: 
\begin{equation}\label{eq:BCM}
  \tag{DW}
  \begin{cases} 
  o_i^{(t+1)}  = o_i^{(t)} + \mu \:  . \: \mathbbm{1}_{[0 , r]}(|d_{ji}^{(t)}|) \:  . \: d_{ji}^{(t)}\\
  o_j^{(t+1)}  = o_j^{(t)} + \mu \:  . \: \mathbbm{1}_{[0 , r]} (|d_{ij}^{(t)}|) \:  .  \: d_{ij}^{(t)} \\
\end{cases}
\end{equation}
in which at any given time the pair $i$ and $j$ are 
chosen randomly and $d_{ji}^{(t)} = o_j^{(t)} -  o_i^{(t)}$. 
The parameter $\mu$ is called the learning parameter 
or the convergence parameter and \emph{usually} is 
taken in the interval $(0, 0.5]$ to avoid crossover.
Please note that for the basic DW model,
\begin{itemize}
\item The system is homogeneous in learning rate $\mu$.
\item The system is homogeneous in confidence radius $r$.
\item The system updates are pairwise.
\end{itemize}

An example of a bounded-confidence model interaction in which
each person has a different level of openness/closedness would be the following:
 Assume Bob and Alice participate in an interaction. If the opinion of Bob is close enough
 to the opinion of Alice, i.e., Bob's opinion falls in the 
 blue region in~\ref{fig:homoAsymConf}, then Alice learns from him and
 her opinion gets closer to that of Bob. But Bob is not very
 open-minded toward those whose opinions are far smaller
 than his own opinion, i.e., Alice's opinion does not fall in the 
 blue region on the left side of Bob in Fig.~\ref{fig:homoAsymConf},
 therefore Bob will not learn and his opinion will remain unchanged.
\item The most general form of the 
Hegselmann-Krause~\cite{Hegselmann2002} (HK) model is given by:
\begin{equation}\label{eq:HK}\tag{HK}
\mathbf{o}^{(t+1)} = \mathbf{A}(t, \: \mathbf{o}^{(t)}) \mathbf{o}^{(t)} 
\end{equation}
where $\mathbf{A}(t, \: \mathbf{o}^{(t)})$ is an arbitrary function of time
and opinion. This is the HK model in its most complex and flexible form, 
which is too complicated to study without simplification:
``In this generality, however, one cannot hope to get an answer, 
neither by mathematical analysis nor by computer simulations.'' 
Because Eq.~\eqref{eq:HK} is too general to be suitable for direct analysis, 
Hegselmann inevitably studied simplifications of it. Please note that the set of models 
studied in Ref.~\cite{Hegselmann2002}, 
regardless of the direction they take are all referred 
to as the HK model in other research papers.

Note that by fixing the matrix 
$\mathbf{A}(t, \mathbf{o}^{(t)}) := \mathbf{A} $ 
the Eq.~\eqref{eq:HK} collapses to DeGroot. 
Please also note that fixing the matrix takes away the 
bounded confidence idea and the dynamic 
changes dramatically in nature.

Setting $\mathbf{A}(t, \mathbf{o}^{(t)}) := \mathbf{A}(\mathbf{o}^{(t)})$ 
where the adjacency matrix is only dependent on
the current profile, following the bounded confidence 
rule, results in a synchronous bounded confidence
 model. 
\end{enumerate}

In Sec.~\ref{degrootianModelSec} we study models that are based on DeGroot and FJ models 
, defined above, and in Sec.~\ref{BCMSection} bounded 
confidence model extensions is discussed. An important difference 
between DeGrootian and bounded confidence model is 
that the former is linear and latter is nonlinear and more complex.
\section{DeGrootian models and their applications}\label{degrootianModelSec}
The \ref{eq:DeGroot} model is the simplest method used 
for representing opinion dynamics. 
Such a simple scenario
is traceable in time and opinion 
space, enabling researchers
to produce analytical results. 
Over the years, different modifications
of and additions to the DeGroot model have been used
to investigate a variety of real human traits such as 
stubbornness (the study of which gave rise to the \ref{eq:FJmodelEq} model in 1990); since then,
the \ref{eq:FJmodelEq} model has undergone further developments, and here
we look at some of the newest results: the evolution of the social power of agents in the 
DeGroot and FJ models over a sequence of topics, the
co-evolution of expressed and private opinions, the evolution of 
opinions given sequentially dependent topics,
and the evolution of an agent's susceptibility to influence.


\subsection{Power evolution}
In any society, whether it be a colony of ants, 
a pride of lions, or the US house of representatives, any
given member will have power over some and will be submissive to others.
As time passes and issue after issue is addressed by the society, the
type of hierarchy governing the individual agents will become more
distinguishable and distinct. Depending on the particulars of the society
in question, an autocrat may arise or a democracy may develop.
The phenomena of power evolution has been studied by
researchers~\cite{YeLatest, MirTabatabaei, Tian2019, Askarzadeh2019}
and we will devote following two 
subsections to this type of scenario.

\subsubsection{Evolution of social power in the DeGroot model over a sequence of topics}\label{DeGrootSocialPower}

Let us start this section 
with a proposition about the 
DeGroot model that may enlighten 
the motivation for the rest of the section.

\begin{prop}\label{DeGrootConvergence80} [Ref.~\cite{Berger1980}]
The DeGroot model will reach consensus if and 
only if there exists a power $t$ of the adjacency matrix 
for which $\mathbf{A}^t$ has a strictly positive column.
\end{prop}

This proposition basically states that if there 
comes a time that everyone in the community
listens to an agent (directly or indirectly), i.e., takes his opinion 
into account, then the community will come to consensus eventually.

\begin{prop}\label{DeGrootConvergence74}[Ref.~\cite{DeGroot1974}]
Let $\mathbf{A}$ be an adjacency matrix for 
which the DeGroot model reaches consensus. 
Then the final state of the system is $\mathbf{o}^* := o^*\mathbf{1}$, where:
\begin{equation}\label{eq:lefteig}
o^* = \langle \bm{\ell}_\mathbf{A},  \mathbf{o}^{(0)}  \rangle  
\end{equation}
\noindent and $\mathbf{1}_N = [1, 1, \dots, 1]^T \in \mathbb{R}^N$ is the vector of size $N$.
 $ \bm{\ell_A^\text{T}}$ is the left eigenvector of $\mathbf{A}$ 
 associated with 1, i.e., $\bm{\ell_A^\text{T}} \mathbf{A} = 1 \bm{\ell_A^\text{T}}$,  
 constrained to $\langle \bm{1}_N, \bm{\ell_A} \rangle = 1$. Since the entries of 
 $\bm{\ell_A^\text{T}}$ are non-negative, the $o^*$ is a convex combination of 
 the initial opinions.
\end{prop}

Jia et al.~\cite{MirTabatabaei} introduced a very realistic 
scenario involving the evolution of social power in 
which individuals become aware of their power to
 influence and control others and the outcome of 
 debate on each topic under negotiation in a sequence of topics.
 
Let $\mathbf{0}_N = [0, 0, \ldots, 0]^T$
 be the vector of zeros with length $N$. 
 Let the simplex $\Delta_N$ 
 be the set of points in 
 $\{ \mathbf{x} \in \mathbb{R}^N\ | \mathbf{x} \ge 0, \langle \mathbf{1},  \mathbf{x} \rangle = 1 \}$. 
 A nonnegative matrix $\mathbf{M}$ is irreducible 
 if its associated digraph is strongly 
 connected. A nonnegative matrix $\mathbf{M}$ is reducible 
 if its associated digraph is not strongly connected.
 
In the~\ref{eq:DeGroot} model the weight matrix is static 
and does not change over time for the given topic under 
discussion. Here we consider a sequence of topics or subjects $s=0, 1, 2, \dots$ 
where each topic is represented by the DeGroot model, i.e., the weight matrix $\mathbf{A}$ 
does not change over time for a given 
topic, although it changes from topic to topic. The changes 
depend on the outcome of the previous topic, i.e., 
$\mathbf{A}(s+1)$ depends on the outcome of topic $s$:
\begin{equation}\label{eq:DeGrootFriedkin}
\mathbf{o}^{(t+1)}(s) = \mathbf{A}(s) \mathbf{o}^{(t)}(s)
\end{equation}
Just as in the DeGroot model, each weight/adjacency 
matrix is stochastic. The diagonal entries, $a_{ii}$, 
determine the degree of openness to change 
and the off-diagonal entries, $a_{ij}$, determine the degree 
to which agent $i$ is influenced by agent $j$. The off-diagonal 
entries can be decomposed and
written as $a_{ij} = (1 - a_{ii} ) c_{ij}$ where the $c_{ij}$ values
 are referred to as \emph{relative interpersonal weights}. 
 Define the matrix $\mathbf{C}:=[c_{ij}]$ with diagonal 
 entries equal to zero. Then $\mathbf{C}$ is 
 stochastic and we refer to it as the relative interaction 
 matrix. Note that the self-weights $a_{ii}(s)$ 
 are topic dependent, however the matrix $\mathbf{C}$ 
 is static and does not depend on the topic.

We can write 
\begin{equation}
\mathbf{A}(s) = diag([a_{11}(s), a_{22}(s), \dots, a_{NN}(s)]) + (\mathbf{I} - diag([a_{11}(s), a_{22}(s), \dots, a_{NN}(s)] )) \mathbf{C}
\end{equation}

From now on we assume that matrix $\mathbf{C}$, which 
is stochastic with zero diagonals, is irreducible unless 
otherwise stated. Based on this assumption, the 
influence matrix $\mathbf{A}(s)$ has a unique 
left eigenvector $\bm{\ell_A}$ with non-negative 
entries, normalized so that $ \langle \mathbf{1}, \: \bm{\ell_A} \rangle = 1$, 
and associated with eigenvalue $\lambda = 1$, i.e., $\bm{\ell_A} \in \Delta_N$. For a large variety of 
the self-weight vectors $[a_{11}, a_{22}, \dots, a_{NN}]$, the 
eigenvector $\bm{\ell_A}$ satisfies 
$\lim_{t\to\infty} \mathbf{A}^t = \mathbf{1} [a_{11}, a_{22}, \dots, a_{NN}] $, 
which explains 
consensus in the DeGroot model: 
$\lim_{t\to\infty} \mathbf{o}^{(t)} = \langle \bm{\ell_A}, \mathbf{o}^{(0)} \rangle \mathbf{1}$.
 Hence, the entries of the left eigenvector $\bm{\ell_A}$ 
determine the contribution of each individual to the 
final state of the system, that is, this eigenvector 
defines the agents' power. This fact, also mentioned in Props.~\ref{DeGrootConvergence80} and ~\ref{DeGrootConvergence74}, motivates the 
definition of the evolution of social 
power for a sequence of topics: 
\begin{equation}
[a_{11}(s+1), a_{22}(s+1), \dots, a_{NN}(s+1)] = \bm{\ell}_{\mathbf{A}}(s)
\end{equation}

In other words, the self-weights for topic 
$s+1$ are equal to the agents' power contributions 
to the final state of the system for topic $s$. 
This leads us to the definition of the DeGroot-Friedkin model.
\begin{definition}{}
Let a group of $N$ agents discuss a sequence of topics 
$s \in \mathbb{Z}_{\ge 0}$ and the matrix $\mathbf{C}$ be 
the relative interaction matrix. The DeGroot-Friedkin model is given by 
\begin{equation} [a_{11}(s+1), a_{22}(s+1), \dots, a_{NN}(s+1)] = \bm{\ell_A}(s) \end{equation}
\noindent where $\bm{\ell_A}(s) \in \Delta_N$ is the dominant left eigenvector of the adjacency/weight matrix:
\begin{equation} \mathbf{A}(s) = diag([a_{11}(s), a_{22}(s), \dots, a_{NN}(s)]) - (\mathbf{I} - diag([a_{11}(s), a_{22}(s), \dots, a_{NN}(s)])) \mathbf{C} \end{equation}
\end{definition}

The following proposition is a bridge that connects the 
DeGroot-Friedkin model to dynamical systems theory, 
enabling the application of this model to dynamical 
systems and the establishment of results such as 
proof of the existence and uniqueness of fixed points. 
Before introducing the proposition let us introduce 
$\bm{\ell_C}$ as the left eigenvector of the
 relative interaction matrix $\mathbf{C}$ that 
 corresponds to the eigenvalue $1$ such that 
 $\langle \mathbf{1} \: , \: \bm{\ell_C} \rangle = 1$. 
 The $i^{th}$ entry of $ \bm{\ell_C} $ is called the
 eigenvalue centrality score of agent $i$.
\begin{prop}\label{dynamicDFModel}
Let $( \mathbf{1} , \bm{\ell}_\mathbf{C} ) $ be 
the eigenpair of relative interaction $\mathbf{C} \in \mathbb{R}^{N \times N}$. 
The DeGroot-Friedkin model is equivalent 
to \[ diag([a_{11}(s+1), a_{22}(s+1), \dots, a_{NN}(s +1 )]) = F(diag([a_{11}(s), a_{22}(s), \dots, a_{NN}(s)]))\] 
where $F: \Delta_N \rightarrow \Delta_N$ 
is a continuous map given by
\begin{equation}\label{eq:DeGrootEquivToDynamEq}
F(x) = 
\begin{cases} 
\vspace{.2in}
\mathbf{e}_i  & \text{if $\mathbf{x} = \mathbf{e}_i$,} \\
\left (\frac{c_1}{1- x_1}, \frac{c_2}{1- x_2}, \dots, \frac{c_N}{1- x_N} \right )^T \big / \sum_{i=1}^N \frac{c_i}{1-x_i}& \text{o.w.} \\
\end{cases}
\end{equation}
\noindent and where $c_i$ is the $i^{th}$ entry of the left eigenvector $\bm{\ell_C}$.
\end{prop}

If the relative interaction matrix $\mathbf{C}$ is doubly 
stochastic, then the left eigenvector associated with 
eigenvalue $1$ is $\bm{\ell_C} = \mathbf{1}/N$ and Eq.~\ref{eq:DeGrootEquivToDynamEq} simplifies to
\begin{equation}\label{eq:DeGrootEquivToDynamEq}
F(x) = 
\begin{cases} 
\vspace{.2in}
\mathbf{e}_i  & \text{if $\mathbf{x = e}_i$,} \\
\left (\frac{1}{1- x_1}, \frac{1}{1- x_2}, \dots, \frac{1}{1- x_N} \right )^T \big / \sum_{i=1}^N \frac{1}{1-x_i}& \text{o.w.} \\
\end{cases}
\end{equation}
\begin{prop}\label{dynamicDFModelDoublyStochProp}
If the relative interaction matrix is doubly stochastic, then the following two properties hold true: 
\begin{enumerate}
\item The equilibrium points of the dynamical system given by $F$
are $\{ \mathbf{1}/N, \: \mathbf{e_1,\: e_2, \dots, \: e_N}\}$

\item For all initial conditions $\mathbf{x}(0) \in \Delta_N \backslash \{  \: \mathbf{e_1, \: e_2, \dots, \: e_N} \}$
we have $\lim_{s\to\infty} \mathbf{x}(s) = \mathbf{1}/N$.
\end{enumerate}
\end{prop}

The second property in Prop.~\ref{dynamicDFModelDoublyStochProp} 
indicates that the DeGroot model results in consensus with
 the final opinion being an average of the initial opinions, thereby indicating equal social ranking among agents.

The authors~\cite{MirTabatabaei} presented 
conclusions and results (which are not included here) 
for interaction networks, including a star graph 
with propositions similar to those above. We close this subsection with the following theorem:
\begin{theorem}\label{MirTabThm}
Let there be $N \ge 3$ nodes in an interaction network 
(that is not a star graph) with a relative interaction matrix 
$\mathbf{C}$. Let $\bm{\ell_C}$ be the left eigenvector 
of $\mathbf{C}$ associated with eigenvalue $1$. 
Then, for the dynamical system defined by
\begin{equation}
[ a_{11}(s+1), a_{22}(s+1), \dots, a_{NN}(s+1) ]  = F([ a_{11}(s), a_{22}(s), \dots, a_{NN}(s) ] )
\end{equation}
\noindent we have the following:
\begin{enumerate}
\item The set of points of $F$ is $\{ \mathbf{x^*, e_1, e_2, \dots, e_N}\}$
where $\mathbf{x}^*$ lies in the interior region of $\Delta_N$ and 
the ordering of the entries in $x^*$ is the same as that in $\bm \ell_\mathbf{C}$.

\item For all initial conditions 
$\mathbf{x}^{(0)} \in \Delta_N \backslash \{  \mathbf{ \: e_1,\: e_2, \dots, \: e_N} \}$
we have:
\begin{equation}
\lim_{s\to\infty} [ a_{11}(s), a_{22}(s), \dots, a_{NN}(s)] = \mathbf{x}^*
\end{equation}
\end{enumerate}
\end{theorem}

According to Thm.~\ref{MirTabThm}, if the network does not 
establish an individual with total power as a consequence 
of its graph topology (i.e., the graph is not a star graph) 
or if in the initial system the social ranking of individuals 
is not set up so that one has power over all others, 
then the social ranking amongst agents will converge 
to an egalitarian state in which all individuals have the same power.

Later Ye et al.~\cite{Ye2018} showed that the convergence discussed above is 
exponentially fast. And they also studied the case of  dynamic topology 
in which the matrix $\mathbf{C}$ changes along the sequence of topics,
i.e. $\mathbf{C}(s)$ is a function of $s$, and show the conditions under which
the same results hold true.
\subsubsection{Evolution of social power in FJ model over sequence of topics}\label{friedkinMirtab}
Let us now consider the idea of the evolution of power over 
the course of a sequence of topics in the FJ model studied 
in Ref.~\cite{Tian2019}. The FJ model is an extension of the 
DeGroot model in which each agent has a memory and 
is attached to its initial opinion at time $t=0$ and cannot 
completely let go of it. This model is given by Eq.~\ref{eq:FJmodelEq} 
in~\cref{sec:definitions}. To refresh the readers' memory, we repeat:
\begin{equation}\label{eq:eq11}
\mathbf{o}^{(t+1)} = \mathbf{D} \mathbf{A o}^{(t)} + ( \mathbf{I} -  \mathbf{D}) \mathbf{o}^{(0)}
\end{equation}
where $\mathbf{D} = diag([d_1, d_2, \dots, d_N])$ 
with entries that are individuals’ susceptibility to \
influence, i.e., $(1-d_i)$ is the level of stubbornness of 
agent $i$. Moreover, since in this subsection we are 
applying the FJ model to a sequence of topics where 
each adjacency/weight matrix depends on the topic 
$s$ currently under discussion, we can rewrite the equation above as
\begin{equation}
\mathbf{o}^{(t+1)}(s) = \mathbf{D} \mathbf{A}(s) \mathbf{o}^{(t)} + ( \mathbf{I} -  \mathbf{D}) \mathbf{o}^{(0)}(s)
\end{equation}

As before, we can write the adjacency matrix as:
\begin{equation}\label{eq:adjAgain}
\mathbf{A}(s) := diag([ a_{11}(s), a_{22}(s), \dots, a_{NN}(s)]) +  ( \mathbf{I} - diag([a_{11}(s), a_{22}(s), \dots, a_{NN}(s) ])) \mathbf{C} 
\end{equation}
where $\mathbf{C}$ is a stochastic matrix with zeros on the 
diagonal; however, in this section we drop the irreducibility 
of $\mathbf{C}$. Let us define a few concepts followed by 
a definition of the FJ model modified to handle a sequence of topics.
\begin{definition}{}
A directed graph is said to be \emph{strongly connected} 
if every node is reachable 
from every other node, i.e. if there is a path between 
any node to any other node.
A \emph{strongly connected component (SCC)} 
of a graph $\mathcal{G}$ is a strongly connected subgraph of $\mathcal{G}$
and is maximal in the sense that no additional edge or vertex from $\mathcal{G}$
can be added to it without breaking the strong connectivity property.
\end{definition}
\begin{definition}{}
An SCC of graph $\mathcal{G}$ is called a \emph{sink SCC} 
if there are no directed edges from it to the nodes outside of it.
\end{definition}

To avoid repetition, we list a few assumptions 
here and refer to them later, as needed.
\begin{assumption}\label{ass1}
Every sink SCC of $\mathcal{G}(\mathbf{C})$ 
has at least one stubborn agent, and $d_i < 1$ if for the self-weight 
vector we have $[a_{11}(0), a_{22}(0), \dots, a_{NN}(0)] = \mathbf{e}_i$.
\end{assumption}

\begin{assumption}\label{ass2}
$\forall i \: d_i < 1$, $\exists j, \: s.t. \: d_j > 0$.
\end{assumption}

Now we are ready to define the FJ 
model for a sequence of topics.
\begin{definition}{}
Let $s= 0, 1, 2, \dots$ be a sequence of topics, 
and assume Asm.~\ref{ass1} holds. Let 
$\mathbf{C}$ be the relative influence matrix, and 
let $\mathbf{D} = diag([d_1, d_2, \dots, d_N])$ be 
the susceptibility matrix. Then the FJ model for a sequence of topics (FJS) is given by:
\begin{equation}\label{eq:FJSUpdateRule}
[a_{11}(s+1), a_{22}(s+1), \dots, a_{NN}(s+1)]^T = (\mathbf{I} - \mathbf{D}) ( \mathbf{I} - \mathbf{A}^T(s) \mathbf{D})^{-1} \frac{\mathbf{1}}{N}
\end{equation}
where the adjacency matrix, $\mathbf{A}(s)$, is given by Eq.~\ref{eq:adjAgain}.
\end{definition}
\noindent \textbf{Derivation of Eq.~\eqref{eq:FJSUpdateRule}}: By Eq.~\eqref{eq:eq11}
we have

\begin{equation}\label{eq:allInOne}
\begin{array}{lllllll}
\mathbf{o}^{(t+1)} &=& \mathbf{DA} {\color{ForestGreen}{\mathbf{o}^{(t)}}} + (\mathbf{I - D}) \mathbf{o}^{(0)}\\
 &=& \mathbf{DA} \left[{\color{ForestGreen}{\mathbf{DA}  \mathbf{o}^{(t-1)} +  (\mathbf{I - D}) \mathbf{o}^{(0)}}}  \right] + (\mathbf{I - D}) \mathbf{o}^{(0)}\\
&=& (\mathbf{DA})^2 \mathbf{o}^{(t-1)} +\mathbf{DA} \mathbf{(I-D)} + (\mathbf{I - D}) \mathbf{o}^{(0)}\\
&=& (\mathbf{DA})^2 {\color{Cyan}{\mathbf{o}^{(t-1)}}} + \left[ \mathbf{DA} + \mathbf{I} \right] (\mathbf{I - D}) \mathbf{o}^{(0)}\\
&=& (\mathbf{DA})^2 \left[{\color{Cyan}{ \mathbf{DA} \mathbf{o}^{(t-2)}  + (\mathbf{I-D}) \mathbf{o}^{(0)} }}\right] + \left[ \mathbf{DA} + \mathbf{I} \right] (\mathbf{I - D}) \mathbf{o}^{(0)} \\
&=& (\mathbf{DA})^3 \mathbf{o}^{(t-2)} + \left[(\mathbf{DA})^2 + (\mathbf{DA}) + \mathbf{I}\right] (\mathbf{I - D}) \mathbf{o}^{(0)} \\
&\vdots& \\
 &=&
 (\mathbf{DA})^{t+1}\mathbf{o}^{(0)} + \left[(\mathbf{DA})^{t} + (\mathbf{DA})^{t-1} + \cdots + (\mathbf{DA})^1 + \mathbf{I} \right] (\mathbf{I - D}) \mathbf{o}^{(0)}
\end{array} 
\end{equation}
Therefore, since $\mathbf{DA}$ is strictly sub-stochastic,
we have $\lim_{t\to\infty}(\mathbf{DA})^{t+1} = \mathbf{0}$
and 
\begin{equation}
\lim_{t\to\infty}  \left[ (\mathbf{DA})^{t} + (\mathbf{DA})^{t-1} + \cdots + (\mathbf{DA})^1 + \mathbf{I}  \right]  = \mathbf{(I - DA)}^{-1}
\end{equation}
\noindent consequently,
\begin{equation}
\lim_{t\to\infty} \mathbf{o}^{(t+1)} = \mathbf{(I - DA)}^{-1} (\mathbf{I-D}) \mathbf{o}^{(0)}
\end{equation}
Assumption~\ref{ass1} implies that the system comes to consensus for each topic:
\begin{equation}
\lim_{t\to\infty} \mathbf{o}^{(t)} = ( \mathbf{I} -\mathbf{D}  \mathbf{A}(s) )^{-1}  (\mathbf{I} - \mathbf{D}) \mathbf{o}^{(0)}
\end{equation}

We refer to the matrix 
$( \mathbf{I} -\mathbf{D}  \mathbf{A}(s) )^{-1}  (\mathbf{I} - \mathbf{D})$ 
as the final state matrix. The mean of the $i^{th}$ 
column of the final state matrix given by the $i^{th}$ 
element of 
$ [(\mathbf{I} -\mathbf{D}  \mathbf{A}(s) )^{-1}  (\mathbf{I} - \mathbf{D})]^T \frac{\mathbf{1}}{N}$ 
is the relative control agent $i$ has on the final opinions of 
other agents, i.e., the social 
power of agent $i$. The transpose of the final state 
matrix is stochastic, and it defines a continuous map 
from $\Delta_N$ to itself and hence has a fixed 
point in $\Delta_N$. Lets denote this map by $F(\mathbf{x})$:
\begin{equation} \label{eq:FJSMapping}
F(\mathbf{x}) =  [(\mathbf{I} -\mathbf{D}  \mathbf{A}(\mathbf{x}) )^{-1}  (\mathbf{I} - \mathbf{D})]^T  \frac{\mathbf{1}}{N}  =
(\mathbf{I} - \mathbf{D}) (\mathbf{I} -\mathbf{A}^T(\mathbf{x})  \mathbf{D}  )^{-1}    \frac{\mathbf{1}}{N}
\end{equation}

Furthermore, this map indicates that the final 
state of the system for topic $s$, and hence the social 
power of agents with respect to topic $s+1$, is dependent 
on stubbornness, i.e., stubbornness is equivalent to social power.

Please note that in the definition above 
$\mathbf{x} \in \mathbb{R}^N$, where in 
Eq.~\ref{eq:FJSUpdateRule}, for simplicity and 
to avoid the introduction of new notation, we use 
$\mathbf{A}(s)$ to indicate that the adjacency matrix 
depends on each topic and can be written as Eq.~\ref{eq:adjAgain}, 
which is obtained by representing each element of 
$\mathbf{A}(s)$ as the product $\mathbf{A}_{ij}(s) := (1- a_{ii}(s)) \mathbf{C}_{ij}$, 
where the $a_{ii}(s)$ are the self-weights. 
So, $\mathbf{A}(s)$ emphasizes the 
dependence of the adjacency matrix on topic $s$, 
which in turn depends on the self-weights 
$[a_{11}(s), a_{22}(s), \dots, a_{NN}(s)] \in \mathbb{R}^N$.
\begin{prop}\label{mapFProp1}
For the map $F(\mathbf{x})$ given by Eq.~\ref{eq:FJSMapping} we have:
\begin{itemize}
\item F is continuous on $\Delta_N$ and is differentiable in its interior region.
\item $\forall x \in \Delta_N,  \: F_i(\mathbf{x}) \in [\frac{1-d_i}{N} , \frac{1 + \zeta}{N}]$
where $\zeta =  \: \langle [d_1, d_2, . . . , d_N ]^T , \mathbf{1} \rangle - \: d_{min}$, and $d_{min} = \min \{d_1, d_2, . . . , d_N \}$.
\end{itemize}
\end{prop}
\begin{theorem}\label{Thm1OfFJS}
Consider the dynamical system given by 
Eq.~\ref{eq:FJSUpdateRule}, and let $[a_{11}(0), a_{22}(0), \dots, a_{NN}(0)] \in \Delta_N$. 
Denote the set of 
fully stubborn agents, (i.e., $d_i = 0$) with $\mathcal{V}_f$ 
and the set of partially stubborn agents (i.e., $d_i > 0$) 
with $\mathcal{V}_p$. WLOG, assume $\mathcal{V}_f = \{ 1, 2, \dots, r\} $ and $\mathcal{V}_p = \{ r+1, r+2, \dots, N\} $. Then,
\begin{enumerate} [ label = (\arabic*) ]
\item $\exists \mathbf{x}^* \in \Delta_N$ satisfies:
    \begin{enumerate} [ label = (\roman*) ]
       \item $\forall i \in \mathcal{V}_f, \mathbf{x}^*_i \ge \frac{1}{N}$, \text{and}, $\mathbf{x}^*_i = \frac{1}{N} \: \text{iff} \:\: \text{ for any }  j \in \mathcal{V}_p,\mathbf{C_{ji}}=0$

       \item $\forall i \in \mathcal{V}_p, \: \mathbf{x}^*_i > \frac{1-d_i}{N}, \text{and } \mathbf{x}^*_i<\frac{1}{N} \: \text{if} \:\: \mathbf{C_{ji}}=0 \text{ for any } j \in \mathcal{V}_p$

       \item $\max_i \:\: \mathbf{x^*_i} < \frac{1}{N} (1 + \langle {[d_1, d_2, \dots, d_N]^T, 1}\rangle) $
    \end{enumerate}

\item $\mathbf{x}^*$ is unique if $\max_i \:\: d_i < \frac{N}{N + 2(1+\zeta)}$.
\end{enumerate}
\end{theorem}

Theorem~\ref{Thm1OfFJS} shows that an autocracy 
is not a possible outcome for the system defined 
above, as constrained by the associated assumptions. 
Moreover, if two agents can influence a third one, 
then the more stubborn agent of the two will 
have more social power at the end.
\begin{theorem}\label{Thm4OfFJS}
Consider the dynamical system given by 
Eq.~\ref{eq:FJSUpdateRule} for which Asm.~\ref{ass2} 
holds, and let $\zeta$ be defined as it was in 
Prop.~\ref{mapFProp1}. If $\max_i \:\: d_i < \frac{N}{N + 2(1+\zeta)}$, 
then all trajectories of the dynamical system 
converge exponentially fast to the unique 
equilibrium point given in Thm.~\ref{Thm1OfFJS}.
\end{theorem}

In Ref.~\cite{Tian2019} the author establishes a number 
of the properties for the system associated with 
a star graph. Moreover, the evolution 
of social power is considered for a single topic, as 
opposed to the evolution occurring over a sequence of 
topics with social power fixed for a given topic. The 
author shows that the two approaches have 
similar behavior and equivalent properties.

The idea of the evolution of social power over a sequence 
of topics has been empirically studied in Ref.~\cite{Friedkin2016socialPower}. 
In this model, for a strongly connected network with assumptions such 
as the ones outlined above, one dominant agent with maximal 
influence will typically emerge, with the rest of the agents having 
minimal influence. An example in which the above scenario does
 not happen is a fully connected graph with all individuals having 
 the same level of influence at $t=0$. The findings of Ref.~\cite{Friedkin2016socialPower} 
 involve mostly artificial experiments in which people are represented
 unnaturally as interacting simultaneously. However, one might 
 consider a simultaneous interaction as equivalent to a pairwise 
 interaction with a different influence matrix. For example, agent 
 $i$ may be influenced by agent $j$ at some time $t$, and agent 
 $j$ may have been influenced by agent $k$ at some earlier time. 
 Hence, agent $i$ is influenced by agent $k$ indirectly, which one 
 might consider a simultaneous interaction with agent $i$ allocating 
 different influence weights to $j$ and $k$ in two pairwise and 
 simultaneous interactions. It is the combination of influence 
 weight and the frequency of interaction that matters, really.
  
\subsection{Susceptibility evolution}
A new line of thought in opinion dynamics is 
considered in Ref.~\cite{Abebe2018}, followed 
by Ref.~\cite{Chan2019}; while the idea upon which it 
is based‚ a dynamic susceptibility to 
persuasion‚ has a long(er) history in social psychology, 
the mathematical study of it in opinion 
dynamics is novel. Susceptibility to persuasion 
denotes the extent to which a given agent is 
willing to change its opinion. It is the opposite 
of stubbornness. In the FJ model:
\[\mathbf{o}^{(t+1)} = \mathbf{DAo}^{(t+1)} + \mathbf{(I - D)o}^{(0)}\]

Recall that $\mathbf{D} = diag([d_1, d_2, \cdots, d_N])$, 
where the level of stubbornness of 
agent $i$ is $1 - d_i$; therefore, by definition, $d_i$ is agent $i$'s 
susceptibility to persuasion. Abebe et al.~\cite{Abebe2018} 
built on the FJ model by studying the effects of manipulating the $d_i$'s. It 
seems reasonable to assume that the susceptibility 
to persuasion of the agents in a network can be 
influenced by different tools. 
Abebe ran simulations with the goal of determining 
how the opinion dynamic could be altered by changing 
the susceptibility to persuasion in order to maximize (or minimize) 
the sum of opinions at equilibrium; such 
optimization translates into the network 
being pushed toward one or the other of the 
two extreme points in opinion space, 1 or 0. 

To follow Abebe's work, take the FJ model 
and a simple undirected graph 
(where simple means there is no self-loop in the 
network, i.e., $\mathbf{A_{ii}} = 0 $, and  $N_i = N_{\bar i}$). 
Let every agent put equal weight on all of its 
neighbors' opinions, i.e., $\mathbf{A}_{ij} = 1/|N_{\bar i}|, \: i \neq j$. 
Then the update rule for any given agent at time $t$ can be written as:
\[ o_i^{(t+1)} =  (1 - d_i) o_i^{(0)} + \frac{d_i}{|N_{\bar i}|} \sum_{j \in N_{\bar i}} o_j^{(t)}  \]

The system is known to have an equilibrium solution given by:
\[ \mathbf{o^*} = [\mathbf{I - DA}] ^ {-1} (\mathbf{I-D}) \mathbf{o^{(0)}}\]

Therefore, the objective, maximization of the sum of opinions, 
can be written as a maximization of 
\begin{equation}
f(\mathbf{o^{(0)}}, [d_1, d_2, \cdots, d_N]) = \langle \mathbf{1},  \mathbf{o^*} \rangle
 \end{equation}

 Abebe showed that this problem can 
 be solved in polynomial time provided that the 
 susceptibility to persuasion of all agents can be 
 modified. However, if the number of agents is 
 limited, the problem is NP-hard for which the 
 author provides a greedy algorithm. Chan et al.~\cite{Chan2019} 
 continued Abebe's line of work and claimed 
 that one of his findings is wrong. Chan also suggested an algorithm for use 
 with large graphs. Although the aforementioned 
 works examine the problem from an algorithmic 
 point of view using computer science, the approach 
 of changing the susceptibility to persuasion of 
 the agents in a network existed in psychology 
 before being utilized in the opinion dynamics community.
 
\subsection{Sequentially dependent topics}
Let us start this section with an example. Suppose 
you have built a machine that produces gears. 
The completed machine is making the gears, and 
now you want to use the gears to manufacture 
mechanical wrist watches. Now suppose that back 
at the beginning, before building the gear-making 
machine, you had not carefully considered the size 
of gears that would be necessary for watches, 
but you went ahead and built the machine anyway. 
Now, because you do not want to re-do everything, 
you keep the machine running, even though the 
gears it produces means the wrist watches have 
to be extra-large and clunky. This type of phenomena 
is known as \emph{path-dependency}. 

In reality, as in the FJ model, consensus may not occur
over a single topic. However, in psychology and path-dependence
theory it has been shown that consensus can occur over a sequence of
topics that are dependent (or if a topic is arising repeatedly.) 
 It has been shown that the connectivity of a social network is enhanced when a sequence
of dependent topics is considered by a network. The connection
between the \emph{influence network} and 
the \emph{network of initial opinions for successive topics}
can illuminate why consensus can occur for topics later in the sequence.
The quest for understanding was the motivation of Tian and Wang~\cite{TIAN2018213} 
when considering a sequence of successively dependent 
topics and the agents' related opinions/decisions (like the example above) 
and studying the conditions under which a community 
can come to consensus or form clusters of 
opinions for a sequence of topics.

Moreover, in the FJ model stubborn agents are unwilling to change their
opinions on a single topic. In path-dependence theory \emph{cognitive inertia} 
is defined as people's unwillingness to change their opinions over
a sequence of chain-dependent topics. Hence, the FJ model
for each topic is employed to study the opinion dynamics
of sequentially dependent topics, where the stubbornness factor
is equated with cognitive inertia. In a sequence of dependent
topics an agent’s initial opinion for topic $s+1$ is a function of 
(or a trade-off between) the agent's ``cognitive inertia'' and ``being social''.
In other words, each agent forms its initial opinion for topic
$s+1$ by making a tradeoff between its initial opinion for topic $s$
and the initial opinions of others about topic $s+1$.

\subsubsection{The model}

Let us start by pointing out that Tian~\cite{TIAN2018213} uses the
term \emph{interdependent} in his work. However, we prefer the term
\emph{sequential dependency} because the terms
\emph{interrelated, interdependent and coupled topics} have been used earlier,
and we believe that \emph{sequential dependency} or perhaps \emph{chain-dependency}
are more accurately descriptive for this scenario; 
Parsegov et al.~\cite{Parsegov7577815} discussed and 
defined topics that are interrelated or interdependent by: 
``Dealing with opinions on interdependent topics, the
opinions being formed on one topic are influenced by 
the opinions held on some of the other topics, so that 
the topic-specific opinions are entangled. \dots Adjusting 
his/her position on one of the interdependent issues, 
an individual might have to adjust the positions on several related issues 
simultaneously in order to maintain the belief system’s consistency.''
In addition, Noorazar~\cite{Noorazar2016} provided 
the following definition for coupling: ``Change of 
opinion about topic $s_k$ as a result of change 
of opinion about topic $s_\ell$ is called coupling.'' 

In accordance with the definitions given above, if an 
agent discusses topic $s_k$ (with another agent), and 
as a result changes its opinion about topic $s_k$, that agent will also change 
its opinion about coupled topic $s_\ell$ as well, even 
though topic $s_\ell$ was not discussed during the 
interaction with the other agent. 
For example, Alice and Bob may talk only about 
education, but as a result, their opinions about gun 
control may also change even though they did not 
mention gun control at all during their discussion. 
In this case it is as though the agent is moving on 
a manifold or surface, such that if the agent moves in the 
$x$ direction, it must also move in the $y$ direction 
in order to stay on the manifold. 
For this reason, we use \emph{sequential dependency} 
for the model proposed by Tian~\cite{TIAN2018213}.

Suppose there is a sequence of topics $s = 1, 2, 3, \cdots$, 
and we apply the FJ model to each topic:
\begin{equation}
\mathbf{o}^{(t+1)}(s) = \mathbf{DA}\mathbf{o}^{(t)}(s) +( \mathbf{I - D})\mathbf{o}^{(0)}(s)
\end{equation}

As before, $\zeta_i = 1-d_i$ will be the level of 
stubbornness, or \emph{cognitive inertia}, and 
$d_i$ will be the susceptibility to influence. This 
is where the FJ model and the path-dependency theory, 
in which agents exhibit stubbornness over a 
sequence of dependent topics, collide. For this reason, 
``cognitive inertia'', $\zeta_i$, is taken to be the 
same as the level of stubbornness, $1-d_i$. 

When presented with a topic $s+1$, agents will 
form an initial opinion about it; in order to do 
so, a given agent $i$ will minimize the following:
\begin{equation}\label{eq:tianOptimization}
C_i(\mathbf{o}) = \zeta_i ({o}_i - o_i^{(0)}(s)) + (1 - \zeta_i) \sum_{j=1}^N \mathbf{A}_{ij} ({o}_i  - {o}_j )^2
\end{equation}
The optimal solution, $x^\dagger$, satisfies
\begin{equation}\label{eq:tianOptimal}
(\mathbf{I - DA}) x^\dagger = \mathbf{(I-D)}\mathbf{o}^{(0)}(s)
\end{equation}

We need the following definition to present the results of
this section.

\begin{definition} 
In an SCC of digraph $\mathcal{G}$, if there exists a 
vertex that has parents belonging to 
other SCCs, we say that the in-degree 
of this SCC is nonzero; otherwise, 
we call it an independent strongly connected 
component (ISCC).
\end{definition}

\begin{prop}\label{tianFirstProp}
Eq.~\ref{eq:tianOptimal} has a unique solution if and 
only if there exists no non-stubborn ISCC in $\mathcal{G}(\mathbf{A})$.
\end{prop}

Proposition~\ref{tianFirstProp} allows 
the use of the initial opinion for topic $s+1$ as the
 limiting opinion of topic $s$, i.e., $\mathbf{o}^{(0)}(s + 1) = \lim_{t\to\infty} \mathbf{o}^{(t)}(s)$. 
 For such a scenario/system, the authors~\cite{TIAN2018213} 
 focused on the sequence of initial opinions of the topics and 
 defined the consensus states of the system in terms of the initial opinions: 
\begin{definition}
The system given by 
\begin{equation}
\begin{cases}
\mathbf{o}^{(t+1)}(s) = \mathbf{DA}\mathbf{o}^{(t)}(s) +( \mathbf{I - D})\mathbf{o}^{(0)}(s)\\\\
\mathbf{o}^{(0)}(s + 1) = \lim_{t\to\infty} \mathbf{o}^{(t)}(s)
\end{cases}
\end{equation}
is said to reach consensus if: $\lim_{s\to\infty}{o}^{(0)}_i(s) = o^* \in \mathcal{O}, \: \forall i$.
\end{definition}

Next we consider a topological condition that ensures the system 
reaches consensus in the sense defined above, 
\begin{theorem}\label{thm4graphdef}
Suppose there exists no non-stubborn ISCC, and there is no 
fully stubborn agent in the network. Then the system outlined 
above will reach consensus if and only if there exists a 
partially stubborn agent who has a directed path to any 
other partially stubborn agent.
\end{theorem}

This result, which hinges on the existence of an agent who 
can influence others directly or indirectly, is similar to what 
we have seen before in other systems. A corollary to the 
theorem above is that the system will form clusters 
if more than one ISCC is in the network and vice versa.\\

Another reasonable case, considered in Ref.~\cite{TIAN2018213}, 
has the initial opinion of a given agent $i$ for topic $s+1$ 
be a weighted average of those agents whose limiting 
opinions on topic $s$ are close enough to that of agent 
$i$’s opinion on topic $s$. This is the bounded 
confidence adaptation of the system above in 
which the initial opinion of agent $i$ for topic $s+1$ 
is influenced only by those who hold similar opinions 
to him about topic $s$. Results similar to those from 
the system above also hold true for this adaptation 
of the bounded confidence dynamic.

\subsection{Expressed vs. private opinions}

In many situations the expressed opinion of an agent
may be different from its candid belief, e.g., such as when a candidate is trying to capture
voters' attention. Such a
discrepancy between the private and expressed opinions
of people has been studied in psychological fields, for example
by Asch~\cite{Asch1951} whose work has motivated numerous 
researchers~\cite{ShengWen, huang2014, Medina2019} including Ye et al.~\cite{ye2019_EPO}. 
The expressed opinion of an agent is the result of pressure 
to conform to the average expressed opinion of the group 
the agent belongs to (\emph{local public opinion}),
or to conform to the group norm. And the private opinions
of the agents evolve under the influence of other agents,
as a function of their expressed opinions. Of course
agents have ``\emph{resilience}'' in the face of social pressure and therefore
their expressed opinions are different from their private opinions.

\subsubsection{The model}
Let the co-evolution of a 
given agent's private ($o_i^{(t)}$) and expressed ($\tilde o_i^{(t)}$) opinions is given by
\begin{equation}\label{eq:EXPPRU}
\begin{cases} 
\vspace{.1in}
o_i^{(t+1)} = d_i \mathbf{A}_{ii} o_i^{(t)}  +  \left ( d_i \sum_{j \neq i} \mathbf{A}_{ij}  \tilde o_j^{(t)} \right ) + (1 - d_i) o_i^{(0)}& \text{(based on FJ)} \\
\tilde o_i^{(t)} = \phi_i o_i^{(t)} + (1 - \phi_i) \sum_{j \in N_i} m_{ij} \tilde o_j^{(t-1)} 
\end{cases}
\end{equation}
\noindent where we have the following:
\begin{itemize}
\item As before, $\mathbf{A}_{ij}$ is the weight agent $i$ puts on agent $j$'s opinion.
\item $\mathbf{A}_{ii}$ is self-confidence (self-loops are allowed).
\item  $\mathbf{A}$  is row stochastic.
\item Updates are synchronous, i.e., all agents update simultaneously.
\item $(1- d_i)$ is the level of stubbornness and $d_i$ is the susceptibility to influence.
\item $m_{ij} > 0  \Leftrightarrow \mathbf{A}_{ij} > 0$, where $\sum_{j \in N_i}m_{ij} = 1$.
\item $\phi_i \in [0, 1]$ is agent $i$'s \emph{resilience} to the pressure to conform.
\item $\tilde o_i^{(0)} := o_i^{(0)}$
\end{itemize}
For such a system, the conditions under which 
the opinions converge to their limits exponentially 
fast has been studied. The conditions under which 
expressed opinions and private opinions reach 
constant values (i.e., consensus) have been examined as well. 
Ye et al. also considered the interesting 
case in which the expressed opinions and private 
opinions of agents, at the limit, reach a state 
of persistent disagreement at equilibrium that is 
caused by ``\emph{the presence of both 
stubbornness and pressure to conform}.'' 
The paper concluded with the application 
of such a system to Asch’s~\cite{Asch1951} experimental studies.
Let us consider the details more carefully. 

Define the vectors 
$\mathbf{\tilde o}^{(t)} = [\tilde o_1^{(t)}, \tilde o_2^{(t)}, \ldots, \tilde o_N^{(t)}]$, 
$\mathbf{o}^{(t)} = [o_1^{(t)}, o_2^{(t)}, \ldots, o_N^{(t)}]$. 
Re-write the influence matrix 
$\mathbf{A} = \mathbf{\tilde A} +  \mathbf{\hat{A}}$
where $ \mathbf{\hat{A}}$ is obtained 
by setting the diagonal of $\mathbf{A}$ to zero and
$  \mathbf{\tilde A}  = diag([a_{11}, a_{22}, \ldots, a_{NN}]) $. As in the FJ model
$\mathbf{D} = diag([d_1, \ldots, d_N])$, and let $\mathbf{\Phi} = diag([\phi_2, \ldots, \phi_N])$. 
We can write Eq.~\ref{eq:EXPPRU} in matrix form:
\begin{equation}\label{eq:EXPPRUMatrixForm}
\begin{bmatrix} 
\mathbf{o}^{(t+1)}  \\
\mathbf{\tilde  o}^{(t)}
\end{bmatrix}
= 
\mathbf{P} 
\begin{bmatrix} 
\mathbf{o}^{(t)}  \\
\mathbf{\tilde  o}^{(t-1)}
\end{bmatrix} +
\begin{bmatrix} 
(\mathbf{I} - \mathbf{D} ) \mathbf{o}^{(0)}\\
\mathbf{0}
\end{bmatrix}
\end{equation}
\noindent where $\mathbf{P}$ is a block matrix:
\begin{equation}\label{eq:blockMatrix}
\mathbf{P} = 
\begin{bmatrix} 
\mathbf{P_{11}} &  \mathbf{P_{12}}\\
\mathbf{P_{21}} &  \mathbf{P_{22}}
\end{bmatrix} = 
\begin{bmatrix} 
\mathbf{D(\tilde A + \hat A \Phi)} &  \mathbf{D} \mathbf{\hat A} (\mathbf{I - \Phi}) \mathbf{M}\\
\mathbf{\Phi} &  (\mathbf{I} - \mathbf{\Phi}) \mathbf{M}
\end{bmatrix} 
\end{equation}

 In Ref.~\cite{ye2019_EPO}, $\mathbf{\tilde o}^{(0)}$ is set to
$\mathbf{\tilde o}^{(0)} := \mathbf{o}^{(0)}$, 
though other choices are possible of course,
and we get $\mathbf{o}^{(1)} = (\mathbf{DA + I - D}) \mathbf{o}^{(0)}$.

Recall that a directed network is called strongly connected 
if there is a directed path between any pair of vertices.

\begin{definition}
A cycle is a path with equal starting and ending vertices,
and no other repeated vertices. 
\end{definition}

\begin{definition}
An aperiodic graph
is a graph in which the greatest common divisor of the lengths of
all of its cycles is 1.
\end{definition}

\begin{assumption}\label{ass3} Suppose the influence/weight/adjacency matrix 
$\mathbf{A}$ is stochastic, $\mathcal{G}[\mathbf{A}]$ is aperiodic and
$d_{i}, \phi_i \in (0, 1)$.
\end{assumption}
\begin{theorem}\label{Express1stTHM}
Suppose Asm.~\ref{ass3} holds and the agents' (expressed and private) opinions 
evolve according to Eq.~\ref{eq:EXPPRU}.
Then the system will converge, exponentially fast, to
its limit:
\begin{equation}\label{eq:EXPPRU}
\begin{cases} 
\vspace{.1in}
\lim_{t \to \infty} \mathbf{o}^{(t)} = \mathbf{o}^* = \mathbf{Ro}^{(0)}\\
\lim_{t \to \infty} \mathbf{\tilde o}^{(t)} = \mathbf{\tilde o}^* = \mathbf{So}^*
\end{cases}
\end{equation}
\noindent where $\mathbf{R = (I - (P_{11} + P_{12}S))}^{-1}\mathbf{(I - D)}$
and $\mathbf{S = (I - P_{22})}^{-1} \mathbf{P}_{21}.$
\end{theorem}
The theorem above shows that both expressed and private
opinions at the limit depend on the initial private profile. 
The initial expressed opinion
is forgotten. Hence, the choice of initial expressed profile
will change the trajectory while reaching the limit, but not
the final state. The matrices $\mathbf{R}$ and $\mathbf{S}$
are positive and stochastic, which means the final expressed and private
profiles are convex combinations of the initial private profile.
\begin{prop}\label{ExpPriProp1}
Suppose $\mathbf{A}$ is stochastic and the network
given by it is strongly connected and aperiodic.
Also assume $\phi_i \in (0, 1)$ and there is no stubborn
agent in the network ($d_i = 1$). Then the system
given by~\ref{eq:EXPPRUMatrixForm} converges
exponentially fast to a consensus value shared by both
private and expressed opinions: 
$\lim_{t \to \infty} \mathbf{o}^{(t)} = \lim_{t \to \infty} \mathbf{\tilde o}^{(t)} = o^* \mathbf{1}$
where $o^* \in \mathcal{O} = \mathbb{R}$.
\end{prop}

Let us examine the conditions that determine whether a discrepancy
 will exist between the private and expressed
profiles. Let $\mathbf{v}_{min} \:(\mathbf{v}_{max})$ denote the minimum (maximum)
of the vector $\mathbf{v}$.
\begin{theorem}\label{Express2nsTHM}
Suppose the assumptions of Thm.~\ref{Express1stTHM}
hold and the initial private profile is not at consensus, i.e.
$\mathbf{o}^{(0)} \neq \alpha \mathbf{1}$, for some $\alpha \in \mathbb{R}$.
Then we have:
\begin{equation}
\begin{cases} 
\vspace{.1in}
\mathbf{o}^{(0)}_{max} > \mathbf{o}^{*}_{max} > \mathbf{\tilde o}^{(0)}_{max}\\
\mathbf{o}^{(0)}_{min} < \mathbf{o}^{*}_{min} < \mathbf{\tilde o}^{(0)}_{min}
\end{cases}
\end{equation}
and $\mathbf{\tilde o}^{*}_{min} \neq \mathbf{\tilde o}^{*}_{max}$.
Furthermore, the set of initial profiles $\mathbf{o}^{(0)}$ for which 
exactly $m$ agents will have identical expressed and private opinions
at the limit, i.e. $o^{(*)}_k = \tilde o^{(*)}_k$,
lies in a subspace of $\mathbb{R}^N$ with dimension $n-m$.
\end{theorem}

Therefore, as long as stubborn individuals are in the 
network, a discrepancy will exist between expressed
and private opinions. We have seen before that if every agent
is maximally open, i.e. non-stubborn, then consensus will be reached.
Now we can conclude that without pressure to conform ($\phi_i=1$),
the agents' expressed and private opinion would be the same.
An interesting result of theorem~\ref{Express2nsTHM} is that
there is more agreement among the expressed opinions compared
to the private opinions, at the limit.

\subsection{Repulsive behavior in the DeGroot model}
A recent model that incorporates the repulsion property is Ref.~\cite{backfire2019}. 
Dandekar et al. \cite{Dandekar5791} modified the DeGroot model to account 
for bias among agents in the sense that agents will 
learn more from those whose opinions are closer to 
that of a particular agent. The modified equation is given by:
\begin{equation}\label{eq:dandekar}
o_i^{(t+1)} = \frac{w_{ii} o_i^{(t)} + (o_i^{(t)})^{b_i}}{ w_{ii} + (o_i^{(t)})^{b_i}   
\sum_{j \in N_{\bar i}}  w_{ij} o_j^{(t)} + (1 - o_i^{(t)} )^{b_i} ( \sum_{j \in N_{\bar i}}  w_{ij} -  \sum_{j \in N_{\bar i}}  w_{ij} o_j^{(t)})     }
\end{equation}
\noindent where $o_i$ is the level of support for 
opinion 1, $1-o_i$ is (consequently) the level of 
support for opinion 0, and $b_i \ge 0$ is the bias 
parameter. Hence, $(o_i^{(t)})^{b_i}$ is the weight 
given to neighbors supporting opinion 1 and 
$(1 - o_i^{(t)})^{b_i}$ is the weight given to neighbors supporting opinion 0.

Dandekar’s work motivated Chen et al.~\cite{backfire2019} 
to devise a model that supported both bias and 
repulsion, or ``backfire''. In Chen's model the opinion space 
is $\mathcal{O} = [-1, 1]$, and the opinions 
products are used to assign dynamical 
weights to the edges, as opposed to static 
ones. The opinion space is set to include 
negative as well as positive opinions, so that 
products of opinions could be either positive 
(for attraction) or negative (for repulsion). The 
weights on the $e_{ij}$ edge at any given 
time is given by $w_{ij}^{(t)} = 1 +  \beta_i o_i^{(t)} o_j^{(t)}$. 
The larger the value of parameter $\beta_i > 0 $ becomes, 
the greater the strength of both the bias and repulsion. 
The update rule is given by:
\begin{equation}\label{eq:backFire}
o_i^{(t)} = 
\begin{cases} 
\vspace{.2in}
\frac{w_{ii} o_i^{(t)} + \sum_{j \in N_{\bar i}} w_{ij}^{(t)} o_j^{(t)} }{ w_{ii}  + \sum_{j \in N_{\bar i}} w_{ij}^{(t)}  } & \text{,  } w_{ii}  + \sum_{j \in N_{\bar i}} w_{ij}^{(t)} > 0 \\
sgn(o_i^{(t)})  & \text{o.w.} \\
\end{cases}
\end{equation}
For $\beta_i \neq 0$, there are two cases:

\begin{itemize}
\item $w_{ij} < 0 $: In this case repulsion occurs: $\beta_i o_i^{(t)} o_j^{(t)} < -1$,
where $ o_i^{(t)} o_j^{(t)} < 0$, i.e., 
agents hold opinions with opposing signs in $\mathcal{O} = [-1, 1]$.

\item If $w_{ij} > 0 $, then bias assimilation occurs.
\begin{enumerate}
\item $\beta_i o_i^{(t)} o_j^{(t)} >0$: Both agents have 
either positive or negative opinions. In this case, agent 
$i$ takes agent $j$'s opinion more seriously if the 
level of agreement is high between the two.

\item  $-1 < \beta_i o_i^{(t)} o_j^{(t)} < 0$: opinions are 
opposed, but not too strongly. In this case agent $i$ 
assimilates the opinion of agent $j$, but to a lesser extent.
\end{enumerate}
If the update rule stated by Eq.~\eqref{eq:backFire} violates 
the boundaries of opinion space, it is clamped.
\end{itemize}

Consider two agents $i$ and $j$, where agent $j$ does 
not change its opinion. Then depending on the initial 
opinion of agent $i$ and the parameter $\beta_i$, $i$ 
can be attracted to $j$ or be repulsed by it so that $i$’s 
opinion ends up at either of the endpoints or it never 
changes its opinion, an unstable equilibrium similar to 
an unstable fixed point in dynamical systems. Let 
us take a look at a general case below.

\begin{theorem}\label{backfireTHM1}
Let $\mathcal{G}= (V, E)$ be any connected unweighted 
undirected graph. For all $i \in V$, $o_i^{(t)} \in (-1, 0) \cup (0, 1)$, $w_{ii}=1$, $\beta_i = \beta >0$. 
Let $ | \mathbf{o}^{(t)}|$ be the vector whose elements 
are absolute values of the opinion vector and 
$\min(\mathbf{o}^{(t)})$ be the minimum entry of $\mathbf{o}^{(t)}$. Then,
\begin{itemize}
\item If $\beta > \frac{1}{[\min(|\mathbf{o}^{(0)}|)]^2}$, 
then $\forall i \in V: |o_i^*| = 1$, i.e., polarization occurs.
\item If $\beta < \frac{1}{[\max(|\mathbf{o}^{(0)}|)]^2}$, 
then there exists a unique $o^* \in [-\max(|\mathbf{o}^{(0)}|), \max(|\mathbf{o}^{(0)}|) ]$ 
such that, $\forall i: |o_i^*| = o^* $, $o_i^*$ is the 
final opinion of agents as time goes to infinity.
\end{itemize}
\end{theorem}
\begin{prop}\label{prop1BackFire}
Let $\mathcal{G}= (V, E)$. Let $V = V_1 \cup V_2$ 
such that $ V_1 \cap V_2 = \emptyset$ where all 
agents in $V_1$ hold the same initial opinion
 $o_i^{(0)} = o^{(0)} \in (0, 1)$ and all agents 
 in $V_2 $ hold the same initial opinion 
 $o_i^{(0)} = - o^{(0)} \in (-1, 0)$, i.e., 
 the opinions are opposite in 
sign, but equal in absolute value. 
Moreover, let $w_{ii} = 1$ and $\beta_i = \beta > 0 $. Then:
\begin{itemize}
\item If $\beta > \frac{1}{(o^{(0)})^2}$, then, $\forall i \in V, |o_i^*| = 1$.
\item If $\beta = \frac{1}{(o^{(0)})^2}$, then the agents' opinions do not change over time.
\item If $\beta < \frac{1}{(o^{(0)})^2}$, then there exists a unique $o^* \in (-o^{(0)}, o^{(0)})$ s.t. $o_i^* = o^*$.
\end{itemize}
\end{prop}

\subsection{Managing consensus in the DeGroot model}
In the history of opinion dynamics to this day, researchers mostly
have been hunting the conditions under which consensus 
occurs, e.g. Ref.~\cite{Choi2019}. 
One interesting part 
that has been missing up to this point is the question of how to
prevent consensus, or perhaps how to manipulate
the system to reach a desired final state, 
whether that state is consensus or not. Such questions are
especially relevant in the era that we are now witnessing, which 
features the interference of various countries in 
other countries' elections. Because of the importance of the dynamics of interference,
researchers have recently begun to investigate 
such phenomena. The literature on interference is 
in its infancy, however, due to its importance we include a discussion on it here.

Dong et al.~\cite{Dong2017} studied how to 
manipulate a network to reach a desired consensus, 
and, if such manipulation is not possible, then how 
to manage the network to reach a given set of final opinions.
We will start by presenting a definition and looking at the result.

Recall that an agent who can influence
all other agents, directly or indirectly, is called a connected-agent (CA).

\begin{definition}{}
If agent $i$ is not a connected-agent 
it is a follower. 
\end{definition}

Denote the set of connected-agents and the set of followers 
by $V^{CA}$ and $V^{follower}$, respectively. Moreover, let $\mathbf{A}$ be 
the matrix of influence weights that  defines a directed graph. 

Consider the DeGroot model in which each agent 
has a positive self-weight for its own opinion, 
$\mathbf{A}_{ii} \in (0, 1)$, \emph{and} is influenced 
by at least one agent other than itself, and distributes 
equal weight among all of its neighbors; in other words, 
if $N_{\bar i}$ is the set of neighbors of agent $i$ 
(other than itself) who can influence it, then each 
neighbor's influence on agent $i$ is $\frac{1-\mathbf{A}_{ii}}{|N_{\bar i}|}$.
\begin{theorem}\label{dongManageThm1}
In the modified DeGroot model described above, 
agents will reach consensus if there exists at least one connected-agent. 
\end{theorem}

\begin{prop}\label{DongProp1}
If consensus is reached in this modified 
DeGroot model, the final opinion can be 
expressed as a combination of the initial opinions of the connected-agents.
( $o^* = \sum_{v_i \in V^{CA} } \lambda_i o_i^{(0)}$ where $\lambda_i \geq 0$.)
\end{prop}

Theorem~\ref{dongManageThm1} makes it 
clear that it is sufficient to have at least one connected-agent, (i.e., an agent that is reachable by other agents), 
in the network to reach consensus, 
 and the associated proposition suggests that it is possible to 
guide agents towards a particular opinion. The 
goal is then to add the minimal number of 
edges to the network to make the set of connected-agents nonempty. 
To achieve this goal, create a new graph 
$\hat{\mathcal{G}} = (V, \hat{E})$, obtained 
from ${\mathcal{G}} = (V, {E})$, where $E \subset \hat{E}$ such that $V^{CA} \neq \emptyset$.
\begin{equation}\label{eq:optProblem}
\begin{cases} 
\vspace{.2in}
\displaystyle{\minimize_{ \hat{E}} (| \hat{E} | - | E|)} \\
\text{subject to } E \subset \hat{E}\\
\hspace{.8in} V^{CA} \neq \emptyset\\
\end{cases}
\end{equation}

This optimization problem can be solved in two steps:
\begin{enumerate}
\item Form a partition $M$ of the network into 
subnetworks with the following properties:
\begin{itemize}
\item Each subgraph has at least one connected-agent.
\item The union of any pair of subgraphs has no connected-agent.
\end{itemize}

\item Add the minimum number of edges 
among the subgraphs in the partition to form $\mathcal{\hat{G}}$ (see Alg.~\ref{alg:addingEdges}).
\end{enumerate}
\begin{algorithm}[httb!]
\tiny
    \SetKwInOut{Input}{Input}
    \SetKwInOut{Output}{Output}
    \Input{Graph $\mathcal{G}$ and its partition $M$}
    \Output{A new graph  $\mathcal{\hat{G}}$ with connected-agents.}

   Let $\hat{E} = E$ and $\mathcal{G}^{(i)}(V^{(i)}, E^{(i)}) \in M$ be 
   any subgraph, let  $\mathcal{G}^*(V^*, E^*) = \mathcal{G}^{(i)}(V^{(i)}, E^{(i)})$ and $N=M \backslash \mathcal{G}^{(i)}$\\
 
    \While{$N \neq \emptyset$  }{%
    
       let $\mathcal{G}^{(\tau)}(V^{(\tau)}, E^{(\tau)})$ be any subgraph in $N$\\
      
       let $E^{add} = \{ (v_k, v_l)| v_k \in V_{\mathcal{G}^{\tau}}^{\text{CA}}, v_l \in V^* \}
       \cup \{ (v_m, v_n)| v_m \in V_{\mathcal{G}^{*}}^{\text{CA}}, v_n \in V^{\tau} \}
$\\
    let $e$ be any edge in $E^{\text{add}}$ and $\hat{E}=\hat{E} \cup \{e\}$
    (update $\mathcal{G^*}(V^*, E^*) $ as follows)\\
    $V^* = V^{(\tau)} \cup V*$ \\
    $E^* = E^{(\tau)} \cup E* \cup \{e\}$ \\
    $N = N / \mathcal{G^{(\tau)}}$
      }
      
       return $\mathcal{\hat{G}}$
    \caption{\small Adding Edges}
    \label{alg:addingEdges}
\end{algorithm}
\begin{theorem}\label{dongManageThm2}
The $\mathcal{\hat{G}}(\hat{V}, \hat{E})$ obtained via Alg.~\ref{alg:addingEdges} is 
the optimal solution to the optimization problem given by~Eq. \ref{eq:optProblem} 
and $|\hat{E}| - |E| = |M| - 1$, where $M$ is the partitioning of $\mathcal{G}$.
\end{theorem}

The paper~\cite{Dong2017} also proposed a network modification that  
consisted of adding edges so that the final state lies within a target 
interval $o^* \in [o^*_l, o^*_r]$. Perhaps future work will examine 
how to prevent a community from reaching consensus, 
such as what occurred in the 
most recent American presidential election.

\subsection{A general stabilization condition}
Recall that in the DeGroot~\cite{DeGroot1974} model 
each agent trusts the other agents by a fixed amount,
and consequently the adjacency matrix $\mathbf{A}$ is fixed.

Lorenz~\cite{LORENZ2005217} defined a 
model with an evolving weight (adjacency) 
matrix in which the entries (i.e., the level at which agents trust others) 
are a function of both time and current profile.

Fixing the weight matrix of Lorenz's model will
result in the DeGroot model. One can also define the weight matrix
in Lorenz's model so that the model collapses to a bounded-confidence model.
As such, we use his work as a stepping stone to go from
DeGrootian models to the bounded-confidence models in the next section.

The result of Ref.~\cite{LORENZ2005217} 
is given below, after the introduction of some notation.
Let $\mathbf{A}^{(t)}(\mathbf{o}^{(t)})$ be a stochastic adjacency 
weight matrix at time $t$ that is a function of both time and 
the opinion profile at time $t$, $o^{(t)}$. For simplicity we 
refer to this matrix as $\mathbf{A}^{(t)}$. Define the update rule by:
\begin{equation}\label{eq:lorenzUpdateRule}
\mathbf{o}^{(t)} = \mathbf{A}^{(t-1)} \mathbf{o}^{(t-1)} = \mathbf{A}^{(t-1)} \mathbf{A}^{(t-2)} \mathbf{o}^{(t-2)} = \cdots = \mathbf{A}^{(t-1)} \mathbf{A}^{(t-2)} \cdots \mathbf{A}^{(0)}  \mathbf{o}^{(0)}
\end{equation}

 Denote the last term in Eq.~\eqref{eq:lorenzUpdateRule} by 
$\mathbf{A}(0, t) := \mathbf{A}^{(t-1)} \mathbf{A}^{(t-2)} \mathbf{A}^{(t-3)} \cdots \mathbf{A}^{(0)}$ or more
generally 
$\mathbf{A}(t_0, t_1) := \mathbf{A}^{(t_1-1)} \mathbf{A}^{(t_1-2)} \mathbf{A}^{(t_1-3)} \cdots \mathbf{A}^{(t_0)}$. 
Using this notation we can compactly write $\mathbf{o}^{(t)} = \mathbf{A}(0, t) \mathbf{o}^{(0)}$. 
The following theorem is provided by~\cite{LORENZ2005217}.
\begin{theorem}\label{lorenzTheorem}
  Let $A^{(t)}$ be the adjacency matrix of the Lorenz model defined above
  for a given network $\mathcal{G} = (V, E)$. If for any $t$
  the matrix satisfies the following conditions:
   \begin{itemize}
   \item $\mathbf{A}_{ii} > 0$
   \item $\mathbf{A}_{ij} > 0 \iff A_{ji} > 0$
   \item $\exists \: \epsilon \: \forall \mathbf{A}_{ij} \neq 0,  \: s.t. \: \mathbf{A}_{ij} > \epsilon$
    \end{itemize}

\noindent then there exists a time $t_0$ and a pairwise disjoint subgroups of agents $S_i$
    such that
    $\cup_{i=1}^p S_i = V$
    and 
\begin{equation}
\lim_{t \to \infty} \mathbf{A}(0, t) = diag(\mathbf{K}_1, \mathbf{K}_2, \ldots, \mathbf{K}_p) \mathbf{A}(0, t_0)
\end{equation}
\noindent where $diag(\mathbf{K}_1, \mathbf{K}_2, \ldots, \mathbf{K}_p)$  
is a diagonal block matrix with blocks $\mathbf{K}_i \in \mathbb{R}^{|S_i| \times |S_i|}$. Moreover, each $\mathbf{K}_i$ has identical rows.
\end{theorem}

Theorem~\ref{lorenzTheorem} states that if the adjacency matrix that assigns 
weights for the opinions of neighbors satisfies the three conditions, then 
for any initial set of opinions, the network will end up with separate 
clusters in which the agents of each subgroup come to consensus.
 In each of the block diagonal matrices each entry will be greater 
 than zero, i.e., all agents within a subgroup will talk to any other 
 agent in the group. This is one of the early interesting analytical 
 results that motivated additional studies, including 
 a number of the following modifications.


\section{Bounded confidence models}\label{BCMSection}
One of the most famous types of model in the 
field is the so-called ``bounded confidence''‚ 
model in which agents are influenced by those whose 
opinions‚ are close enough to their own. This 
modification is justified by the homophily observation 
and the tendency that ``birds of a feather flock together.''
In the model of Deffuant~\cite{Deffuant2000} (the DW model) 
the interactions are binary, for which Monte Carlo-driven 
results are in abundance. However, due to the 
model's nonlinearity, theoretical results are scarce. 
We could mention Ref.~\cite{Fortunato2004} as an example 
of simulation-driven work in which for a homogenous 
(i.e., all agents share the same level of confidence) 
DW model it has been demonstrated that the 
confidence radius $0.5$ is the limit above which 
consensus occurs for a variety of network topologies. 
Lorenzo~\cite{lorenz2010heterogeneous} has 
shown that if two types of agents‚ open-minded 
and closed-minded‚ with two different confidence 
radiuses are in the same network, then, 
counterintuitively, the final state will be consensus 
for confidence levels below the critical radius 
(of homogenous systems) of 0.5. According 
to simulations, the critical value between 
polarization and consensus is 0.27 for the DW model
and 0.19 for the HK model.

Later, Hegselmann and Krause~\cite{Hegselmann2002} 
established the model for synchronous updates
 for which theoretical results have emerged at a
 higher rate. Recently, the publication rate for 
 analytical results for minor modifications to 
 such models has increased.
 \subsection{Power evolution in a synchronous bounded confidence model}
 We start this section by studying power evolution in a bounded confidence model
 to have a consistent pattern with the last section. However, this study is a simulation based
 for which analytical results do not exist.

New approaches and concepts have been introduced 
in Ref.~\cite{Brede2019}, with agents having the chance to 
change their connectivity in the digraph to maximize their 
influence. For example, when the opinion space is set to $\mathcal{O} = [-1, 1]$, 
the influence of agent $i$ is defined as $I_i = |o_i - \bar o|$ where $\bar o$ 
is the average of all the agents' opinions. $I_i = 0$ 
indicates that the entire network is in total agreement 
with agent $i$, who has maximum influence; likewise, $I_i = 2$ 
indicates total disagreement with agent $i$. In such 
a scenario, agent $i$'s goal would be to maximize 
its influence in the network. The dynamics of the model 
in Ref.~\cite{Brede2019} are given by the repetition of two alternating steps:
\begin{enumerate}
\item Update the opinions of agents synchronously
\item Rewire the network to maximize the influence of a particular agent
\end{enumerate}
The first step is given by:
\begin{equation}\label{eq:bredeEq}
o_i^{(t+1)} =  o_i^{(t)} + s \eta_i + \mu \sum_{j \in N_i} \mathbf{A}_{ij} d_{ji}
\end{equation}
\noindent where $s$ is a parameter that strengthens or weakens 
the effect of random external noise $\eta_i$, and as before $N_i$ 
is the set of agent $j$s whose opinions fall within the confidence 
radius of agent $i$, that is, $|d_{ji}| = |o_j^{(t)} - o_i^{(t)}| < r$. The 
reason for incorporating $ \mathbf{A}_{ij}$ above is that the network 
is not a complete graph, but a directed graph which may be incomplete. 
The second step involves rewiring, which is accomplished in the 
following way: $m$ agents are chosen randomly and each of them 
will consider rewiring (if all do decide to rewire, $m$ rewirings will take place). 
Then each agent (such as agent $i$) from among the $m$ agents,
chooses an agent out of all agents which are not currently its neighbor 
(suppose agent $k$ is chosen by agent $i$ and they are not currently neighbors).
Then agent $i$ predicts its influence (using Eq.~\ref{eq:bredeEq}) in
 the new topology where $i \rightarrow k$, without taking into account
 the external noise. If agent $i$’s influence has increased, then
 rewiring takes place. Simulations for the above scenario can 
 be divided in two sub-categories: \emph{endogenous} and 
 \emph{exogenous}. In the endogenous case, a large fraction 
 of the agents in the network participate in the rewiring step 
 that maximizes influence. In the exogenous case, external 
 sources compete to maximize their influence on the network, 
 while they themselves cannot be influenced by members of 
 the network. Simulations can be run for different confidence 
 radiuses $r$ and for various rewiring schemes. 
 \emph{The main results are that: for small confidence 
 radiuses, the population will be polarized into two 
 sub-communities of comparable size; for medium-sized 
 confidence radiuses, one of the two sub-communities 
 will be significantly larger than the other; and for large 
 confidence radiuses, consensus will be achievable.} For more discussion of randomly 
 changing topologies, please consult the references given in Ref.~\cite{Brede2019}.

\subsection{Convergence and convergence speed of heterogeneous (in confidence level) DW model}

In the work of Ref.~\cite{Chen2019} a version of DW is considered
in which each agent had its own confidence bound. From a probabilistic 
standpoint, it is shown that when the DW model is heterogeneous in
confidence levels, the network will reach a final state in which
any pair of agents either are in agreement, or the distance between them
is greater than the confidence radius of the two. This means
that if there are two communities, $C_1$ and $C_2$ in the network,
then $C_1$ is at the consensus state, $C_2$ is at consensus state,
and the distance between the two clusters, 
(or distance between the opinions of two group), is greater than
the confidence radius of the most open-minded agents
in the two communities. 

\subsubsection{The model}

Let us start this section with a definition.

\begin{definition}{}
The system almost surely comes to consensus if
\begin{equation}
 p \left( \lim_{t \to \infty} o_i^{(t)} = o^* \right) =  1, \: \forall i
\end{equation}
where $o^* \in \mathcal{O}$. Similarly it will almost surely converge if
$ p \left( \lim_{t \to \infty} \mathbf{o}^{(t)} = \mathbf{o}^* \right) =  1$.
\end{definition}
Chen et al.~\cite{Chen2019} studied a 
\emph{heterogeneous (in confidence radius)} 
DW model, i.e., each agent has its own 
confidence interval, and the learning rate 
is homogeneous.

Let $\mathcal{N} = \{(i, j) | \: i, j \in \{1, 2, \cdots, N\}, i<j \}$ 
be the set of pairs that is used to select a 
random pair to interact at any time. Then 
the update rule is defined by
\begin{equation}\label{eq:ChenheterogenousBCM}
\begin{cases} 
\vspace{.2in}
o_i^{(t+1)} = o_i^{(t)} + \mu \:  . \:  \mathbbm{1}_{[0 , r_i]}(| d_{ji}^{(t)} |)  \:  . \: d_{ji}^{(t)}\\
o_j^{(t+1)} = o_j^{(t)} + \mu \:  . \:  \mathbbm{1}_{[0 , r_j]}(|d_{ij}^{(t)}|)  \:  . \:  d_{ij}^{(t)}\\
\end{cases}
\end{equation}
\noindent where $\mu$ is the learning rate 
and $d_{ji}^{(t)} = o_j^{(t)}  - o_i^{(t)} $ is 
defined as before. In Chen et al.'s work 
$\mu=\frac{1}{2}$, and the introduction 
of $\mathbbm{1}_{[0, r_i]}(| d_{ji}^{(t)} |)$ 
follows Chen et al.'s notation and can be dropped
 with the understanding that this is in fact 
 a bounded confidence model with updates 
 that take place only if agents fall within each 
 other's confidence intervals. Order the agents 
 so that the confidence radiuses of the agents are 
 decreasing, i.e. $r_1 \ge r_2 \ge \cdots \ge r_n > 0$. (We will make use of this ordering later.)
\begin{theorem}\label{almostSureConv}
  Consider the heterogeneous BC model whose update rule 
  is given by Eq.~\ref{eq:ChenheterogenousBCM}. Assume each agent
  has a positive confidence radius and let the interactions be performed
  in a randomized pairwise fashion at any given time where all $N$
  agents of the network $\mathcal{G}$ are fully connected.
  Then we have
\[\forall \mathbf{o}^{(0)} \in [0, 1]^N, \: \exists \: \mathbf{o}^* \in [0, 1]^N \:\: \text{such that }\]
\begin{itemize}
\item $\lim_{t\to\infty} \mathbf{o}^{(t)} \xrightarrow {\text{almost surely}} \mathbf{o}^*$
\item $\forall i \neq j, \:o^*_i = o_j^*$ or $\abs{o^*_i - o_j^*} > \max\{r_i, r_j\} $
\end{itemize}
\end{theorem}

\begin{corr}\label{almostSureConsensus} 
If the assumptions of Thm.~\ref{almostSureConv} 
are met and one of the confidence bounds is 
greater than or equal to 1, then the network 
will almost surely reach consensus for any initial profile.
 \end{corr}
  The convergence rate is established by:
\begin{theorem}\label{ChenConvRate}
 Let the opinion dynamic system be given 
 by Eq.~\ref{eq:ChenheterogenousBCM} 
 with positive confidence radiuses. Then 
 for any initial state $\mathbf{o}^{(o)} \in [0, 1] ^ N$ 
 there exist $c \in \mathbb{R}$ s.t. \[ E \left[\sum_i^N (o_i^{(t)} - o_i^*)^2 \right]= O(exp(-ct))\]
\noindent where $E[.]$ denotes the expected value.
\end{theorem}
\subsection{A more inclusive bounded confidence model}
In the DW version of bounded confidence, if Alice and Bob are 
chosen at time $t$, they will interact if their opinions are close enough.
In the HK model, Alice talks to all of her neighbors whose
opinions are close enough. In the two bounded confidence models of 
Zhang and Hong~\cite{Zhang2013} the situations are different. 
In the first version, at a given time Alice chooses several
agents and updates her opinion using a subset of the chosen
agents. The subset includes only the agents whose opinions are close enough to her own.
In other words, the agents whose opinions are too far away are omitted from
the update. The analytical results for this case also
apply to the DW model since the DW model is a special case of the HK model. 
In the second version, Alice chooses several agents and computes
a weighted average of their opinions, and if this value falls
within her confidence level, then she uses it to update her opinion. 
The analytical result for the first scenario is that as time goes
to infinity, any two agents are either in agreement
or the distance between them is greater than the confidence radius $r$
which is shared among all agents. 
In the second scenario it is shown that as the confidence 
radius increases, consensus will occur more often.

Let us define the two scenarios more formally below.
\subsubsection{The model}
The first scenario is called
Short-range Multi-choice DW (SMDW), in which agent $i$
has its own \emph{choice number $c_i$}. At a given time $t$
agent $i$ chooses $c_i$ agents randomly, removes
those whose opinions are too far from that of its own, and then
updates its opinion by a weighted average of the opinions of the agents:
\begin{equation}\label{eq:SRMC}
o_i^{(t+1)} = o_i^{(t)} +  \mu_i \: . \sum_{j=1}^{c_i}\: w_{ij} \mathbbm{1}_{[0 , r]} (|d_{ji}^{(t)}|) \: . \: d_{ji}^{(t)}
\end{equation}
\noindent where $c_i$ is the number of 
agents that agent $i$ selects to learn from. 
The influence weights $0 < w_{ij} \leq 1$ add up to 1 
and they determine how much agent $i$ is
 influenced by each agent $j$'s opinion. Note that in Eq.~\ref{eq:SRMC} 
 only those agents are taken into account whose opinions are close
 to that of agent $i$.
 
The second scenario, the Long-range Multi-choice (LMDW), is introduced below. 
Agent $i$ first chooses several agents to learn from. 
Then the overall weighted average of the opinions of the chosen 
agents are computed, and if this value falls within 
the confidence interval of agent $i$, then an update 
occurs; otherwise nothing happens. This arrangement 
allows agent $i$ to be influenced by some agents 
whose opinions would not have fallen within agent 
$i$'s confidence interval if they had participated in 
private interactions; in other words, opinions that fall
outside of the confidence radius of agent $i$ can be influential now:
\begin{equation}\label{eq:LRMC}
o_i^{(t+1)} = o_i^{(t)} +  \mu_i \: . \: \mathbbm{1}_{[0 , r]} ( | y_i^{(t)} - o_i^{(t)} | ) \: . \: (y_i^{(t)} - o_i^{(t)})
\end{equation}
\noindent where $y_i = \sum_{j=1}^{c_i} w_{ij} o_j^{(t)}$. 
Let us examine the results and implications.

\begin{theorem}\label{zhang2013_theorem_1}
(SRMC Thm.)
Let $\mathcal{G}$ be a fully connected graph with 
$N$ nodes. Let $r$ be the confidence radius for 
all agents whose interactions are governed by 
the Short-range Multi-choice Eq.~\ref{eq:SRMC}; 
then, for any initial profile of the network given 
by $\mathbf{o}^{(0)} \in [0, 1]^N$, one of the 
two following results will almost surely hold true for any pair of agents:
\begin{enumerate}
\item $\lim_{t \to \infty} d_{ij}^{(t)} = 0$
\item $\lim_{t \to \infty } |d_{ij}^{(t)}| > r$
\end{enumerate}
\end{theorem}

\begin{theorem} 
\label{zhang2013_theorem_2}
(LRMC Thm.)
Let $\mathcal{G}$ be a fully connected graph with $N$ nodes.
Let $r$ be the confidence radius for all agents whose interactions
are governed by the Long-range Multi-choice Eq.~\ref{eq:LRMC}. Furthermore,
assume $c_i > 1$, for all $i$. If \[ r \ge \max_{1 \leq i \leq N} \min_{1 \leq j \leq c_i} w_{ij} \] 
then consensus can almost surely be reached.
\end{theorem}

Another paper that derives theoretical properties of 
opinion dynamics that are similar to those of Ref.~\cite{Chen2019} 
which is a minor modification of Ref.~\cite{Zhang2013} 
is Ref.~\cite{Zhang2019Plos}. In Ref.~\cite{Zhang2019Plos}, 
the LRMC Eq.~\ref{eq:LRMC} is modified so that 
all agents choose the same number of agents, $c$, to 
look up to and the weights $w_{ij}$ are all equal to $1/c$. 
The agents chosen to learn from can be chosen with 
replacement. Formally, the update rule is given as follows:
\begin{equation} \label{eq:Zhang2019-rule}
o_i^{(t+1)} = o_i^{(t)} + \mu \cdot \mathbbm{1}_{[0 , r]}(|y_i^{(t)}|)  \:  y_i^{(t)} 
\end{equation} 
\noindent where 
\begin{itemize}
\item $y_i^{(t)} = \frac{\mathlarger{\sum_{j=1}^c o^{(t)}_{\mathcal{J}(i, j, t)}}}{c} - o_i^{(t)}$
\item $\mathcal{J}(i, j, t)$ is the index of the agent selected by agent $i$ at its $j^{th}$ selection at
time $t$.
\item $c$ is a constant.
\item The confidence radius $r$ and the learning rate $\mu$ both are in the interval $(0, 1)$.
\end{itemize}

After the following definitions we will be able to 
represent the final result of Zhang et al.'s paper.
\begin{definition}{}
Let $o_{[i]}^{(t)}$ be the opinion of the agent whose 
opinion at time $t$ is the $i^{th}$ largest opinion,
 i.e. the $i^{th}$ opinion when we order 
 opinions: $o_{[1]}^{(t)} \leq o_{[2]}^{(t)} \leq \cdots \leq o_{[N]}^{(t)}$. 
 Then we can define $D_{[i, i+1]}^{(t)} = o_{[i+1]}^{(t)} - o_{[i]}^{(t)}$ 
 and the opinion range at time $t$ by $\Delta^{(t)} = o_{[N]}^{(t)} - o_{[1]}^{(t)}$.
\end{definition}

The interesting result of this generalized model is 
obtained by stepping foot into the world of probability . 
The following theorem puts a lower and upper bound 
on the probability of consensus as a function of network 
population, $N$, confidence radius, $r$, and the selection parameter, $c$.
\begin{theorem}\label{zhang2019_theorem}
For the model defined by Eq.~\ref{eq:Zhang2019-rule}, let the 
population of the network be $N$, and the confidence 
radius $r$ be smaller than $\frac{1}{c}$, where $c$ is the
 selection parameter. Then the lower and upper bounds 
 for the probability of convergence are given by:
\begin{equation}
  N(c r)^{N-1} - (N-1)(cr)^N \leq p(\lim_{t\to\infty} \Delta^{(t)}=0) \leq 
  \begin{cases} 
     N!(cr)^{N-1} & \text{if $N \leq \ceil{\frac{1}{cr}} $,} \\
    N!(cr)^{N - \floor{\frac{1}{cr}}} &\text{o.w.} \\
  \end{cases}
\end{equation}
\end{theorem}

The convergence of $\Delta^{(t)}$ to zero is the convergence 
of the population to consensus, and if the assumptions outlined 
above are met, then we have $0 <  \ell \leq p(consensus) \leq u < 1$, 
where $\ell$ and $u$ are the lower and upper bounds, 
respectively, defined in Thm.~\ref{zhang2019_theorem}. 
Although computing the exact probability is impossible, a simple 
experiment is used to support the result. 
\subsection{A somewhat different bounded confidence model}
\begin{figure}[httb!]
  \centering
  \includegraphics[width=.4\linewidth]{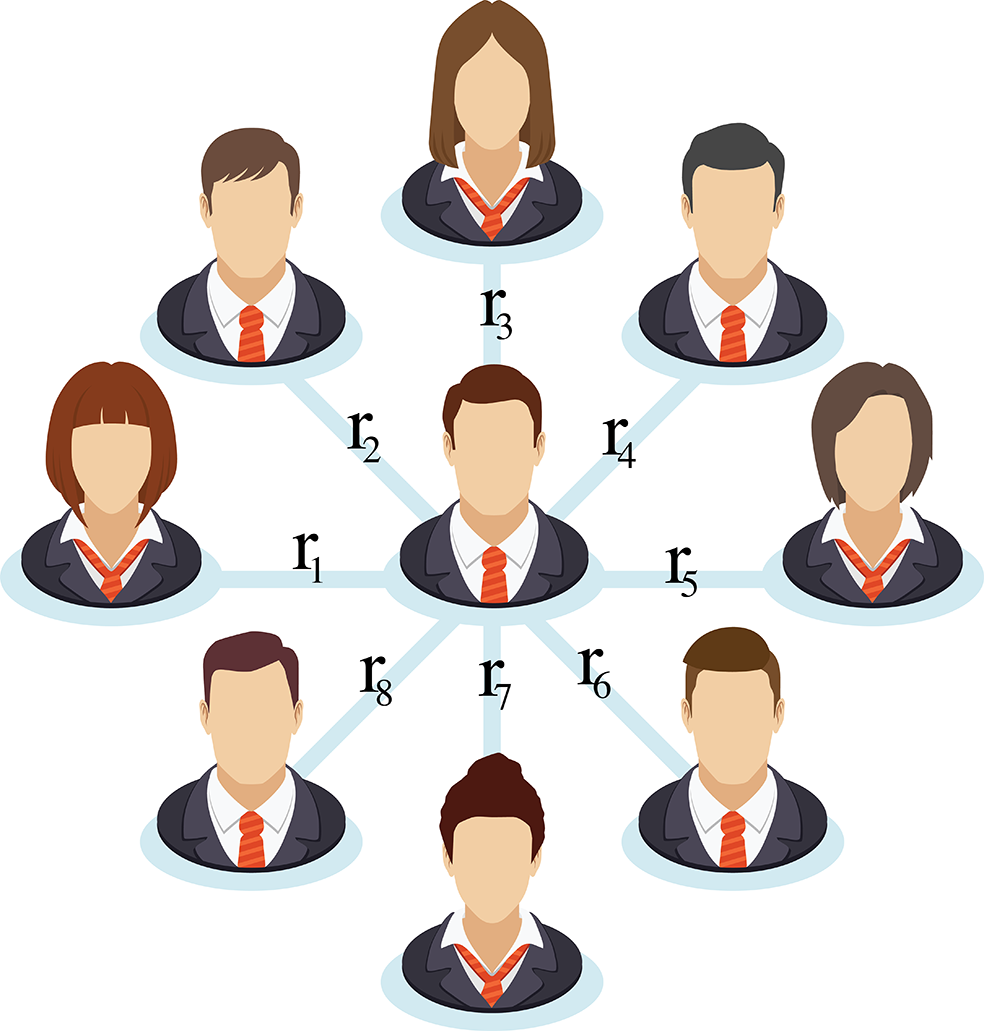}
\caption{Edge dependent confidence intervals. 
In this case a given agent, like Alice, has more than 
one confidence interval—she has one per neighbor. 
She trusts different people differently.}
\label{fig:edge_conf}
\end{figure}

In what we have seen before in BC models, 
any agent, like Alice, has \emph{only} one confidence radius,
Alice trusts everyone equally.
However, in this section we have a DW model in which
a given agent has more than one confidence radius.
More precisely, Alice trusts her friends by different amounts (See Fig.~\ref{fig:edge_conf}).
These confidence radiuses are assigned to edges like $e = (i, j) = (\text{Alice, Bob})$, i.e., relationships,
via a random (Poisson) process. The confidence assigned
to the edge, $e$ connecting Alice and Bob is denoted by $r_e =  r_{\text{Alice-Bob}}$. 
The paper~\cite{Shang2014} that introduces this model starts by 
presenting analytical results for a 1-dimensional lattice (i.e., each
agent only has two friends); this is followed 
by simulation results obtained from 
applications to ring and Barab{\'a}si-Albert networks.
The (analytical) result based on the 1D lattice is 
that $0.5$ is a critical value, in the sense that 
if the expected value of the confidence 
radiuses is below $0.5$, then, with a probability of 
one, any two adjacent agents are either at consensus 
or their opinions are far apart (by more than the confidence 
radius assigned to their relationship). 
And if the expected value of the confidence levels is 
more than $0.5$, then with probability one
all agents will come to consensus with all agents 
having opinions of $0.5$.
The simulations for ring and BA networks also suggest
that $0.5$, as the expected value of the 
confidence radius, is a critical boundary
 between disordered and ordered phases 
 of the system: the ordered phase of the 
 network is in a consensus state, and the 
 disordered phase lacks consensus.
 
 The details are formally presented below:

\subsubsection{The model}
In this section we discuss a somewhat different
bounded confidence model~\cite{Shang2014} that is based 
on the DW model. However, in this case 
\begin{itemize}
\item The system is homogeneous in learning rate $\mu$.
\item In this system confidence intervals are handled differently:
every edge can have a different confidence interval.
\item The system updates are pairwise.
\end{itemize}

In spite of this model using a 1-dimensional lattice network,
 whose agents sitting on the $\mathbb{Z}$ line, 
this modification of the DW model is interesting 
and will be discussed below. 
Originally Lanchier~\cite{Lanchier} proposed the model 
and geometrically proved that if the confidence radius 
is greater than 0.5 consensus will be achieved. Later 
H{\"a}ggstr{\"o}m~\cite{Haggstrom2012} re-proved 
the conjecture, in a probabilistic framework, 
that was put forward using simulations. 
The aforementioned works are based on a 
homogenous network in which all agents share 
the same confidence level; however, in this section 
we focus on a heterogenous version~\cite{Shang2014} 
where the heterogeneity is implemented via a Poisson process.
 
Let $\mathbf{o}^{(0)}$ be the initial profile. For 
a given edge $e$, a random unit rate Poisson 
process is assigned that governs the interaction 
time as well as an i.i.d random variable $r_e$ 
whose values belong to $(0, 1)$. Moreover, 
let the confidence radius $r$ be a random 
variable with the same distribution as $r_e$. 
Denote the opinion of a given agent, right 
before an interaction at time $t$, by 
$o_i^{(t_-)} = \lim_{t \to t-} o_i^{(t)}$. 
Then the update rule is given as follows 
(similar to the discrete-time DW model):
\begin{equation}\label{eq:shangUpdate}
\text{if} \: \:  |o_j^{(t_-)} - o_i^{(t_-)}| < r_e : 
\begin{cases} 
o_i^{(t)} = o_i^{(t_-)} + \mu (o_j^{(t_-)} - o_i^{(t_-)}) \\
o_j^{(t)} = o_j^{(t_-)} + \mu (o_i^{(t_-)} - o_j^{(t_-)}) \\
\end{cases}
\end{equation}

Please note that the confidence radius $r_e$ is 
assigned to an edge, i.e., each agent will have a 
different confidence radius for each neighbor.
For a given edge $e_{ij} = \{i, j\}$ 
we denote $r_i = r_e$. The graph $\mathcal{G}$ is 
the set of integers $\mathbb{Z}$ unless otherwise stated. 
\begin{theorem}\label{shangTheorem}
Let an opinion dynamic be given by Eq.~\ref{eq:shangUpdate}, 
with learning rate $\mu \in (0, 0.5]$ and with $r_i$ 
assigned as random confidence radiuses for all $i \in \mathbb{Z}$, then:
\begin{itemize}
\item if $E[r] < 0.5$ then $\forall i, \: p( \lim_{t \to \infty } | d_{i, i+1}^{(t)} | \in S ) = 1$, 
where $d_{i, i+1} = o_i^{(t)} - o_{i+1}^{(t)}$ and $S = \{ 0 \} \cup [r_u, 1]$.
\item if $E[r] > 0.5$ then $\forall i,  \: \lim_{t \to \infty } o_i^{(t)} = 0.5$.
\end{itemize}
\end{theorem}

The above theorem states that neither the final profile of 
the network nor the critical confidence radius depends 
on the distribution of $r$ except for its expectation. 
Similar to the homogenous case, if $E[r] > 0.5$ 
consensus is observed, and below that threshold fragmentation 
occurs, with the distance between neighboring communities 
connected via edge $e=e_{i, i+1}$ which is greater than $r_e$.

The extensive simulations presented in Ref.~\cite{Shang2014}, 
show that these results hold for graphs that include the 
ring- and scale-free topology of Barab{\'a}si with 
confidence radius parameters drawn from truncated 
normal distributions and from Beta distributions, 
each with a few parameters. 
These results agree with homogenous cases~\cite{Fortunato2004}.

\subsection{Noise in Bounded Confidence Models}
Even though humans are 
capable of rational thought, 
they are still essentially emotional beings. 
A person may accept or reject the exact 
same idea depending on who is promoting it. 
In the previous sections of this paper the exchange of opinions among agents 
has taken place as though the agents are 
following a set of well-defined rules in a closed and 
isolated space, unaffected by external forces,
 or even by internal thoughts. However, such 
 a scenario cannot accurately represent how 
 opinions are exchanged in real life, where 
 people are capable of changing their minds 
 based on their own internal thought processes 
 (without one-to-one interactions with others), 
 or by reading (similar to a unilateral interaction 
 in which one party does not change its mind). 
 Moreover, in a single interaction the transmission 
 of opinions is not absolutely perfect. 
 Individuals usually find their own ways to be 
 unique and distinct from the population as a 
 whole. Such disintegrating actions can as a first approximation be 
 represented by noise.

Hence, despite the lack of research on 
disintegrating forces or noise in social networks, 
especially with an analytical focus, we include 
such sections to emphasize its importance. 
The variation seen in opinion dynamics models 
is a testament to the difficulty of deriving specific 
results while using a given model; therefore, 
researchers modify the models of others to 
obtain desired results or to apply the models
 to scenarios that arise in their own fields. 
 Therefore, not all modifications or papers 
 presented in this section follow the 
 same line of work. Let us start with 
 analytical results for the case of 
 noise added to the DW model.
 
\subsubsection{The Noisy DW model}

In order to eliminate the unrealistic sharp boundary 
between interacting and being indifferent in 
the bounded confidence of the DW, 
Grauwin and Jensen~\cite{Grauwin2012} introduced a 
probabilistic interaction schema to the DW model 
in which two agents interact with some probability 
that depends on their difference of opinion, 
\emph{interaction noise}. This allows an agent 
the potential to interact with an agent whose 
opinion falls outside its confidence radius and 
to ignore an agent whose opinion falls within its confidence radius.
A second type of noise is also considered by
Grauwin and Jensen~\cite{Grauwin2012} that models death and birth of
humans; at time $t$ an agent's opinion randomly changes to a
random number, they refer to it by \emph{turn over}.

It has also been shown that \emph{dynamic and stable clusters} can emerge
in this model, as opposed to ``frozen'' clusters. Moreover, the authors 
claim that this noise is more natural than the one introduced by 
M{\"{a}}s et al.~\cite{Mas2010} that is ``specifically tailored to prevent consensus.''

It is shown that the introduction of 
interaction noise in the DW model (in the absence of turn over)
causes consensus. The DW model with the additional ingredient
of the turn over (in the absence of interaction noise) shows a different
range of behaviors depending on the death/birth probability, 
changes in which cause a progression from an ordered 
phase to a disordered one. 
In the case that both interaction noise and turnover exist in
the model, a phase transition happens for different combinations
of the two parameters.
\subsubsection{The model}
Grauwin and Jensen~\cite{Grauwin2012} investigated the 
addition of noise to a specific DW model, namely, 
the one in which the learning rate is $0.5$. 
Said differently, an interaction means consensus 
between the two interacting agents. To be more precise, 
the update rule is the same as that of the standard DW 
model, however, the confidence rule is ignored some 
of the time (determined by a probability), which is 
the chosen method for implementing noise. 
In Grauwin and Jensen~\cite{Grauwin2012}, two types of noise were added to the opinion dynamics:
\begin{itemize}
\item \textbf{Type 1} Interaction noise: two randomly chosen agents 
at a given time step $t$ will interact with probability 
\begin{equation}
p_{int} = \left[ 1 + exp(\frac{ -1 + |d_{ij}| / r }{\gamma})\right]^{-1}
\end{equation}
This noise acts as an integrating force, providing the opportunity 
for communities to come to consensus.

\item \textbf{Type 2} At any given time $t$, select a random 
agent and, with probability $\nu$, set its opinion to a random 
number  within the opinion space $\mathcal{O}=[0, 1]$. 
This noise acts as a disintegrating force, causing random behavior.
\end{itemize}

The probability of interaction, $p_{int}$, ensures that 
agents with opinion differences greater than the 
confidence radius $r$ have the chance to interact, 
which models rational choices in the real world. We 
do not just simply ignore all people who have opinions 
that differ from ours by a narrow margin. Note 
that a large $\gamma$ indicates that opinion difference is not very important.

The second type of noise, which is studied in 
the social sciences, is interesting, though to 
the best of our knowledge it has not been 
used before in opinion dynamics. 
The rationale for this type of noise is that some 
agents may die and new ones may be introduced 
to the system. Most current models consider the 
long-time behavior of agents as time goes to infinity, 
although individual human beings cannot exist in the 
system for that long, due to both death and the fact 
that we make and lose friends over the course of our lifetimes. 

In Grauwin and Jensen~\cite{Grauwin2012}, each community/cluster 
of agents is determined as follows: the agents are ordered 
by their opinions; if the difference between opinions of 
two neighboring agents is less than the confidence 
radius $r$, then they belong to the same group, 
chained together by the confidence radius, 
(in other words, if the difference between two 
consecutive agents' opinions is more than $r$, 
then the number of clusters goes up by 1 and the 
two agents are the borders between the two clusters).

The results of adding these two types of noise to the model are described below:
\begin{itemize}
\item Let $\nu=0$; under this condition, we 
add only the probability of interaction between 
agents of two distinct communities separated 
by a distance greater than $r$; in other words, 
$\gamma > 0$. In this case the system will end 
up in consensus because when two agents from 
different communities interact, their new opinions 
will be set to the average (since learning rate $\mu = 0.5$) 
and this causes the communities to drift from their current 
states toward each other. The relationship between the 
parameters of the model and the time needed for the 
communities to merge into one is given in \cite{Grauwin2012}.

\item Let $\gamma = 0$, so that interactions are 
governed by the standard DW update rule, and 
let $\nu > 0$. As expected, when $\nu$ is small, 
the system exhibits behavior similar to that of the 
DW model; likewise, when $\nu$ is large, chaos 
rules the system. In this type of scenario, one can 
define order parameters and apply statistical physics 
routines. The order parameter in~\cite{Grauwin2012} 
is given by 
\begin{equation}
\rho = S_{max} \:  \left (1 - 3\frac{1}{{2 \choose S_{max}}} \sum_{i, j \in g_{max}}|d_{ij}| \right )
\end{equation}
\noindent where $g_{max}$ is the largest cluster, 
and $S_{max}$ is the number of agents in $g_{max}$.

\item Finally, it has been observed through the use 
of simulations that for a range of the pair of 
parameters $(\gamma, \nu)$, lasting communities will form.
\end{itemize}
\subsubsection{The Noisy HK model}
In this section we consider an extension of the HK model with noise added to it.
In this section all agents share the same confidence radius $r$.

Su et al.~\cite{Su2017} have shown that if the noise strength
is smaller than $r/2$ then the noisy HK model will almost surely reach
\emph{quasi-consensus}. They also state that ``HK dynamics is known 
to explain the divergence of opinions. However, our results 
reveal that the fragmentation behavior of the HK model 
fails to exhibit robustness against arbitrary weak random 
noise and that the mechanism of opinion 
divergence requires further study''. By means of simulation
it has been shown that if the initial opinions of agents are identical 
and noise strength is larger than $r/2$, then agents will diverge at 
some point. If $r$ itself is too large, then consequently 
the noise will be too large as well, which 
causes fluctuations. Hence, the simulation
that derived the aforementioned divergence in the presence of
strong noise was conducted for $r=0.01$. 

\subsubsection{The model}
Recall that the HK model is given by the following:
\begin{equation}
o_i^{(t+1)} = \frac{1}{|N_i^{(t)}|} \sum_{j \in N_i^{(t)}} o_j^{(t)}
\end{equation}

Su et al.~\cite{Su2017} added noise to this equation to obtain the following:
\begin{equation} \label{eq:HKnoise}
o_i^{(t+1)} = \xi_i^{(t)} + \frac{1}{|N_i^{(t)}|} \sum_{j \in N_i^{(t)}} o_j^{(t)}
\end{equation}

Please note that $N_i^{(t)}$ is the set of 
agents whose opinions lie within the 
confidence interval of agent $i$ at time 
$t$, i.e., $| o_i^{(t)} - o_j^{(t)}| < r$ 
(here all individuals have the same 
symmetric confidence interval). 
The random noise $ \xi_i^{(t)}$ can 
violate the boundaries of opinion 
space, in which case we apply the clamp function to it:
\begin{equation}
clamp(x) = 
\begin{cases} 0 & \text{if $x < 0$,} \\
x &\text{if $0 \leq x \leq 1$,} \\
1 &\text{if $x > 1$.}
\end{cases}
\end{equation}
\noindent so, the update rule in the violating cases is:
\begin{equation}
o_i^{(t+1)}  = clamp\left(\xi_i^{(t)} + \frac{1}{|N_i^{(t)}|} \sum_{j \in N_i^{(t)}} o_j^{(t)}\right)
\end{equation}

We will now look at some definitions specific to this section 
and in the following subsections we will present 
experimental work related to noise as a disintegrating force.
\begin{definition}{}
Let the diameter of opinions associated with the 
graph $\mathcal{G}$ at a given time $t$, and the limit diameter be given, respectively, by:
\[  d_{\mathcal{G}}^{(t)} = \max_{i, j \in V}|o_i^{(t)} - o_j^{(t)}| \: \: \mathrm{and} \: \: 
d_{\mathcal{G}} = \lim_{t\to\infty}  d_{\mathcal{G}}^{(t)} \]
\end{definition}
\begin{definition}{}
Define quasi-consensus (as before $r$ is the confidence radius):
\begin{itemize}
\item The system will reach a state of quasi-consensus if $d_{\mathcal{G}} < r$
\item The system is said to almost surely reach quasi-consensus if $p(d_{\mathcal{G}} < r) = 1$.
\item The system will not reach a state of quasi-consensus if $p(d_{\mathcal{G}} < r) = 0$.
\item Let $t_{min} = \min \{ t \: | \:  d_{\mathcal{G}}^{(t')} \le r, \forall t' \ge t \}$.
If $p(t_{min} < \infty) = 1$, then the system will almost surely reach quasi-consensus 
in finite time.
\end{itemize}
\end{definition}

In the standard HK model, when consensus is
achieved all agents share the same opinion. 
In this modified version, once quasi-consensus 
is achieved the maximum difference of opinions 
cannot exceed $2 \delta := 2  \sup_{i, t} |\xi_i^{(t)}|$.

The following theorems assume the disintegrating
 force, i.e. noise, is randomly and independently 
 chosen. However, in the work of M{\"{a}}s~\cite{Mas2010, Mas2014}, 
 to be presented after this section, this force is 
 actually a function of the level of uniformity 
 among an agent's neighbors, the more similar 
 an agent is to its neighbors, the greater its willingness to be different.
\begin{theorem}\label{quasi-1}
Let $r \in (0, 1]$ and 
the noise $\xi_i^{(t)}$ be independent while 
satisfying $p(\xi_i^{(t)} \le \delta) = 1$, where 
$\delta \in (0, r/2]$, while also satisfying 
$p(\xi_i^{(t)}  \ge a) \ge p $ for some $a \in (0, \delta)$ 
and $p \in (0, 1)$. Then, for any initial state 
$\mathbf{o}^{(0)} \in \mathcal{O}^N$, the 
opinion dynamics given by Eq.~\ref{eq:HKnoise} 
will almost surely reach a state of quasi-consensus in 
finite time, and almost surely $d_{\mathcal{G}} \le 2\delta$.
\end{theorem}

\begin{theorem}\label{quasi-2}
Let $r \in (0, 1/3]$ and let the random 
noise $\xi_i^{(t)} $ have zero mean and 
be i.i.d. with $E[\xi_i^{(t)}] < \infty$, or 
independent with $\sup_{i, t} |\xi_i^{(t)}| < \infty$ almost surely. 
If there exists an $m>0$ such that 
$p(\xi_i^{(t)} > r/2) \ge m$ and $p(\xi_i^{(t)} < - r/2) \le m$, 
then, almost surely, the opinion dynamics given by 
Eq.~\ref{eq:HKnoise} cannot reach quasi-consensus.
\end{theorem}

The next theorem follows from the previous two:
\begin{theorem}\label{quasi-3}
Let the noises in the model be i.i.d with a mean of 
zero and be non-degenerate, and 
let $E[(\xi_i^{(t)})^2]$ be finite. Then the following statements hold:
\begin{itemize}
\item If $p(|\xi_i^{(t)}| \le r/2) = 1$, then 
almost surely the network will reach quasi-consensus in finite time.
\item If the confidence radius 
$r \le 1/3$ and if $p(\xi_i^{(t)} > r/2) > 0$ and $p(\xi_i^{(t)} < -r/2) > 0$, 
then almost surely the network cannot reach a 
state of quasi-consensus.
\end{itemize}
\end{theorem}

The above derivations indicate that when 
the amplitude of the noise is not too great, 
the addition of noise can help lead to consensus, which is an intuitive result. 
A small amount of noise may cause agent $i$ to
 jump into an area of the opinion space in 
which many other agents are present, and hence,
 the opinion of agent $i$ and the other agents 
 would be averaged for the next time step. 
 Next, we will focus on experimental studies 
 of the uniqueness tendency, i.e., disintegrating forces. 
 
\subsubsection{Related models}
 Interested readers can see Refs.~[\cite{Pineda2011,Pineda2013,Carro2013}]
for further examples of the use of noise in opinion dynamics. 
For example, Ref.~\cite{Pineda2013} studied the effect of noise 
on a modified version of the HK model in which updates 
are done for a random selection of agents, and the noise 
is different from what we saw previously. Furthermore, 
this case could be treated like temperature, and 
standard statistical and physics procedures 
(such as dis/order parameter measurements, computation of critical values, etc.) 
could be applied.

Quattrociocchi et al.~\cite{Quattrociocchi} studied a system 
in which there are two networks; a network of agents 
and a network of media. At a given time, agents could 
choose one from among $k$ different media to have 
interactions with. The agent-agent interactions as well 
as agent-media interactions follow the bounded confidence 
rules. Since the media compete to attract a larger audience, 
a good portion of agents can be influenced by a given media. 
Along the same line of reasoning, an interaction force other 
than the pairwise interactions between agents can exist, in 
which the entire network of agents is under the influence of 
a constant external force~\cite{Droz2019}. This external 
force can be thought of as media that does not change its 
opinion and is connected to everybody, or like a magnetic
 field that influences a system of ferromagnetic atoms or 
 particles. The update rule in Ref.~\cite{Droz2019} is given by 
\begin{equation} 
o_i^{(t+1)}  = o_i^{(t)} + \frac{1}{|N_{\bar i} + 1 | } \left[ \sum_{j \in N_{\bar i}} o_j^{(t)} w_{ij} + h \right]
\end{equation}
\noindent where $h$ is the external field constant. 
Quattrociocchi studied the effect of such an 
external force acting in favor of a minority 
community in a society with two (political) 
parties to see if the external force could help 
the minority group take power, while also looking 
at the impact of other parameters such as the 
relative initial population of the minority community.

\subsection{Managing Consensus in a Heterogenous (in confidence radius) DW}
Pineda and and Buend{\'{i}}a~\cite{Pineda2015}, studied the effect of media
on the system. 
This simulation-based paper considered two cases of
a heterogeneous (in confidence radius) DW model:
in the first scenario, agents are 
divided into two groups where each group shared an identical 
confidence radius; in the second
scenario, each agent had its 
own confidence radius. Since the paper
is simulation-based, we will 
highlight only the main conclusions 
drawn from the experiments. 
Let us start by presenting the general form
of the model. 

Let the fully connected graph $\mathcal{G}$ consist of $N$ agents,
and let the opinion of the external media be denoted by $o_M \in [0, 1]$. 
\begin{itemize}
\item At a given time an agent $i$ is chosen randomly.
\item The chosen agent $i$ interacts with the media 
with probability $p_M$ and interacts with another 
randomly chosen agent with probability $1 - p_M$ 
($p_M$ is the probability of agent-media interaction, 
and $1-p_M$ is probability of agent-agent interaction.) 
The larger the value of $p_M$, the greater the probability
 of interaction with the media, or alternatively, the higher/stronger 
 the \emph{media intensity}, a term first used in Pineda et al.~\cite{Pineda2015}.

\item If agent $i$ interacts with 
the media and if $| o_i^{(t)} - o_M| < r_i$
then the agent updates its opinion
according to $o_i^{(t+1)} = o_i^{(t)} + \mu_i(o_M - o_i^{(t)})$ 
although the opinion of the media does not change;
if the interaction is between 
two agents, then each agent 
updates its opinion if the opinion 
of the other agent is within its confidence radius,
i.e.,
\begin{equation*}
\begin{cases} 
\vspace{.1in}
\text{if } | o_i^{(t)} - o_j^{(t)}| < r_i \text{, then}& o_i^{(t+1)} = o_i^{(t)} + \mu_i(o_j^{(t)} - o_i^{(t)}) \\
\text{if } | o_i^{(t)} - o_j^{(t)}| < r_j \text{, then}& o_j^{(t+1)} = o_j^{(t)} + \mu_j (o_i^{(t)} - o_j^{(t)}) \\
\end{cases}
\end{equation*}
\end{itemize}

In the experiments of Ref.~\cite{Pineda2015} 
the opinion of the media
did not change as a result of 
interactions and is set to $o_M := 1$, with all agents 
share the same learning rate $\mu=0.5$. Let us look at 
the results for the two scenarios.

\subsubsection{Heterogenous system with two confidence radiuses}
In this subsection we look at a system where individuals 
are divided into two groups of equal size 
$N_1 = N_2 = N/2$, where agents in each group share
the same confidence radius, which is different from that of the 
other group, $r_1 \neq r_2$. 
Experiments were conducted
for two cases, the first with the media absent and the second with the
media present, and results were compared.
\begin{enumerate}
\item \textbf{No media} (${p_M=0}$) In the homogenous case
where all agents shared the same confidence interval, it is
observed that $r \approx 0.27$ is the critical point
of phase transition from polarization to consensus. The main conclusion
for the experiments of Pineda et al. in this section, for the heterogeneous
system with two confidence 
radiuses, was that a consensus may occur for 
confidence radiuses below $0.27$.
\item \textbf{With media} (${p_M > 0}$) In this case the experiments
suggested that the final 
state of system heavily depends
on the initial profile. 
Moreover, ``in most of the cases and 
provided that the confidence 
levels are not too large, the 
mass media is unable to form a majority around its opinion 
when the system is too 
homogeneous.'' That is,``too homogeneous'' in the sense that 
$r_1$ and $r_2$ are close to 
the diagonal line in the $r_1r_2-$plane. 
Another conclusion suggested by 
the simulations is that
the probability that the media 
will attract more than half of the agents
increases when the system is heterogeneous.
\end{enumerate}
\subsubsection{Heterogenous system with agent specific confidence radiuses}
Let us present the results for the system where 
agents' confidence radiuses 
are chosen randomly for each agent. More specifically
let $r_i = r_0 + \alpha sign(y_i) |y_i|^\beta$, where $r_0$ is a constant,
 $y_i$ is distributed in $[-1, 1]$, the parameter $\alpha \in [0, r_0]$ ``represents the
 range of heterogeneity'', and $\beta \in [0, 9.9]$ ``characterizes the width of distribution.''
If $\beta = 0$ then agents can have either $r_0 - \alpha$ or $r_0 + \alpha$
and if $\beta > 0$ then $r_i \in [r_0 - \alpha, r_0 + \alpha]$, the larger the $beta$
the tighter the interval is, i.e., the larger the $beta$ 
value the more the heterogeneity is reduced.
\begin{enumerate}
\item \textbf{No media} (${p_M=0}$)
An interesting result is that there are intermediate
values of $\beta$ for which 
the chances of obtaining consensus
in the vicinity of one the 
extreme points can be improved by tuning
the parameter $\alpha$.
\item \textbf{With media} (${p_M>0}$)
In this case the interesting 
result, which is also counter 
intuitive, is that when $p_M$ 
is too large the media fails to attract 
agents to its opinion.
In the experiments $r_0:=0.35$, 
and a large $p_M$ along with $r_0$
caused early fragmentation of 
the system which prevented the success
of the media. 
\end{enumerate}

For more details see Ref.~\cite{Pineda2015}. 
A variant of the HK model with different bounded 
confidence levels was considered in Ref.~\cite{Kou2012}, 
where it was shown that the number of opinion clusters
 increased with the number of individuals who had 
 a very low confidence radius. The effect of the 
 media was also studied in Refs.~[\cite{Quattrociocchi,Carletti2006}]
 under the bounded confidence assumption.

\begin{figure}[httb!]
  \centering
  \includegraphics[width=.5\linewidth]{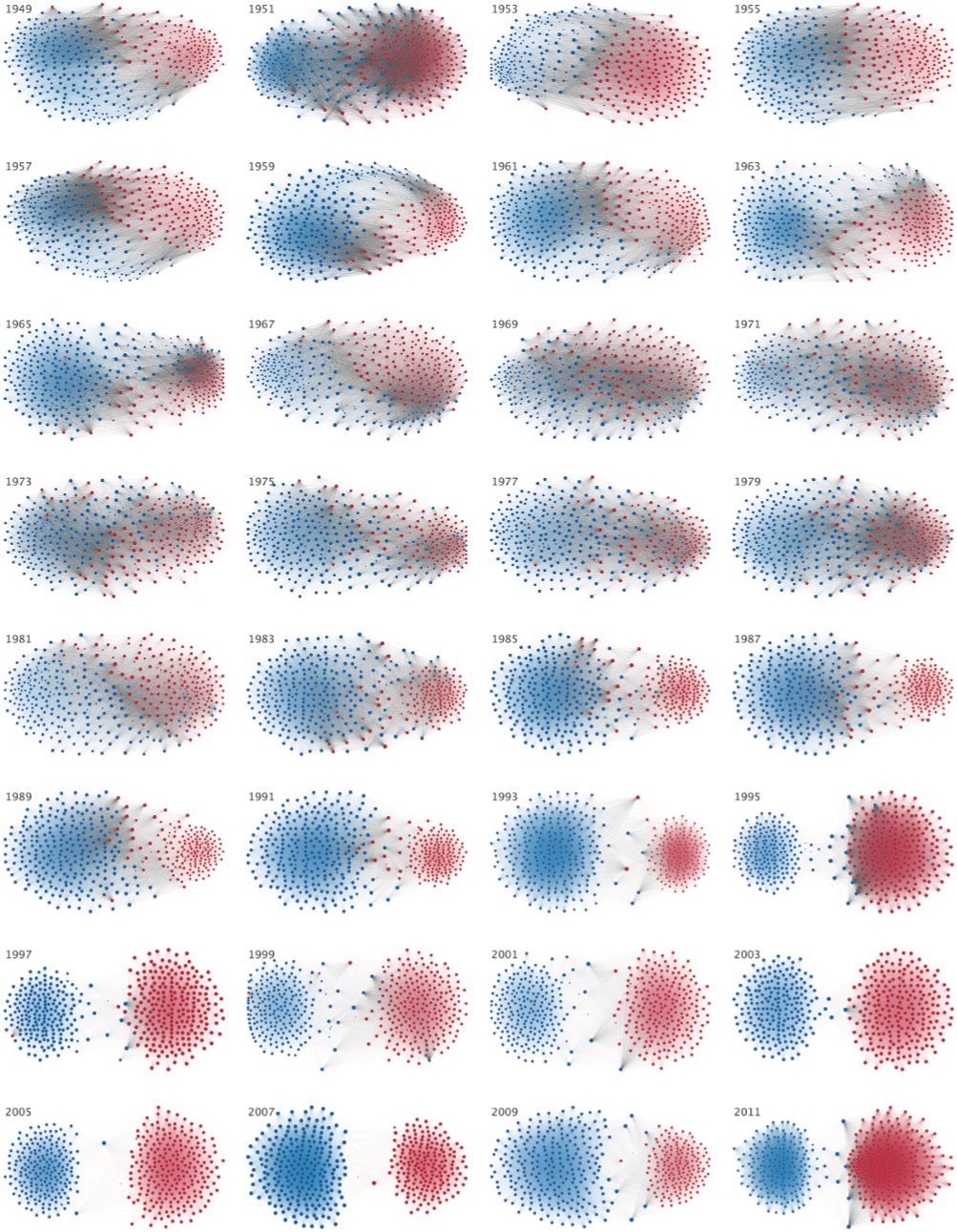}
\caption{Division of Democrat and Republican party 
members over time~\cite{Andris2015}. \small{``Each member 
of the U.S. House of Representatives from 1949-2012 is 
drawn as a single node. Republican (R) representatives are 
in red and Democrat (D) representatives are in blue; 
party affiliation changes are not reflected. Edges are 
drawn between members who agree above the 
Congress’ threshold value of votes. The threshold 
value is the number of agreements where any pair 
exhibiting this number of agreements is equally 
likely to be comprised of two members of the same 
party (e.g. D-D or R-R), or a cross-party pair 
(e.g. D-R). Each node is sized relative to its total 
number of connections; edges are thicker if the pair 
agrees on more votes. The starting year of each 2-year 
Congress is written above the network''}}
\label{fig:plosOneFig}
\end{figure}

Before introducing more references for this section, we would like
to point out, again, that consensus need not be reached
in a natural environment, even if 
agents are assumed to live forever. The Fig.~\ref{fig:plosOneFig}
is taken from the work of 
Andris et al.~\cite{Andris2015} and illustrates how the Democrat and Republican 
members of U.S. House of 
Representatives have been polarized over time.

\subsection{Mini discussion}
Although the original bounded confidence model 
was built using the assumption of homophily to 
explain social interactions, its sharp cut off point, 
which determines who may interact with whom, 
does not exist in real life. However, recent 
technological and cultural changes and the 
rapidly increasing speed of life have resulted in 
an environment in which vast quantities of information are 
sent and received via online media whose algorithms
 are optimized to learn from the actions of users 
 and to show them topics or ideas in which they 
 are already interested and with which they already agree. 
Hence, the news items, tweets, posts and likes 
that are seen by individuals will be overwhelmingly
 likely to support and agree with their current 
 opinions, and homophily is the side effect of such 
 algorithms. This idea is presented in Ref.~\cite{Sirbu2019}
  along with simulation results. In paper~\cite{Sirbu2019}, the 
  DW version of the bounded confidence model 
  (i.e., with pairwise interaction) is modified so 
  that the probability of interaction between two 
  agents, whose opinions lie within the confidence 
  radius, is dependent on the difference of their 
  opinions. The smaller the difference, the greater
   the probability of interaction between randomly 
   chosen pairs. In this model, after agent $i$ is chosen 
   randomly, the interaction partner is then chosen from 
   among agents $j$ for whom $|d_{ji}| = |o_j - o_i| < r$ 
   with probability $p_i(j) = \frac{|d_{ij}|^{-\gamma}}{\sum_{k \neq i} |d_{ik}|^{-\gamma}}$. 
   The parameter $\gamma$ tunes the degree of homophily or 
   bias. The effects of such modification are the following:
   \begin{itemize}
\item The system will end up with fragmented groups of agents, 
whereas in the original model, the same settings resulted in consensus.

\item The time needed to reach steady state or 
equilibrium is significantly increased.

\item The average distance between opinions is also increased.
\end{itemize}
An interesting idea is presented in Pilyugin and Campi~\cite{Pilyugin} 
 where the ``reinforcement theory'' of the social sciences meets 
 opinion dynamics. He used the update rule 
 $o_i^{(t+1)} =  o_i^{(t+1)} + \frac{\mu}{ |N_i|}  \sum_{j\in N_i } o_j^{(t)}$ 
 in the bounded confidence fashion in opinion 
 space $\mathcal{O} = [-1, 1]$. He remarked the fact that 
 since $\frac{\mu}{ |N_i|}  \sum_{j\in N_i } o_j^{(t)}$ 
 includes the agent $i$ itself, ``$i$ in absence of 
 counter-arguments, tends to strengthen her/his own 
 initial opinion'' and ``drift towards a higher level 
 of belief in the absence of opposite voices,'' which is in agreement 
 with reinforcement theory. Of course the update 
 rule can violate the boundaries. 
 (Imagine agent $i$ with no one in its confidence interval. 
 It will strengthen its own opinion.) So, the new opinions 
 will have to be clamped. To be precise, the update rule 
 actually is $o_i^{(t+1)} =  clamp(o_i^{(t+1)} + \frac{\mu}{ |N_i|}  \sum_{j\in N_i } o_j^{(t)})$. 

 The operator that maps $\mathbf{o^{(t)}} = [o_1^{(t)}, o_2^{(t)}, \cdots, o_N^{(t)}]$ to $\mathbf{o^{(t+1)}}$ 
 was considered in Ref.~\cite{Pilyugin}
  and it is shown that the operator's 
 basic fixed points are asymptotically stable 
 and the nonbasic fixed points are unstable. 
 Basic fixed points are of the form 
 $[-1, -1, \cdots, -1, 1, \cdots, 1]$. 
 For a discussion of nonbasic fixed points, 
 please see Ref.~\cite{Pilyugin}. It was also 
 shown that if the confidence radius is $r \le 0.5$ then the 
 trajectories of the operator above will go to 
 fixed points. Pilyugin viewed this 
 continuous dynamic model as a 
 tool for analyzing the voting process 
 in a system in which only one outcome, 
 either 1 or -1, is allowed; by the use
  of simulations, he arrived at the conclusion 
  that the outcome of elections can vary depending 
  on the level of interaction of society, i.e. 
  depending on the size of the confidence radius.

Another work that includes a novel modification 
of the DW model is Ref.~\cite{Huang2018}, in which 
the learning rate is a function of opinion differences
 in single interactions. Using this scenario. Reference~\cite{Huang2018} studied convergence 
 time and the probability of consensus via Monte Carlo simulations.

We would like to flashback to the question posed in 
the Sec.~\ref{intro} about the limitations that prevent 
people from interacting synchronously and introduce 
two works that address this limitation.
Perra and Rocha~\cite{Perra2019} 
states, ``We are bounded by cognitive and temporal 
constraints. Our attention is limited.'' Because of these limitations, 
which are ignored in synchronous models, our attention 
is valuable. Commercial companies or political parties 
must compete to gain and hold our attention. 
Social medias, such as Facebook, are utilized to 
obtain the attention of users for various reasons, 
such as to increase revenue or to influence voting 
decisions. Therefore, the information revealed to 
each user is filtered by machine learning algorithms 
to increase the likelihood of the desired outcome.
Perra and Rocha~\cite{Perra2019} makes use of three types 
of filtering; random filtering, time ordering and 
the accumulation of past ideas, likes, etc., 
expressed by the user‚ to study the filtering 
effect on opinion dynamics. 
Perra demonstrated that the distribution of 
opinions is affected by the filtering algorithm, 
especially when the filtering is based on the 
past behaviors of users. The simulations were 
then applied to three different networks; a random network, 
a Watts-Strogatz network and a 2D lattice‚ and it 
was also shown that if social media sends a 
particular opinion to a fraction of users regularly, 
such messages can affect their behavior and users 
can be manipulated toward a given opinion. 
This result is similar to what we have already 
seen in other cases for fully-stubborn 
agents.

The fact that synchronous interactions are 
unrealistic was also noted by 
Patterson and Bamieh~\cite{Patterson}. 
They modified the DeGroot model, 
$\mathbf{O}^{(t+1)} = \mathbf{A O}^{(t)}$, 
to consider the interaction frequency 
between agents in the following way:
\begin{equation}\label{eq:PattersonEQ}
\mathbf{O}^{(t+1)} = \left(\mathbf{I} -\sum_{(i, j) \in E} \delta_{ij}^{(t)} \mathbf{A}_{ij} \mathbf{L}_{ij} \right)\mathbf{O}^{(t)}
\end{equation}

Please note that $\mathbf{A}_{ij}$ 
are scalars, (i.e., entries located at the $(i, j)$ 
position of the weight matrix $\mathbf{A}$), 
and $\mathbf{L}_{ij}$ is a matrix associated 
with the edge $(i, j) \in E$ of the graph 
$\mathcal{G}$. More precisely, 
the matrix $\mathbf{L}_{ij}$ is 
associated with the subgraph $\mathcal{G}_{ij}$, which 
only has the edge $(i, j)$ in it and it is referred to as 
the \emph{weighted Laplacian matrix of the 
$\mathcal{G}_{ij}$}, and one can write 
$\mathbf{A} = \mathbf{I} - \sum_{(i, j) \in E} \mathbf{L}_{ij}$. 
Furthermore, the $\delta_{ij}^{(t)}$'s are independent random 
variables taking a value of 1 with probability $p_{ij}$ 
and a value of 0 with probability $1-p_{ij}$. The interaction 
frequency is captured by a probability of communication.
Furthermore, the consensus conditions and efficiency 
(i.e., convergence rate) were analyzed along with 
the network modification to improve the efficiency of
both the classical DeGroot model and the altered 
DeGroot model described above.

\subsection{Extensions and related models}
Before moving on to the next section we list some 
interesting bounded confidence models.

A recent bounded confidence model is 
found in Ref.~\cite{Nguyen2019}, in which opinions lie in $\mathbf{o_i} \in \mathbb{R}^d$. 
The interactions are pairwise; a random agent $i$, selected from 
a fully connected graph, chooses a random neighbor, 
and if $||\mathbf{o_i}^{(t)} - \mathbf{o_j}^ {(t)}||_2 < r $, 
then both agents will converge to the mean of their opinions in 
each dimension, 
i.e., $\mathbf{o_i}^{(t+1)} = \mathbf{o_j}^{(t+1)} = \frac{1}{2} (\mathbf{o_i}^{(t)} + \mathbf{o_j}^{(t)})$, 
i.e., the learning rate is $0.5$. Therefore, if the matrix of 
opinions of all $N$ agents is given by 
$\mathbf{O} \in \mathbb{R}^{N \times d}$, where each row 
represents an agent, then at each time step 
two rows will become identical. Let the graph be 
a fully connected one, and define 
$\mathbf{W}(i, j, t) = \mathbf{I} - \frac{1}{2} (\mathbf{e_i} - \mathbf{e_j}) (\mathbf{e_i} - \mathbf{e_j})^T$, 
which is a matrix with $0.5$ at positions $(i, i), \: (i, j), \: (j, i), \: (j, j)$ 
and zeros elsewhere, 
so that $\mathbf{O}^{(t+1)} = \mathbf{W}(i, j, t) \mathbf{O}^{(t)}$ 
defines the update after the interaction between $i$ and $j$. 
Then the equilibrium point of the system, $\mathbf{O^*}$, 
satisfies $\mathbf{O^*} = \mathbf{W}(i, j, t) \mathbf{O^*}$
 for any $t \ge T^*$. For the system mentioned above at 
 equilibrium, any two agents $i$ and $j$ will either be at 
 consensus or separated by a distance greater than the 
confidence radius $r$:
\begin{center} 
$ \mathbf{o_i}^* = \mathbf{o_j}^* $ \hspace{.2in} or  \hspace{.2in}  $|| \mathbf{o_i}^* - \mathbf{o_j}^* ||_2 \ge r $ 
\end{center}

It has also been shown that the bounded confidence 
mentioned above, with learning rate $0.5$, for any 
initial opinion in $\mathbb{R}^d$, will almost surely 
reach such an equilibrium state. This intuitive result 
is also obtained in Ref.~\cite{LORENZ2005217} 
and Refs.~\cite{Fortunato2005,bennaim2003,Chen2019} 
for different versions of bounded confidence models.

In the above system the confidence radius is fixed. Now 
consider an iterative version of the system, whereby the 
game is repeated as follows: 1. initialize $\mathbf{O}^{(0)}$, 
and run the game until the system reaches its equilibrium; 2. 
at this point, use the equilibrium state as the initial state 
of the next round, with the confidence radius increased 
by $\Delta r$, so that for the second round we have 
$r := r + \Delta r$. The iterations can be repeated until 
a predefined maximum confidence $r_{max}$ is reached. 
This approach allows the number of clusters to be reduced 
as the confidence radius is increased.

Two interesting ideas that are related to HK-type dynamics were introduced 
in this model in~\cite{Chazelle2017}. For the first novel idea, let the opinion 
space be $\mathbb{R}^k$, let agents $i$ and $j$ communicate whenever 
$|| o_i - o_j||_2 \le 1$, and define the update rule by 
$o_i^{(t+1)} = (1- \lambda_i ) o_i^{(t)} + \frac{\lambda_i}{|N_{ i}|} \sum_{j \in N_{i}} o_j^{(t)}$, 
where $\lambda_i \in [0, 1]$ 
is called \emph{inertial} 
(note that self-loops exist in the above equation). 
It has been shown that such a system converges asymptotically 
when $\lambda_i \in \{0, 1\}$, and in one dimension the convergence 
is exponentially fast. The second novel idea involves an \emph{anchored HK system}. 
In this system each agent is identified by $ z_i = (o_i^{(t)}, y_i)$, 
where $o_i^{(t)} \in \mathbb{R}^k$ is the moving part of agent $i$ and 
$y_i \in \mathbb{R}^{\hat k}$ is the fixed part. Two agents interact 
whenever $ || z_i^{(t)} -  z_j^{(t)}||_2 \le r$, for some value of $r$. 
The fixed part plays a role only in the communication graph and 
determines whether $i$ and $j$ are neighbors at any given time. 
The authors of ~\cite{Chazelle2017} show that an anchored system 
like this has a symmetric heterogeneous HK equivalent. 
(A symmetric heterogeneous HK system is a system in which each 
edge is assigned its own confidence radius; in other words, 
agents $i$ and $j$ each have their own confidence radius 
when they interact with each other.)
We would also like to mention that in almost all 
models/papers, the equilibrium of the system when 
the interactions stop is determined as if agents have global knowledge of the 
system. Xie et al.~\cite{Xie2018} built on the work of 
others~\cite{Yadav2007DistributedPF, Manitara} 
and considers a local stopping criteria in a system 
where agents are only aware of their neighbors. 
In most of the literature, the dynamics and the simulations 
stop with a knowledge of the global state of the network, or 
after a large number of steps have been taken. 
Such results are questionable when modeling actual 
systems with real-life agents that need not know the 
global state of the network they live in. In addition, 
the asymptotic state of the network also depends on topology 
and the initial state of the system, and computing the 
needed number of iterations may be hard, impossible 
or unrealistic for time varying networks.

\section{Further Work and Further Questions}\label{FurtherWorkQuestions}
In this section we provide more models worthy of notice and end the
paper with questions that are not studied yet.
\subsection{Other works}\label{sec:other}
In the following 
subsections we include newly developed models,
influential works that are not variation of DeGroot or bounded-confidence models
and provide pointers so more works of researchers.


Let us start with models supporting the repulsive behaviors 
that exist in real life--whether in human interactions,
the interactions of gas molecules or birds flying 
together. 

Noorazar et al.~\cite{Noorazar2016} introduced a rich and
flexible opinion dynamic model inspired by energy functionals
from physics. In this model two interacting agents have an energy
between them that is a function of their opinion difference.
The update rule is given by
\begin{equation} \label{eq:updateNoWeights}
\left\{
\begin{array}{lr}
    o_i^{(t+1)} &= o_i^{(t)}  -\frac{\mu}{2} \: \psi' ( |d_{ij}^{(t)}| ) \: \frac{d_{ij}^{(t)}}{|d_{ij}^{(t)}|}	\vspace{.1in}\\
    o_j^{(t+1)} &= o_j^{(t)}  +\frac{\mu}{2}  \psi'  ( |d_{ij}^{(t)}| ) \: \frac{d_{ij}^{(t)}}{|d_{ij}^{(t)}|}
  \end{array}
\right.
\end{equation}
\noindent where $\psi$ is the energy function and $\mu$ is the learning rate. 
In this model, potential functions can be agent specific.
Assigning potential functions to edges (i.e., defining 
interaction-partner-dependent potential functions) will 
make the system highly complicated and flexible. 

\begin{figure}[httb!]
\hspace{.5in}
\begin{subfigure}{.3\textwidth}
  \centering
  \includegraphics[width=1\linewidth]{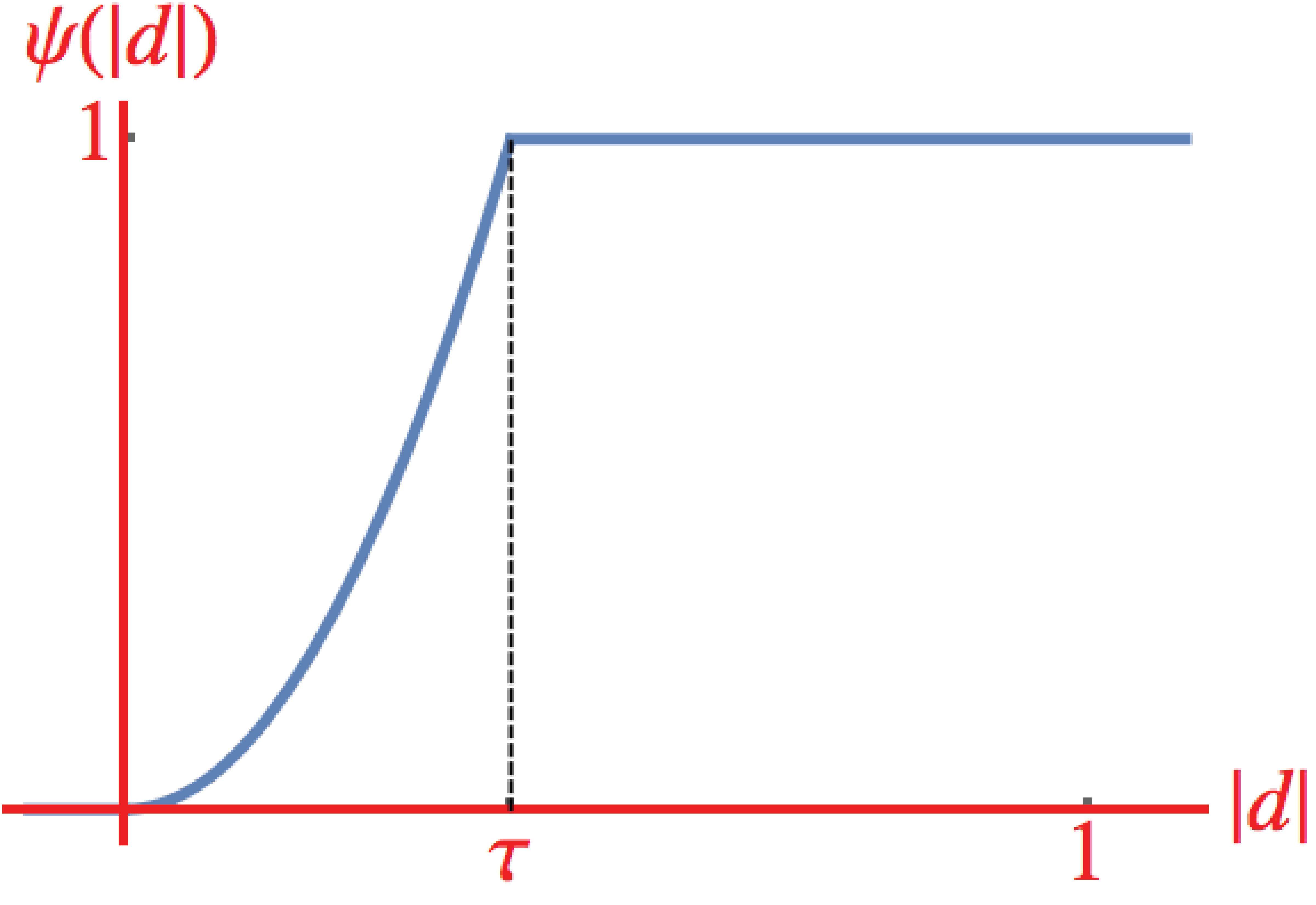}
  \caption{BCM potential function} 
   \label{fig:BCMpoten}
\end{subfigure}
\hspace{1in}
\begin{subfigure}{.3\textwidth}
  \centering
  \includegraphics[width=1\linewidth]{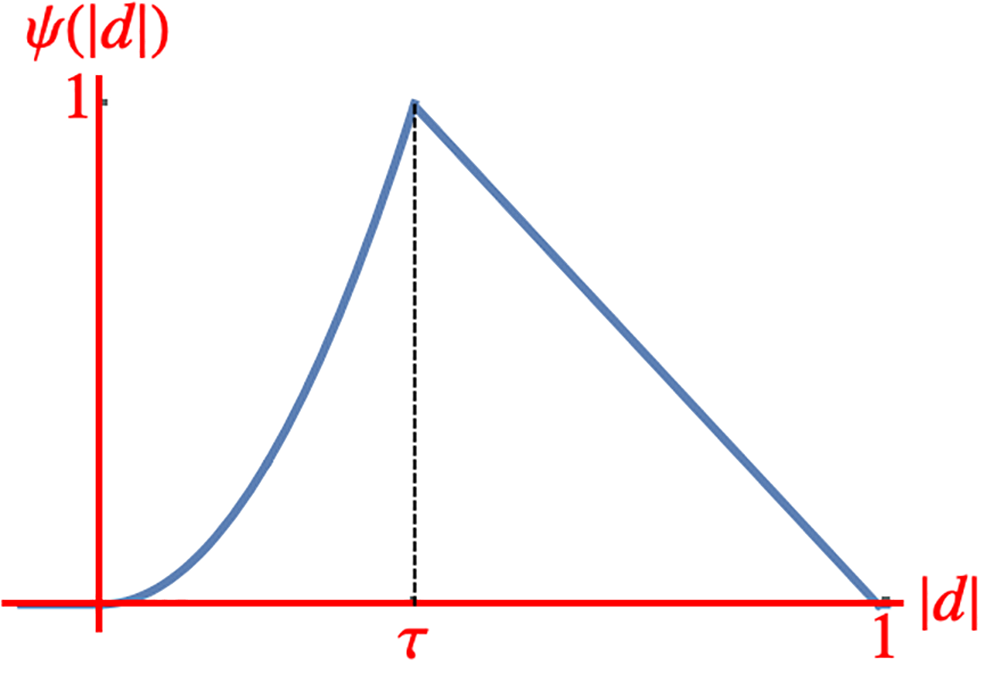}
  \caption{BCM repulsive}
  \label{fig:BCMrepulse}
\end{subfigure}
\caption{BCM potential function and an alternative. \small{Using~\ref{fig:BCMpoten} as 
the potential function in Ref.~\cite{Noorazar2016} 
induces the bounded confidence model and 
using~\ref{fig:BCMrepulse} will cause the 
regions of indifference in BC models be replaced with repulsive behavior.}}
\label{fig:BCMPotentials}
\end{figure}

It has been noticed that, for example, the DW model 
is a special case of this model if the $\psi$
is given by Fig.~\ref{fig:BCMpoten}, in 
which case we have $\psi(x) = x^2$ when $x<0.5$, 
and $\psi'(x)=0$ beyond that. By changing the flat 
part of Fig.~\ref{fig:BCMpoten} to a decreasing 
function, as shown in Fig.~\ref{fig:BCMrepulse},
we can easily obtain repulsive behavior. We also note that:
One can also take the BCM potential function, 
shown in Fig.~\ref{fig:BCMpoten}, and smooth 
out the function at $\tau$ so that it is differentiable, 
thus solving the problem of the sharp 
transition between acceptance and rejection in the BC model.

Two other simple potential functions are 
given below in Fig.~\ref{fig:example-potential}. 
The tent potential function, Fig.~\ref{fig:tentfig1}, 
supports both attraction and repulsion. Whether attraction 
or repulsion occurs depends on the ‚ opinion difference‚ 
which can be thought of as a time-varying topology 
that is not arbitrarily random, as it was in 
Ref.~\cite{Proskurnikov2016}. 
The attraction or repulsion, i.e., friendship and antagonism, 
at any given time is governed by the potential function, 
but the randomness of the relationship between a pair of agents 
is due to random pairwise interactions 
that have led the two agents to their current positions. 
Finally, Fig.~\ref{fig:skewtentfig1}, is the flat top tent potential 
function, which gives an agent the option of behaving 
in one of three modes; attraction, indifference and 
repulsion, making the model even richer and more flexible.
\begin{figure}[httb!]
\hspace{.5in}
\begin{subfigure}{.3\textwidth}
  \centering
  \includegraphics[width=1\linewidth]{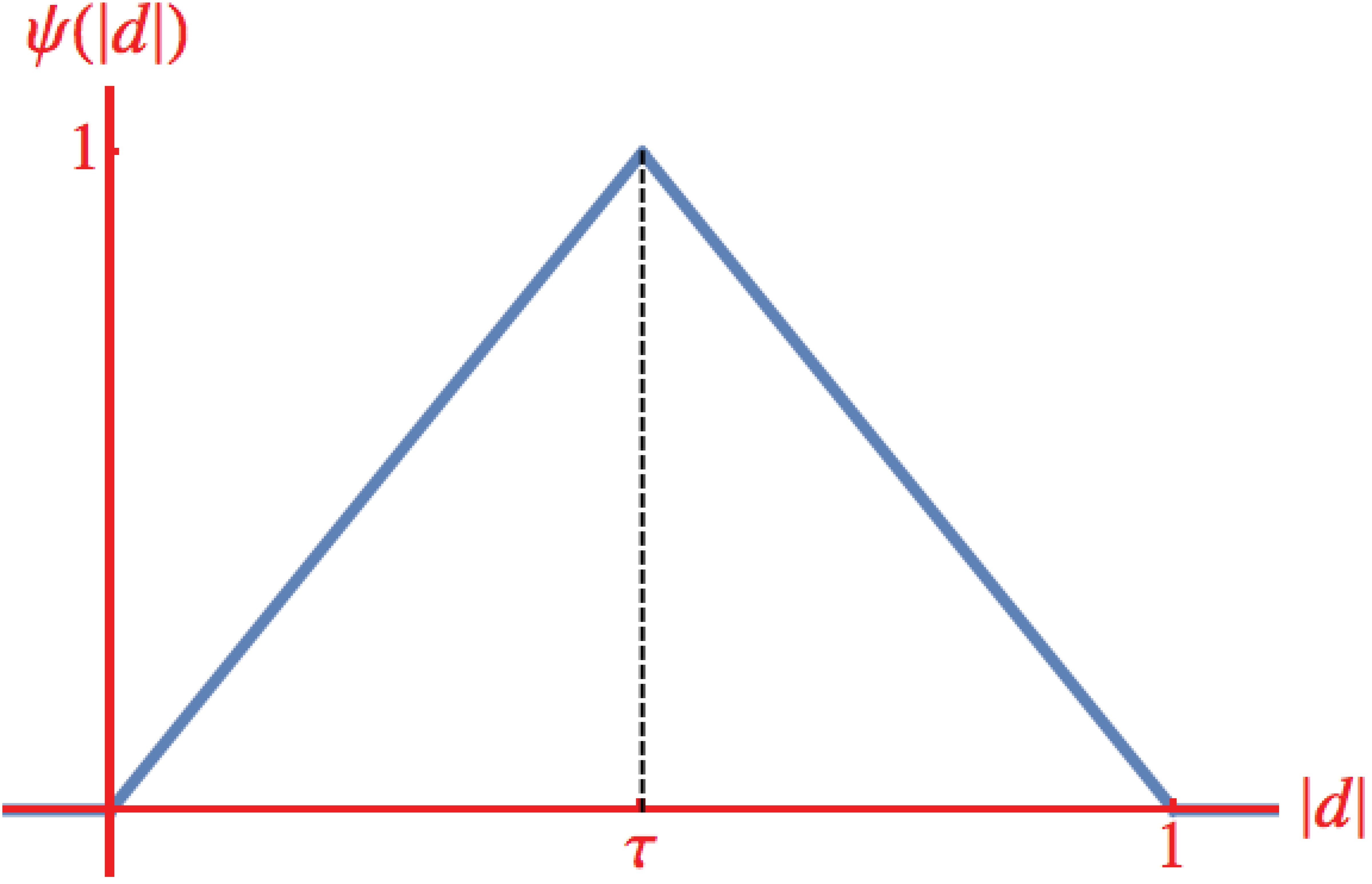}
  \caption{tent function} 
   \label{fig:tentfig1}
\end{subfigure}
\hspace{1in}
\begin{subfigure}{.3\textwidth}
  \centering
  \includegraphics[width=1\linewidth]{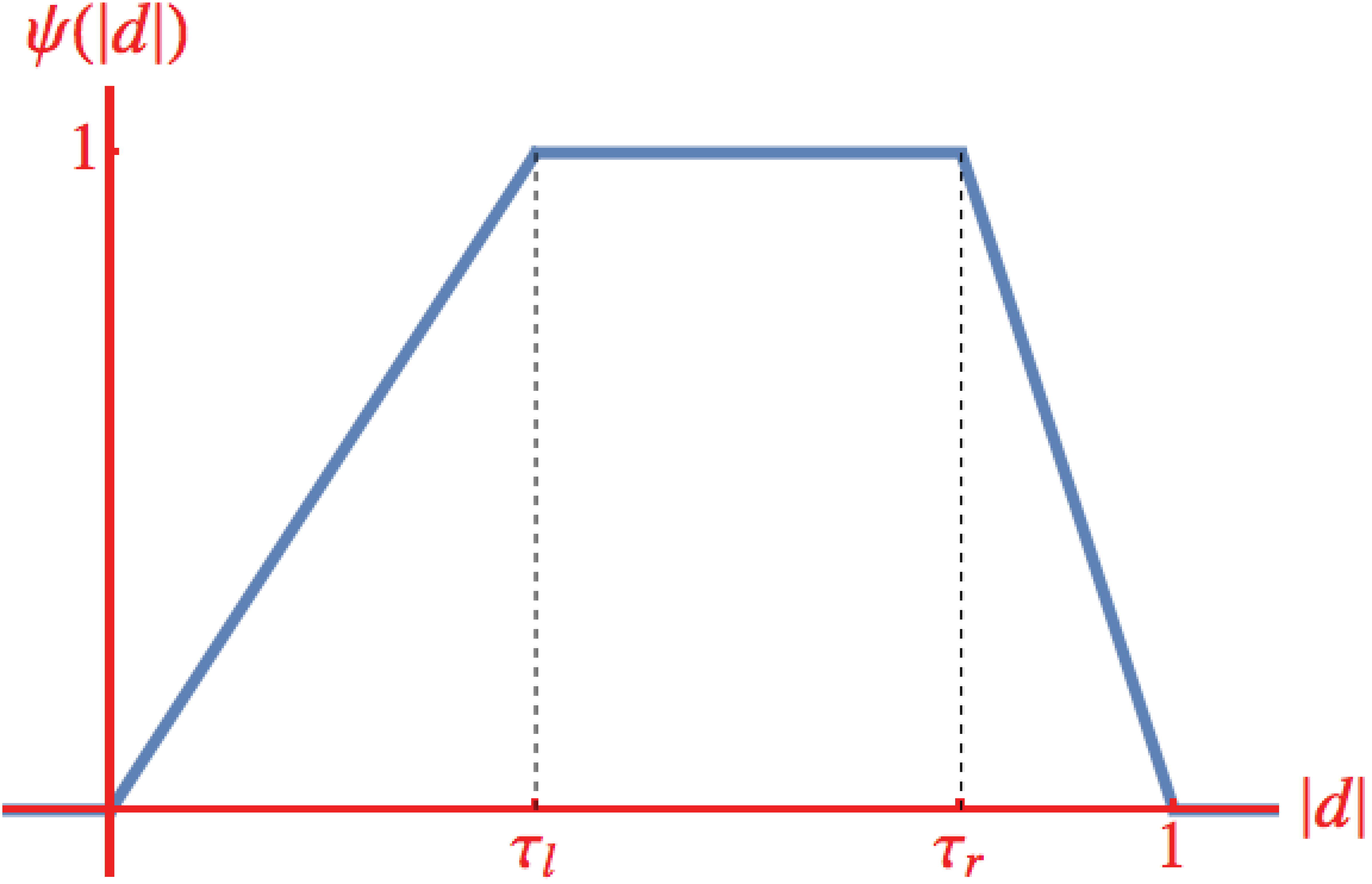}
  \caption{skewed flat top tent}
  \label{fig:skewtentfig1}
\end{subfigure}
\caption{Potential function examples}
\label{fig:example-potential}
\end{figure}
Besides discussing repulsive behavior, Ref.~\cite{Noorazar2016} 
also covers the modeling of interrelated topics in both discrete and continuous \emph{topic} space.


The second class of models that support repulsive behavior is based
on the idea of balanced networks. Let us start with the definition of structurally balanced networks.
\begin{definition}{}
Consider a fully-connected network of 3 agents
where each edge between them is assigned a relationship 
status of either \emph{friend (+)} or \emph{adversary (-)} (See Fig.~\ref{fig:figbalance}.)
The 3-agent fully-connected network is said to be structurally balanced if it has either 
1 or 3 positive signs (the product of the 3 signs is positive in these cases). 
A network with more than $N=3$ agents is said to be structurally balanced
if every fully connected 3-node subgraph in it is structurally balanced.
\end{definition}

An immediate consequence of the above definition 
is that a structurally balanced network can be divided into
two subgroups where the relationships within subgroups
are friendly and the relationships between subgroups (any two agents
from different subgroups) are adversarial. Moreover, it is assumed that
any pair of connected agents are aware of their relationship, and,
their relationship is either friendly or adversarial.
\begin{figure}[httb!]
\begin{subfigure}{.5\textwidth}
  \centering
  \includegraphics[width=.5\linewidth]{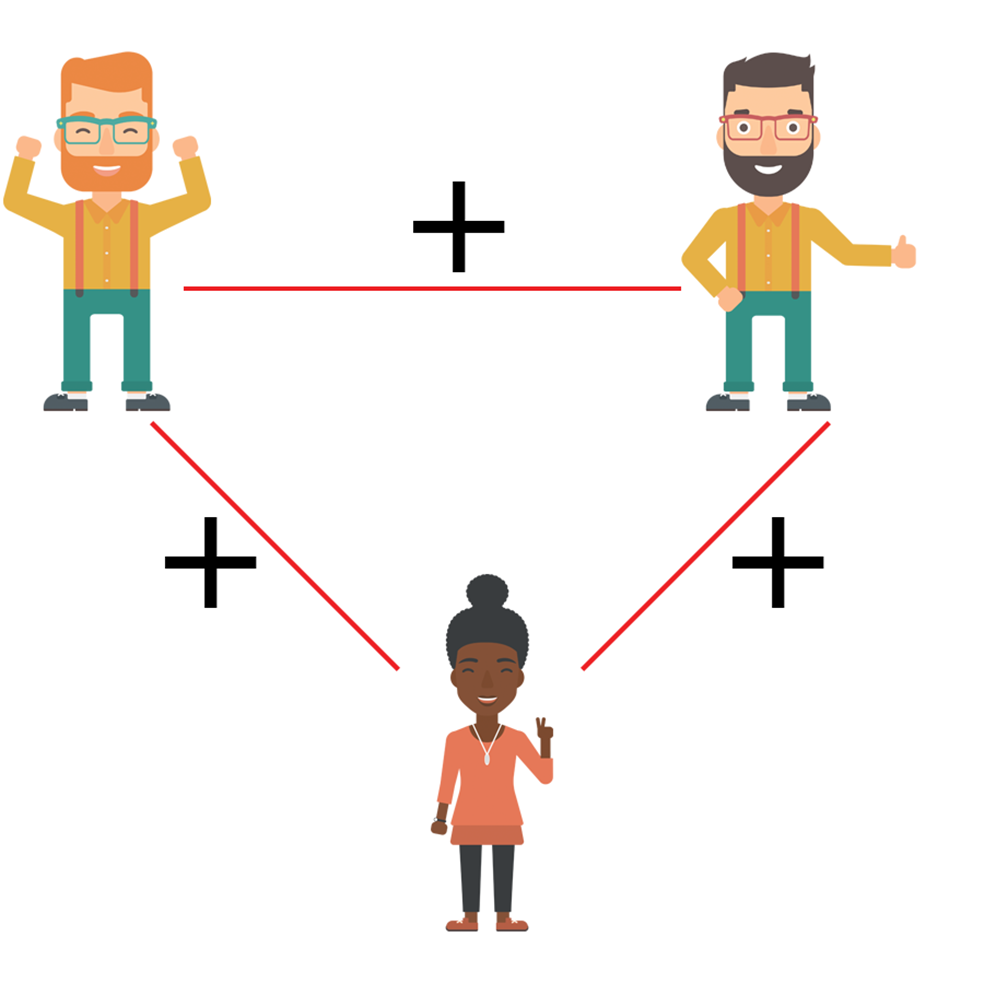}  
  \caption{Everyone is friendly: balanced}
  \label{fig:balanced1}
\end{subfigure}
\begin{subfigure}{.5\textwidth}
  \centering
  \includegraphics[width=.5\linewidth]{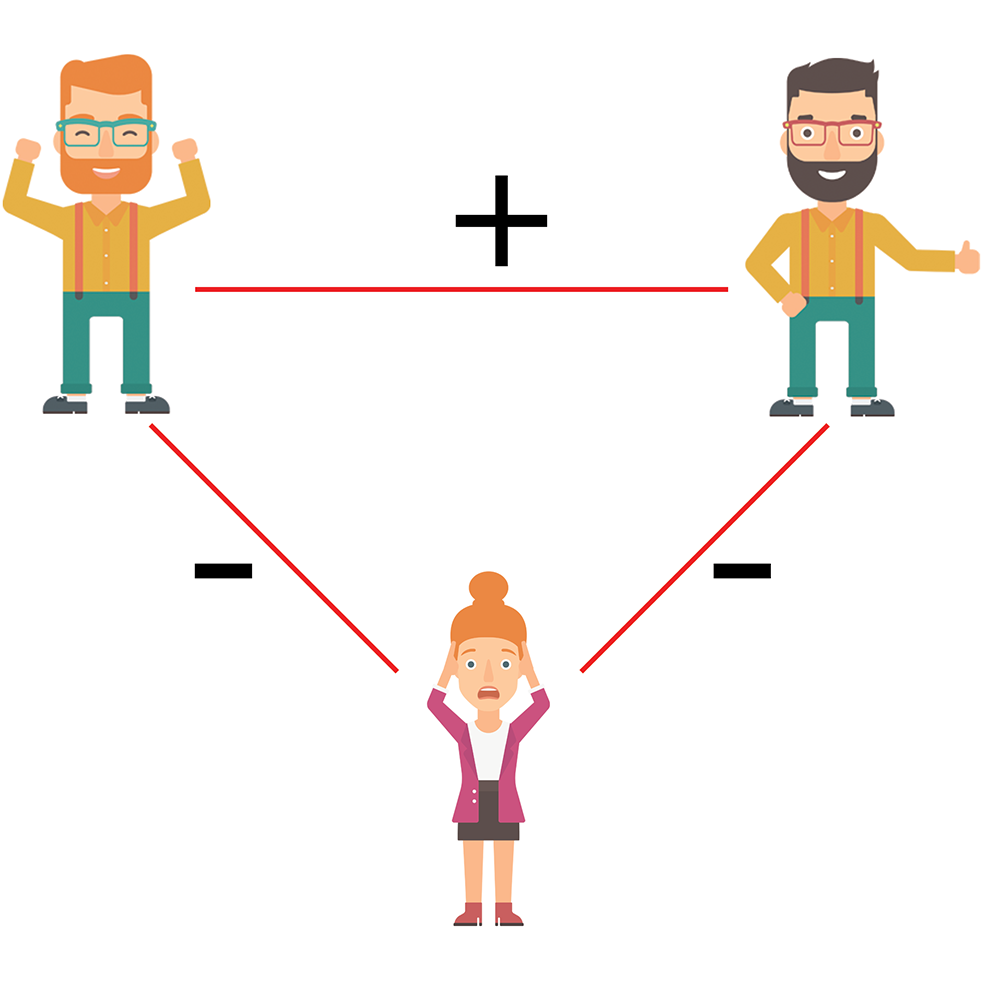}  
  \caption{Two friends have a mutual adversary: balanced}
  \label{fig:balanced2}
\end{subfigure}

\begin{subfigure}{.5\textwidth}
  \centering
  \includegraphics[width=.5\linewidth]{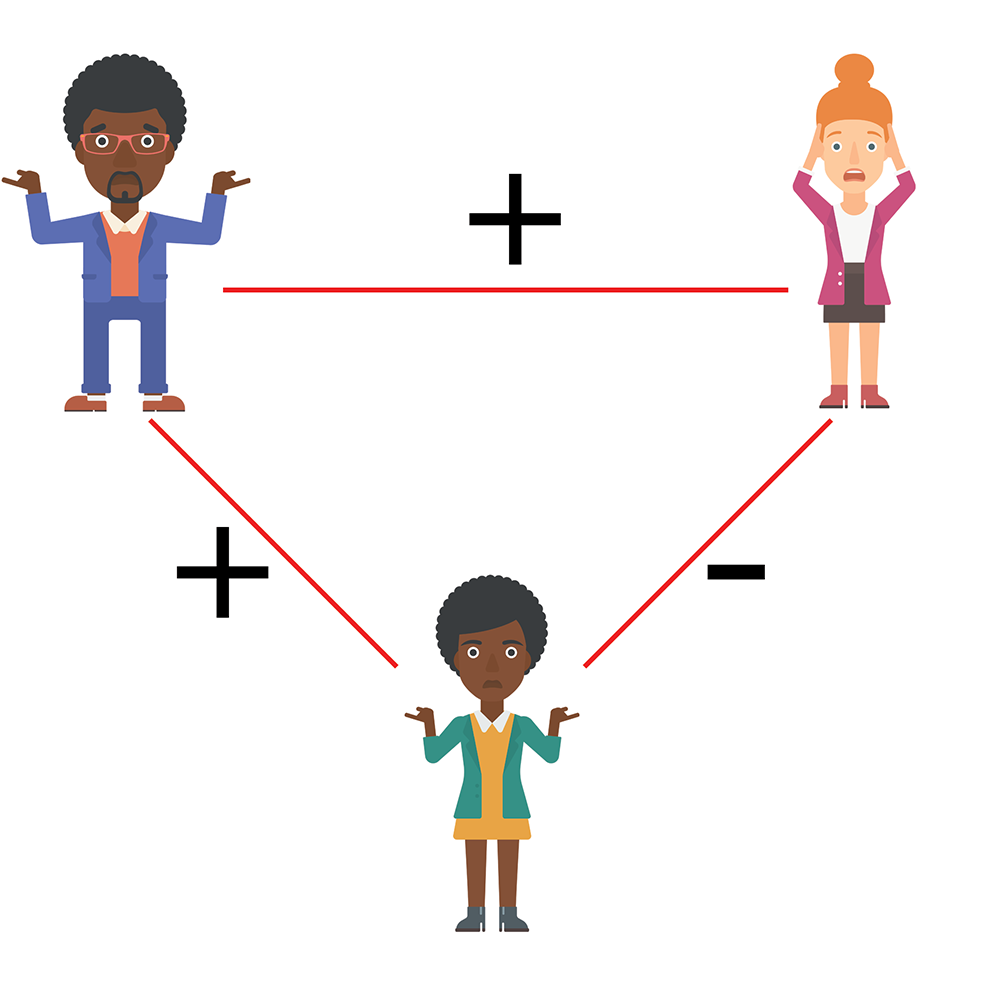}  
  \caption{The antagonistic relationship may effect the friendship relationships: imbalanced}
  \label{fig:imbalanced1}
\end{subfigure}
\begin{subfigure}{.5\textwidth}
  \centering
  \includegraphics[width=.5\linewidth]{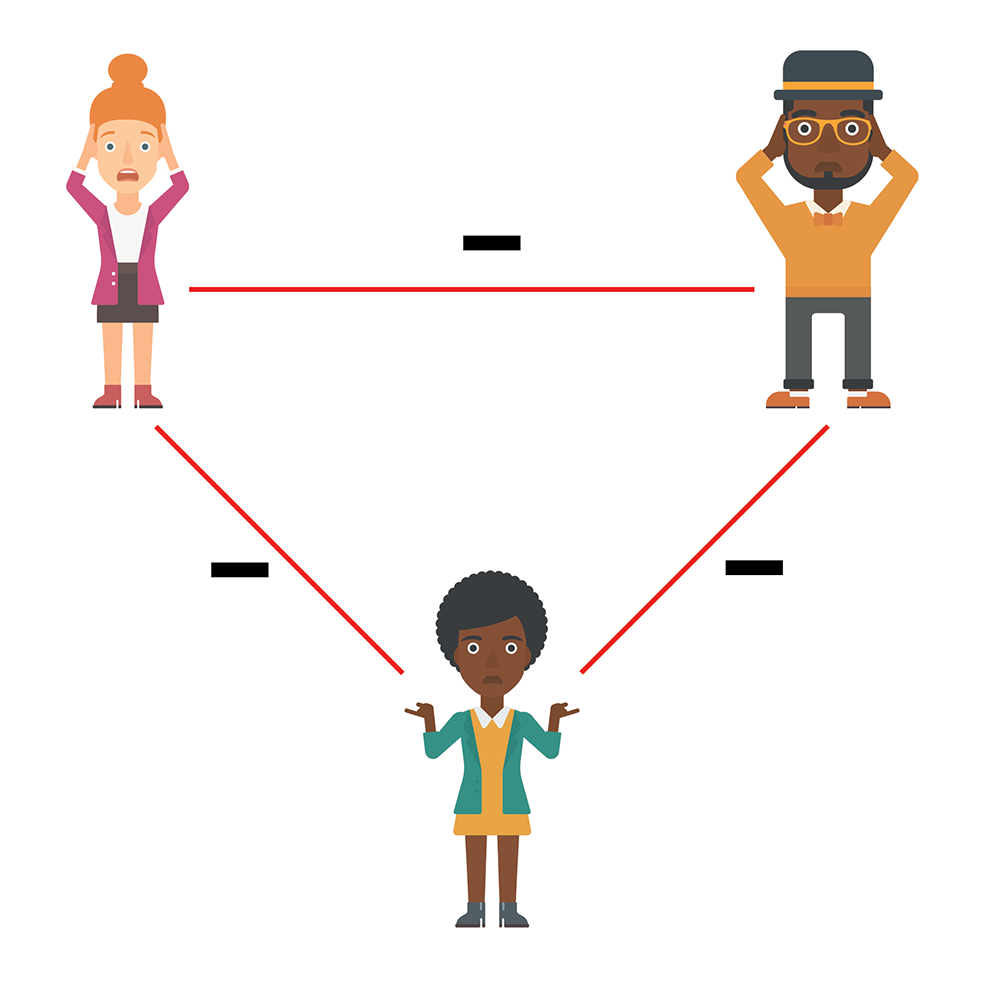}  
  \caption{All enemies: imbalanced}
  \label{fig:imbalanced2}
\end{subfigure}
\caption{Structural balance among three agents. Balance 
requires either 1 or 3 friendship relationships, 
otherwise the structure is imbalanced. A graph with more 
than $N=3$ agents is balanced if all of its fully connected 3-agent subgraphs 
are balanced.}
\label{fig:figBalande}
\end{figure}

Altafini~\cite{Altafini2012} used theories 
presented in Refs.~[\cite{Cartwright1956,wasserman1994}] 
to model a structurally balanced community with a 
dynamical system approach; namely, a monotone 
dynamical system, with its well-known properties, 
was used to model this type of social community. A 
structurally balanced network is a community that is 
divided into two antagonistic sub-communities, and 
agents within a given sub-community are friendly/cooperative 
and have a positive influence on each other, whereas 
agents from different sub-communities are adversaries/antagonistic. In 
this model the relationships among agents remain 
constant‚ they are either on good terms or they are 
adversaries. Relationships are also independent of 
opinions or opinion differences; it is therefore possible 
for two agents with widely different opinions to attract each other, and agents 
with close opinions need not attract each other. The nature 
of these relationships is expressed via constant signs on the 
edges: a positive sign indicates friendship and a 
negative sign, antagonism. Friends, who by 
definition belong to the same sub-community, are 
connected to each other by edges with positive weights 
assigned to them, and inter-community edges are 
assigned negative weights to model antagonism. 

Clearly, the structure of the graph makes the final 
state of the system predictable: polarization is inevitable 
unless all members of a given 
sub-population decide to join the other party.

The edge weights that define the relationships between
 pairs of agents are translated into directional derivatives 
 of the functional defining the dynamics of the system. 
See Ref.~\cite{Altafini2012} for more details. Later 
 the model is extended to explain how it is possible 
 for agents in such a system to reach opinions of 
 equal size but opposite sign~\cite{Altafini2013}. 
 In the works mentioned so far in this section, 
 opinions belong to $\mathcal{O} = \mathbb{R}$. 
 Later, Altifani~\cite{Altafini2013} studied the consensus
  problem using this model and Proskurnikov et al.~\cite{Proskurnikov2016} 
  investigated the use of an arbitrary time-varying signed graph with this model.

Proskurnikov et al.~\cite{Proskurnikov2016} built on Altafini's 
model using a network that is not static; specifically it is
not required to be structurally balanced at the beginning. 
Such a system can reach a structurally balanced state 
at some time $t > 0$. For further reading on the origins 
of evolving networks and the conditions for reaching a 
structurally balanced topology, please see Ref.~\cite{Proskurnikov2016}.

Altafini and Ceragioli~\cite{Altafini2018} extended his work to include 
the case of repulsive behavior as well. 
In what follows, we briefly explain his recent 
work. Consider an opinion dynamic with opinion 
space $\mathcal{O} = \mathbb{R}$ or $\mathcal{O} = [-1, 1]$, or 
any opinion space, really, which contains a point of 
neutrality (in the cases mentioned above, zero
 would be the neutrality point). Positive opinions 
 would be the strength of agreement on a given 
 topic and negative ones would be the strength 
 of opposition to it. Altafini and Ceragioli~\cite{Altafini2018} 
 argued that agents with opinions near zero might 
 fall within each other’s confidence radiuses. 
 However, moving from one side of zero to the 
 other is not easy in the real world, and does not 
 happen frequently. Hence, they added three 
 separate components to his earlier work; incorporated 
 into the bounded confidence model, these additions 
 resulted in three new models. In
 the first model (which is similar to his earlier work) 
 if agents $i$ and $j$ have close opinions that differ in 
 sign, they will be attracted to the opposite of 
 the opinion of the other. In the second model, 
 agents with opinions of opposite signs ignore 
 each other, and in the third, agents whose 
 opinions are close enough and of opposite sign, 
 repel each other. Therefore, opinions in 
 these three models always retain their initial sign 
 during time evolution, and the models are
 referred to as \emph{signed bounded confidence models}. 
 These results were motivated by and compared 
 with the results of the ordinary bounded 
 confidence model with continuous time 
 (the derivation of this model is shown in 
 interesting works referenced in Ref.~\cite{Altafini2018}. 
 Zhang and Chen~\cite{Zhang2014Control} built on 
 earlier work of their own~\cite{Zhang2011Control} 
 along with Altafini's~\cite{Altafini2013} work 
 and designed an \emph{``output feedback control law''} 
 to study consensus in networks that included antagonistic 
 interactions in conjunction with strongly connected graphs, 
 spanning tree graphs and graphs with spanning trees. 
 His work provided results for such scenarios when 
 the network is structurally balanced or unbalanced.

Some recent interesting work related to antagonism 
includes the following: Yang and Song~\cite{Yang2019} 
mapped social networks onto electrical networks, 
with each interaction node between agents being 
mapped to a link in an electrical network and with 
the resistance (or rather conductivity) of each link
 representing the interaction/influence weight of the 
 agents on each other. She used the effective 
 conductance (EC) concept to measure the direct and 
 indirect relationships of a given pair of agents whose 
 interactions are defined by a DeGroot-type update rule: 
$o_i^{(t+1)} = o_i^{(t)} + \lambda \sum_{j \in N_{\bar i}} w_{ij} (o_j^{(t)} - o_i^{(t)})$.
In the update rule it is assumed that 
$w_{ij} = w_{ji} \in \mathbb{R}$. Hence, this 
model also assumed antagonistic interactions 
with no assumption on the structure of the graph, 
and a positive EC that would indicate, overall, the 
direct and indirect nature of the interactions 
between a pair of agents. The consensus criteria for 
this model were considered as well; for more details
 please see \cite{Yang2019} and references 
 therein. Meng et al.~\cite{Meng2019} considered an 
 antagonistic dynamic on a network with agents 
 that are separated into two groups, and with a topology that 
 switched between $M$ finite digraphs. 
 Meng studied the behavior of such 
 systems by lifting restrictions such as the 
 structural balance we saw in previous 
 models. For further discussion please see Ref.~\cite{Meng2019}.
 
 Antal et al.~\cite{Antal2005,Antal2006} considers a network
with two types of relationship, friendly and antagonistic, in which
relationships change in order to turn imbalanced triads into balanced ones.
Refs.~\cite{Antal2005,Antal2006} are not about opinion dynamics, 
but perhaps this line of work can be utilized with opinion dynamics to
explore novel questions.\\




To the best of our knowledge, Ref.~\cite{Mas2010} 
is the first to introduce the individuality tendency to 
the literature. When the individuality 
tendency is added to the model of Durkheim, the update rule becomes:
\begin{equation}\label{eq:masUpdateRule}
o_i^{(t+1)} = o_i^{(t)} + \xi_i^{(t)} + \frac{\sum_{j \neq i} ( o_j^{(t)} - o_i^{(t)}) w_{ij}^{(t)}}{\sum_{j \neq i} w_{ij}^{(t)}}
\end{equation}
\noindent where the influence weights are adaptive 
and are functions of the opinion difference between two given agents:
\begin{equation} 
w_{ij}^{(t)} = e^{-\frac{|d_{ij}^{(t)}|}{\gamma}}
\end{equation}
\noindent where $d_{ij}^{(t)} = o_i^{(t)} - o_j^{(t)}$. 
The parameter $\gamma$ specifies the level of 
confidence each agent has in its own opinion. 
Small $\gamma$ values imply high confidence in 
the current opinion and that agents are influenced 
mostly by those with whom they hold similar opinions. 
The \emph{adaptive} noise is drawn from a normal 
distribution with zero mean and its variance is 
given by ($\xi_i^{(t)} \sim N(0, \sigma_i^{(t)}) $), where
\begin{equation}\sigma_i^{(t)} = s \sum_j e^{-|d_{ij}^{(t)}|} \end{equation}

The parameter $s$ is used to manipulate the 
uniqueness tendency in the model, and the 
variance $\sigma_i^{(t)}$ is larger for agents 
who find themselves similar to a greater number of other agents.

The M{\"{a}}s~\cite{Mas2010} model is motivated by the Durkheim theory of social 
interactions~\cite{Durkheim}, in which individuals conform to 
society's norms while also tending to be unique and different, 
which fosters the co-existence of many different opinions. 
The homophily assumption of confidence models is missing 
here, as an agent is influenced by all other agents. There is 
also no repulsion or negative influence. And please note that 
in Eq.~\ref{eq:masUpdateRule}, agent $i$ is the only one 
updating its opinion, that is, in the simulations a random 
agent is chosen to update its opinion. 
Such updates are neither pairwise nor synchronous. In the 
M{\"{a}}s~\cite{Mas2010} paper, simulations start from a 
state of total consensus in which all agents hold the exact 
same opinion and in the 2-dimensional space of integrating 
and disintegrating parameters, $( \gamma, s)$, there is an 
extensive area for which different meta-stable clusters
 co-exist, while in the other areas various types of 
 quasi-consensus or total chaos is observed.
 
The M{\"{a}}s's results were obtained from a simulation 
on a fully connected network to explain clustering, i.e., to 
explain the lack of complete consensus at the end of long 
simulation runs. The motivation was the need to explain the
 co-existence of different opinions in a network where all 
 agents may interact with each other, as opposed to networks 
 that contain artificial constraints on the topology of the network 
 through disconnections or loosely connected subgraphs~\cite{Hegselmann2002}. 
 M{\"{a}}s’s modification more closely approximates connectivity in 
 the real world, which now includes the internet and social media.\\
 
We mentioned before, humans do not possess a sharp decision boundary beyond 
which they ignore others, as seen in the DW model. Baccelli et al.~\cite{Baccelli2017}
 tried to address this issue by incorporating the effect of probabilistic opinion 
 exchange between agents; in other words, they attempted to smooth out the 
 sharp transition between interacting with and ignoring others in the DW model. 
 The model includes a random internal thought for each agent. The random internal
thought has an expanding effect.
 Their modifications included allowing agents to ignore those whose opinions 
 are close to their own as well as to learn from those who did not think like 
 them. The update rule for these modifications is given as follows:
\begin{equation}\label{eq:stochasticDW}
\begin{split}
\begin{cases}
o_i^{(t+1)} = 
\begin{cases} o_i^{(t)} +  \xi_i^{(t)} + w_{ij} ( o_j^{(t)} - o_i^{(t)} )& \text{if $U_{i,j}^{(t)} = 1$,} \\
o_i^{(t)} +  \xi_i^{(t)} &\text{o.w.} \\
\end{cases},&\\
\\
o_j^{(t+1)} = 
\begin{cases} o_j^{(t)} +  \xi_j^{(t)} + w_{ji} ( o_i^{(t)} - o_j^{(t)} )& \text{if $U_{j, i}^{(t)} = 1$,} \\
o_j^{(t)} +  \xi_j^{(t)} &\text{o.w.} \\
\end{cases}&\\
\\
\hspace*{.1cm} o_k^{(t+1)} = o_k^{(t)} +  \xi_k^{(t)}, k \not \in \{i, j\}
\end{cases}
\end{split}
\end{equation}
\noindent where we must note that:
\begin{itemize}
\item $U_{i,j}^{(t)} \in \{0, 1\} $ is a random variable 
indicating whether agent $i$ is influenced by 
agent $j$ or not; it is a function of their opinion 
difference. So, when agent $i$ is not influenced 
by agent $j$, agent $j$ is simply ignored. 
This is where stochasticity comes in: 
\begin{equation}
p(U_{i,j}^{(t)} = 1) = f_{i,j}(|o_j^{(t)} - o_i^{(t)} |) : \mathbb{R}^+ \rightarrow [0, 0.5)
\end{equation}
\item $\xi_k^{(t)}$ is the noise (``endogenous belief or bias'').
\item $w_{ij} \in (0, 0.5]$ is the influence 
weight of agent $j$ on agent $i$.
\item All agents subject to Eq.~\ref{eq:stochasticDW}, 
regardless of whether they are participating 
in an interaction or not, will have internal 
thoughts, i.e., noise that is added to all agents at any given time $t$.
\item The opinion space in the Eq.~\eqref{eq:stochasticDW} is $\mathcal{O} = \mathbb{R}$.
However, in Ref.~\cite{Baccelli2017} the model is restricted to $\mathcal{O} = \mathbb{Z}$,
and consequently the update rules given by Eq.~\eqref{eq:stochasticDW} are
modified by a rounding method like ceiling or flooring or else.
\end{itemize}

Baccelli et al.~\cite{Baccelli2017} included a \emph{stability} 
or \emph{weak consensus} definition in their paper that 
will not be covered here, due to lack of space. (The idea is that
 ``if all agents move to infinity while remaining 
 close to each other, the society is stable''.)
 However, their work showed that it is sufficient to have an agent in 
 the network with high influence on all other agents to 
 reach a weak consensus. Similarly, it is possible to 
 reach a weak consensus if there is a pathway of 
 strong influences from one agent to another.


Managing consensus or preventing a community from reaching a consensus 
has attracted the attention of researchers.
A recent paper~\cite{Bauso2018} with a basis in control 
theory employs a novel idea to managing 
and controlling consensus. The opinions in this paper are 
in $\mathbb{R}^n$, and the update rule is given by 
$o_i^{(t+1)} = u_i^{(t)} + \sum_{j=1}^N w_{ij} o_j^{(t)} $, 
where $u_i^{(t)}$ is the \emph{pay-off of a repeated game}
 between agent $i$ and an external source. The second term, 
like any other averaging scheme, made sure the convex hull 
of states shrank as time passed, or as the authors put it, 
the ``space averaging process reduces the total squared 
distance.'' The conditions under which such a system 
converges to a predefined set (which very well may 
consist of only one element) based on the iteration 
of games with vector pay-offs was addressed in 
Ref.~\cite{Bauso2018} with a level of detail that is beyond the scope of this paper.\\

Another relatively recent work~\cite{Dietrich2018} 
focused on opinion dynamics systems controlled 
by a fully-stubborn agent. This paper assumed the influence 
weights depended on both the current state of the 
agents and time. In this work fully-stubborn agents influence 
other agents with an influence function but are not influenced 
in return. 
Unlike previous bounded confidence models, which 
assume the influence stops when the opinion 
distance is greater than some threshold, the 
model in this paper included ``other shapes for the 
dependency between influence 
strength and opinion distance.'' It also generalized 
the time-invariant dynamics associated 
with fully-stubborn agents to a time-dependent dynamic model. 
The model converged because it required agents to 
gather around the fully-stubborn agent in finite time, and the fully-stubborn then influenced the other agents to 
adopt the target consensus value. 
Another work with a novel idea uses one \emph{strategic-agent} to maximize the number 
of agents that fall within a given interval is Ref.~\cite{Hegselmann2015}. 
 Tools such as those presented 
in these papers can be utilized to target retail consumers 
or voters in political elections 
by both spreading misinformation and combating it. \\

In addition to the analytically tractable DeGroot and bounded
confidence models we looked at in detail in this paper, there are
simulation driven approachs to understanding models~\cite{Biccari2019,
  Rajpal, Wang2017, Liang2016, Pan2016, Afshar2010}.  
This allows for more realistic communication patterns (e.g., pairwise
interactions) as well as heterogeneous agent behaviors.  Agent-based
and Monte Carlo simulations are commonly employed in this case.
Agents could have different influence on each
other~\cite{Friedkin2011}, or different topics may be
interrelated~\cite{Noorazar2016, Friedkin2016}.  In order to introduce
a specific kind of agents' internal thoughts, the tendency for
individuality was introduced by~\cite{Mas2010} in which agents try to
be different when uniformity in a group increases, and later this idea
was applied with agents' having a memory~\cite{Noorazar2017}, where
their uniqueness tends to be more in the opposite direction of their
community's movement direction.
 
There are also analytical approaches for studying opinion formation
that are based on well-known Boltzmann-type equations of dilute
gases~\cite{During2009, Helbing1993, Boudin2009, Biswas2011, Chowdhury2011, Pareschi2017}. 
The advantage of this approach is
that well established methods from statistical physics can be used to
study the evolution of densities/distribution of agents/opinion in
regions of opinion space in the time limit.  However, some aspects of
physical systems are removed when applied to agents' interactions.
Some of the simplifications, for example, are: interaction among
agents could happen with very few agents whereas for molecules of
gases to interact they have to have high density, or, in opinion
dynamics, limitations are imposed around opinion space
boundaries~\cite{toscani2006kinetic} whereas gas molecules interact
also with the walls of the environment they live in. The application
of integro-differential equations of Boltzmann type are not limited to
opinion dynamics, they are also applied to other areas such as
economics and wealth distribution for which a gentle introduction
source is~\cite{pareschi2013interacting}.  A relatively recent work in
this area~\cite{alexanian2018anti} shows how nonlinear dynamics of
diffusion and anti-diffusion can create clusters, where formation of
clusters and their attributes are one of the challenges of the field
as of today.\\

Consensus formation has of course attracted quite a bit of attention
due to its practical impact.  Given the invention of the Internet,
social media and such, the connectivity among people has increased and
we observe polarizations frequently.  Hence, consensus might not be
the most interesting limit state that systems may reach after all.

If a decision is to be made by a group of people, the majority vote
models are more suitable where consensus may not occur, but a decision
is made.  Research about group decision making has also recently
received attention~\cite{Dong2018aa, Wu2019}, with focus on
consensus~\cite{Dong2018}, decision making on the
web~\cite{Morente-Molinera2018}, and with presence of non-cooperative
agents~\cite{Zhang2018}, which the aformentioned subject is not
studied much.  Please take a look at Ref.~\cite{Li2019} for a recent survey
of research about decision making.\\

In some models~\cite{Zhang2013, Altafini2012, Altafini2013, Proskurnikov2016, Li2017, Noorazar2017} 
agents do not simply ignore other
agents with opinions that are too far from their own. 
In some works~\cite{Ren2005, Cao2008, Li2017, Nedic2017} evolving
networks are considered and
in others~\cite{MirTabatabaei} evolving influences. Another interesting
update rule, in Ref.~\cite{Li2017}, is based on the difference of opinion
of a given agent and all of its connection, however, the update takes
place, unlike bounded confidence model, when the difference is
\emph{larger} than a threshold $\delta > 0$, due to social
pressure. It is unlike most developed models, and it may intuitively
seem agents will move towards consensus, however, in fact the model
allows existence of a spectrum of opinions,
$(o - \frac{\delta}{2}, o + \frac{\delta }{2})$.\\

A major contributor to the field is Galam whose work goes back
to 1982~\cite{galam1982} and includes binary opinion dynamics, 
group decision making and more. For example, the idea of \emph{inflexible} stubborn agents
in binary opinion dynamics was introduced in Ref.~\cite{Galaminflexible}.
We refer the reader to a survey of his works~\cite{GalamReview}
and his book~\cite{GalamBook} to learn more about opinion dynamics from
a sociophysics perspective. 

Galam~\cite{Galam2017} and Biswas et al.~\cite{Biswas2017}
both attempted, with different approaches, to explain Trump's victory in the US presidential election.\\

While in this work we are focused on real-valued, continuous
opinion space, there are binary version of the model~\cite{Ding2010, Biswas2009}
for which opinions are restricted to boolean values. 
Some of these boolean models are inspired by spin
systems (e.g., the Ising model) commonly studied in statistical
physics~\cite{Herrero2004, Sznajd2000, Dietrich2003}. 
The Ising model is a
simplified model of ferromagnetism in which spins
are restricted to two orientations and
are influenced by their neighbors on a grid. 
In between continuous and boolean valued opinions are
models in which are discrete but
more than two choices are available, for example a recent 
model is suggested by Bolzern et al.~\cite{Bolzern2019219} which
is a Markov chain model. Another model proposed by Martins~\cite{Martins2007}
is defined for a continuous opinion space but agents take discrete actions.

\subsection{New Questions}

We have only covered part of the work in the area of opinion dynamics,
but even if all of the work is considered, much 
opportunity remains for exploration and creativity. Here are a few questions and directions
that seem interesting that, to our knowledge, are mostly open at
this time.

\begin{enumerate}
\item There is no work that we are aware of in which higher order
  interactions are considered, where, for example, collections or
  \emph{cliques} of $k$ people ($k > 2$) are modeled as
  interacting in a way that is more than just $\frac{k(k-1)}{2}$ simultaneous
  pairwise interactions, even though some interactions are not
  reducible to a bunch of pairwise interactions.
  
\item Little work has been done on the coupling of topics to each
  other, even though there is the work on topic sequences. A
  reasonable approach might consider the topics themselves to form a
  graph or network, with nodes that are topics and edges and cliques
  that represent interactions. One would then have an opinion space
  for each agent that is a network of topics. The interaction dynamics between 
  the topic networks for individual agents are mediated by agent-to-agent 
  interactions. While this is much more complicated, it is also much
  more realistic and still much simpler than the true reality we are
  trying to model and understand. 

\item There is little work we are aware of that explicitly tries to
  model the effects of time-to-think and the evolution of opinions
  outside of an interaction. This is related to the effect that deep
  reading has on people who take the time to think 
  (see Ref.~\cite{Baccelli2017} for work that considers opinion-influencing 
  noise occurring outside of interactions, 
  which can be thought of as the effects of internal thoughts. 
  We briefly mentioned it in Sec.~\ref{sec:other} where the update rule is given by Eq.~\ref{eq:stochasticDW})

\item The evolution of the connections in the network of agents has
  been considered, but as far as we know, it has not included
  geophysical movements, normal life cycle changes, and the economics
  of connections. There is also the issue of trust evolution due to
  socially/emotionally impactful interactions. While some of these
  events would be tightly coupled in time with changes in opinions,
  others would be hard to tie to immediate opinion changes.
\item Consensus in continuous opinion dynamics is usually studied as
  time goes to infinity, and humans (or their social groups) do not
  live forever. Is there a better way to define equilibrium given the
  finite number of interactions that are physically possible in a
  fixed time period or lifespan?
\end{enumerate}


\bibliographystyle{unsrt}
\bibliography{survey_refs}

\end{document}